# A Probabilistic Chemical Programmable Computer

Abhishek Sharma, Marcus Tze-Kiat Ng, Juan Manuel Parrilla Gutierrez, Yibin Jiang and Leroy Cronin*

School of Chemistry, The University of Glasgow, University Avenue, Glasgow G12 8QQ, UK,

*Corresponding author email: Lee.Cronin@glasgow.ac.uk

**Abstract:** The exponential growth of the power of modern digital computers is based upon the miniaturisation of vast nanoscale arrays of electronic switches, but this will be eventually constrained by fabrication limits and power dissipation. Chemical processes have the potential to scale beyond these limits performing computations through chemical reactions, yet the lack of well-defined programmability limits their scalability and performance. We present a hybrid digitally programmable chemical array as a probabilistic computational machine that uses chemical oscillators partitioned in interconnected cells as a computational substrate. This hybrid architecture performs efficient computation by distributing between chemical and digital domains together with error correction. The efficiency is gained by combining digital with probabilistic chemical logic based on nearest neighbour interactions and hysteresis effects. We demonstrated the implementation of one- and two- dimensional Chemical Cellular Automata and solutions to combinatorial optimization problems.

**One sentence Summary:** A digital-chemical probabilistic computational machine that can solve combinatorial problems and instantiate a physical programmable cellular automaton has been demonstrated.

**Main text:** The exponential increase in computing power has been driven by the vast growth of transistors on silicon chips *(1)*. This growth was made possible by developments in fabrication technology reducing the feature sizes of the transistors (*2*), but this paradigm is currently approaching the limits imposed by physics as quantum effects become more pronounced (*3*). To overcome these limitations, various novel computational architectures based on physical and chemical processes have been proposed (*4–8*). Quantum computers show the potential to solve problems that are intractable on classic computing machines but currently suffer from scalability issues due to error correction (*9*). Alternative computing substrates and unconventional computation paradigms are being developed based on mapping computational logic to various physical phenomena (*10*). These tend to emulate transistor-based logic gates and other circuit components into the physical domain using architectures based on Boolean circuits (*11*) or discovered using artificial intelligence (*12*). Other classic computational architectures which utilize the true nature of physical phenomena include reaction-diffusion (*13*) and neuromorphic computers (*14*). These architectures and their algorithms have been designed to solve a specific set of abstract mathematical problems (*15–18*). The challenge is to develop new platforms that can take advantage of chemical substrates but can be easily programmable.

Herein, we present a probabilistic computational machine based on a chemical computational architecture that is digitally addressable (*20*). This device is built from a programmable chemical array-based (*21*) around a new type of chemical state machine for information processing. The chemical-computational array operates using the Belousov–Zhabotinsky (BZ) reaction which is an oscillating chemical reaction (*22*, *23*). The cells are arranged in a rectangular array and are individually programmable with a central oscillator driver and has four tuneable gates that allow the oscillations to be coupled between adjacent cells in the grid. By programming cell and interfacial controls, the emerging chemical oscillations of the i$^{th}$ cell (with j$^{th}$ neighbour cells) are defined by the analogue chemical state ($\underline{\boldsymbol{CS}_i^t}$) that can be monitored and mapped to a digitally representable Chemical State ($\boldsymbol{CS}_i^t$) for each cell. The chemical state evolution in the analogue domain occurs via the chemical oscillations, while the digital state is based on a finite state logic. Both the analogue and digital representations of chemical states are synchronized utilizing the chemical oscillations. Thus, this approach creates a hybrid electronic-chemical logic where the digital domain has deterministic logic using $CS_i^t$ and the chemical domain performs analogue computation using chemical oscillations. A generalized representation of the time evolution of the hybrid electronic-chemical logic in the analogue domain using chemical states ($CS_i^t$) can be defined by,

$$CS_i^{t+1} = G\left(F(CS_i^t, CS_j^t), CS_i^t\right)$$

where $F$ represents the digital operation and $G$ represents interactions in the physical system. $F$ can be defined by a digital Finite State Machine (FSM) that reads the current analogue signal from the physical system and performs operations back into the analogue domain and $CS_j^t$ are the neighboring cells to $CS_i^t$. The physical system evolves in a high dimensional space controlled by the complex myriad interactions between the chemical oscillations and their hysteresis effects combined with the effect of localized digital operations and the outcomes are probabilistic. Therefore, the computation is performed by iterating the single-step operation involving analogue and digital states with finite state logic defined for a specific problem.

In our experiments, using a digitally programmable input-output (I/O) system, we showed that it is possible to have an error correction system whereby the oscillatory chemical state can be reinforced, and the effects of phase shift can be eliminated so that the hybrid electronic control system amplifies and stabilizes the chemical states, see Fig. 1. As a proof of computation, we showed that the system could embody a cellular automaton (CA) and increase the connectivity in the configuration space way beyond the digital feature space even for simple rules of the elementary CA (*24*). The probabilistic nature within the hybrid computational architecture was demonstrated by implementing a two-dimensional probabilistic Chemical Cellular Automata (CCA) which shows emergent behaviour similar to that seen in Conway's Game of Life (*25*), as well as by solving combinatorial optimization problems such as number partitioning (26), Boolean satisfiability (27), and the travelling salesman problem (28).

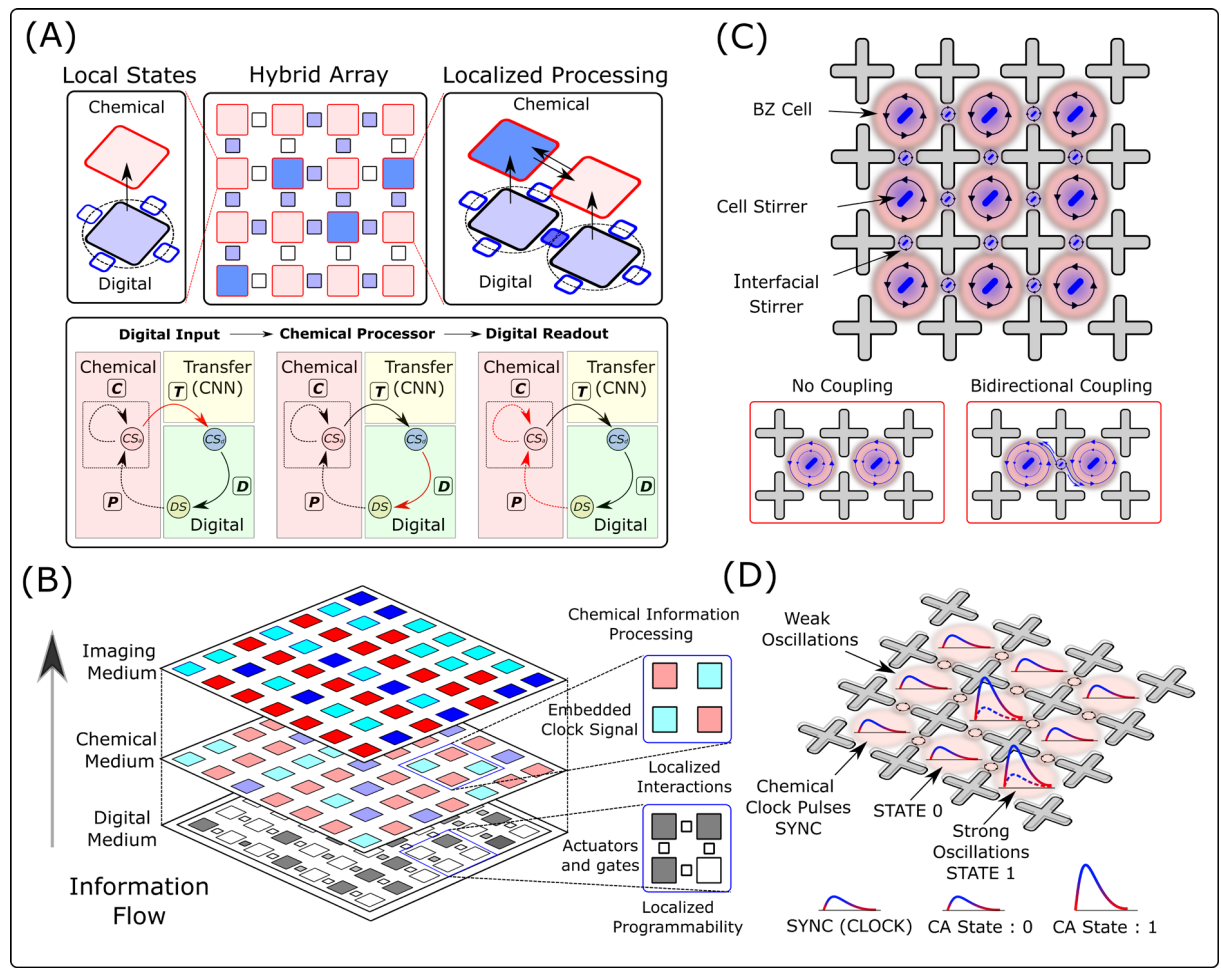

**Fig. 1: Conceptual and schematic design.** (A) Conceptual diagram of the proposed hybrid computation architecture comprising of an array of chemical reactors. The top figure shows hybrid digital-chemical information processing within single and coupled hybrid logical units. The bottom figure demonstrates a single processing step of the hybrid state machine with information looping between digital and chemical domains. Here, $D, P, C$ & $T$ represents digital, physical, chemical and transfer state machines (with $CS_a \equiv \underline{CS_i^t}$ and $CS_d \equiv CS_i^t$). (B) Shows a pictorial representation of how the information propagates together with local information processing in the chemical array. The weakly connected network in the chemical medium provides the global clock (SYNC signal) on which the local interactions process information and perform computation. (C and D) Schematic diagram of the two-dimensional BZ architecture showing how local cellular vortices interact by tuning the speed of the interfacial stirrers. The amplitude of the oscillations is controlled by the speed of the cell stirrers and can be used to define discrete states for information processing. Due to the well-defined periodic behaviour in the weak coupling limit, these oscillations can also be used to create a global clocking SYNC signal for decision making.

**The Chemical Computing Platform**

The chemical array exploits the excitability of the non-linear chemical oscillations of the Belousov Zhabotinsky (BZ) reaction via localized spatial control (*22*, *23*). The BZ reaction is highly excitable, can be maintained far from equilibrium, and it can be both spatially and temporally addressable. To exploit these features, we designed an experimental setup where we can program the addressable medium using an electronically controlled input system, see Fig. 1 (A, B). The experimental architecture consists of a 3D-printed 1D and 2D grid of interconnected reactors supported on an array of motors equipped with magnetic heads. At the centre of each reactor and at the interface of neighbouring cells, magnetic stirrers are placed to match the position of the motor shaft. Each motor is individually addressable, and its speed is controlled using a Pulse Width Modulation (PWM) signal generated using a microcontroller. The schematic diagram and the physical implementation of the two-dimensional experimental setup are shown in Fig. 2 and SI Section 1 and movie SM1. At the start of each experiment, the reagents required to initiate the BZ reaction (solutions of malonic acid, potassium bromate, an iron-based redox catalyst and sulfuric acid) are added into the reactors using an automated liquid handling system interface (see SI section 1). The role of the central cell stirrer is to initiate and then maintain the chemical oscillation and amplitude by varying the stirring speed. The mass transfer due to the hydrodynamic coupling between the neighbouring cells leads to interactions between the chemical oscillations of the BZ reactions, see Fig. 1 (C, D) and movie SM2. By tuning the speed of the interfacial stirrers (PWM levels), we can control and program the strength of intercellular couplings and limit them to their nearest neighbours. The BZ oscillations induced in a single cell are extremely sensitive to the composition of local redox species and the time of the actuation of the stirrer. This extreme sensitivity on the initial state, localized fluctuations, and time of actuation causes the phase of the chemical oscillations in individual cells to show significant drift with time. We observed that these unfavourable phase shifts between individual cells potentially limited the programmability of the system, and an error correction process to prevent decoherence was needed.

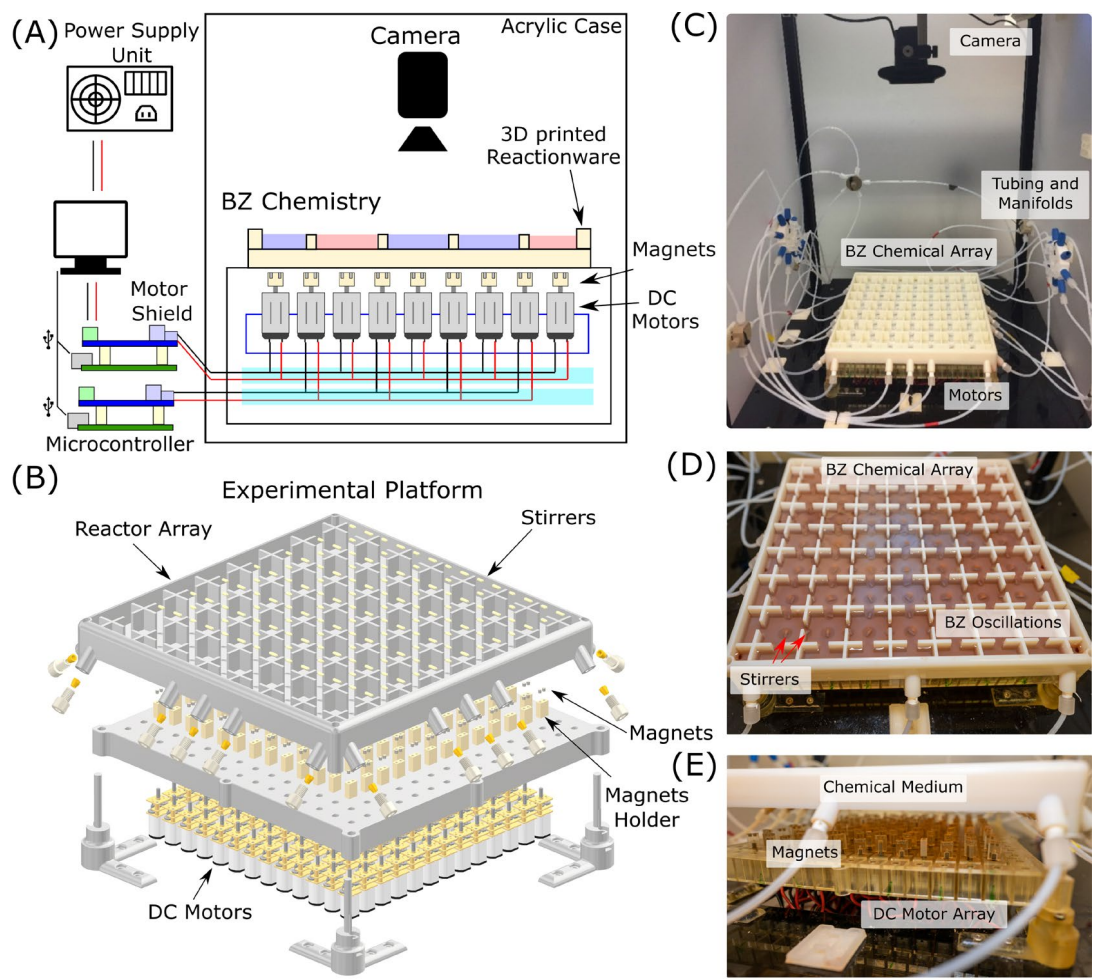

**Fig. 2: Schematic design and physical implementation of the experimental platform.** (A) Schematic diagram of the automated closed-loop experimental setup showing 3D printed reactor, motor control, imaging unit and supporting electronics. (B) Exploded view of the 3D printed reactor array with stirrers mapped with motor array and fluidic connections. (C) Complete experimental setup inside a light controlled acrylic housing with fluidic connections from the pump control unit. (D) Closeup view of the 3D printed reactor array with emerging BZ oscillation patterns. (E) Closeup view of the motor array with magnets connected to motor shafts controlling stirrer actuation. See SI Section 1 for complete details.

The low amplitude Chemical State is defined digitally as ($CS_i^t = 0$) and the high amplitude Chemical state ($CS_i^t = 1$). This state is created in a cell by switching the stirrer from pulsing to continuous mode at a higher stirring rate. To address the potential for errors resulting from state-decoherence in the oscillations, we found that the introduction of a global 'clock' signal (SYNC) between nearest neighbours produces weak oscillations, and these could be used to ensure the cells remained synchronised (see SI Section 2). These global weak oscillations are used for defining clocking signals which allows us to prevent unwanted dephasing, see Fig. 3.

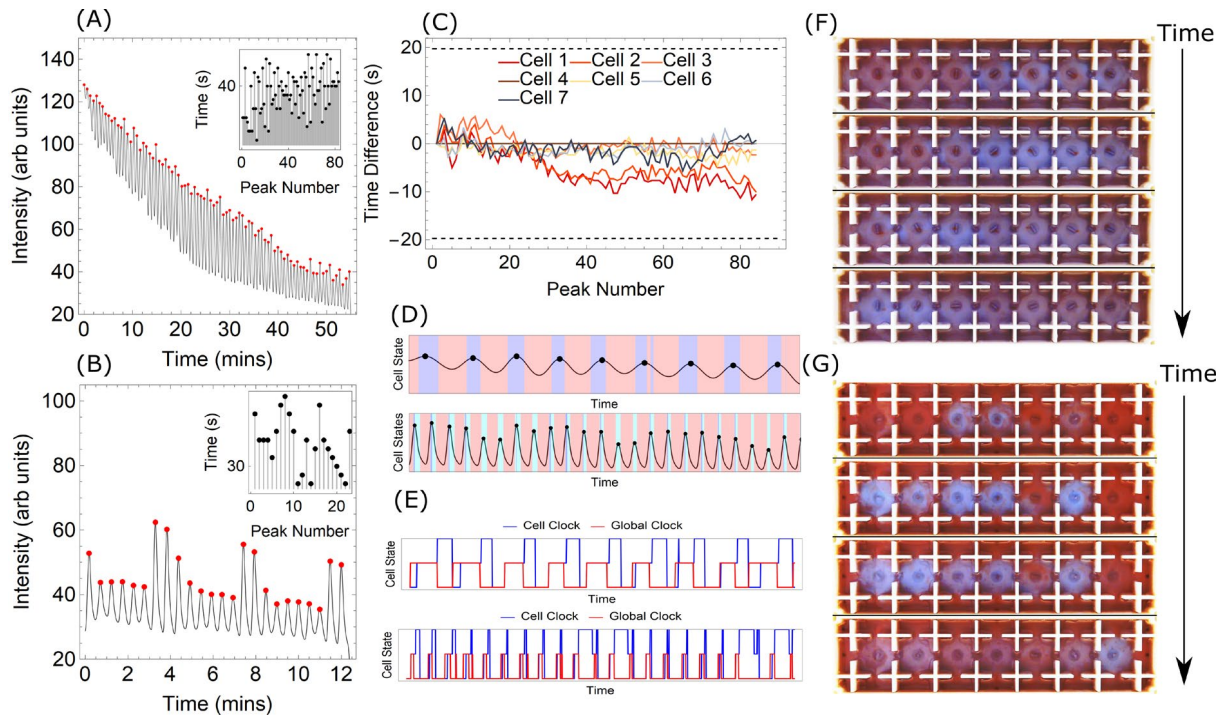

**Fig. 3: Chemical clocking in one-dimensional BZ experiments.** (A) BZ oscillations recorded for 60 mins in a continuously stirred single reactor, (B) BZ oscillations recorded in a single cell with LOW/HIGH states with programmable switching based on global clocking signal, (C) Deviations between the peak oscillations with time (peak number) between different interconnected cells with global clocking, (D) shows chemical oscillations and discrete chemical states by applying CNN for a single BZ cell vs. time for pure clocking tests (top) and cellular automata tests (down). (E) shows two different examples of cell and global clock interpreted from CNN output. (F) and (G) are the temporal snapshots of the actual experiments in a one-dimensional cell array demonstrating clocking wave without and with chemical mapped chemical states.

As the BZ reaction proceeds and the malonic acid 'fuel' is consumed, the amplitudes of the oscillations decrease, see Fig. 3 (A, B). To consistently define discrete chemical states based on the observed colour amplitudes, we trained a Convolutional Neural Network (CNN) on a dataset of time-dependent images labelled with discrete states. The three discrete states based on time-dependent colour classification are red (R), light blue (LB) and blue (B) which are also referred to as CNN states. In the presence of only weak oscillations, two distinct CNN states (Red and Light Blue) are recognized, while in the presence of weak and strong oscillations, three distinct states (Red, Light Blue and Blue) emerge, see Fig. 3, and we use these to describe the pattern of oscillations in a Finite State Machine (rFSM) to define the chemical states. In our current design, the rFSM logic consists of two different states: $CS_i^t = 0$ and $CS_i^t = 1$. The emergence of state $CS_i^t = 0$ occurs when a weak wave pattern is observed (R → LB → R) and the emergence of state $CS_i^t = 1$ occurs when a stronger oscillation is observed (R → LB → B → LB → R). From the chemical state readout, a feedback loop can be completed by implementing a deterministic digital state machine (**D**) that reads out the chemical states of all or a subset of cells as inputs and returns the PWM levels of the stirrers. These new PWM levels were then applied to the cell and interfacial stirrers (see SI Section 3&4).

**Probabilistic Logic and One dimensional Chemical Cellular Automata (1D-CCA)**

A single information processing loop can be represented by a combination of four state machines (**C, T, D, P**) acting on digital ($CS_i^t$) and an analogue representation of chemical states ($\underline{CS_i^t}$), see Fig. 4 (A and B). The state machine **C** is probabilistic and represents the evolution of $\underline{CS_i^t}$ while **D** represents a deterministic digital state machine that reads $CS_i^t$ and updates stirrer states (**DS**). The state machines **T** and **P** represents analogue to digital chemical state conversion ($CS_i^t = \mathbf{T}(\underline{CS_i^t})$) and physical effects of stirrers to the analogue chemical states, respectively. The time evolution of the digital and analogue representation of chemical states in hybrid probabilistic computation can be represented by

$$CS_i^t = \mathbf{K}^{t-1}\mathbf{K}^{t-2}\dots\mathbf{K}^1\mathbf{K}^0(\mathbf{C}(\mathbf{IC}))$$

$$\mathbf{K}^t(\underline{CS_i^t}) \equiv \mathbf{C}(\mathbf{P}(\mathbf{D}(CS_i^t, CS_j^t)), \underline{CS_i^t})$$

where $K$ represents hybrid state machine comprising of four state machines $C, T, D, P$ and i, j represents the central and neighbouring cells and the state machine $T$ in included into $C$ for simplicity. $IC$ defines the initial conditions. The emergence of the new chemical state has both implicit and explicit dependence on the previous chemical states. The implicit dependence comes from the hysteresis effect of the oscillations and the explicit dependence comes from the physical interaction of the stirrer state ($P$) on the oscillations which depends on the previous chemical states via finite state logic $D(CS_i^t, CS_j^t)$.

To demonstrate the working principle of the closed-loop hybrid electronic-chemical logic and the programmability with clocking logic, we implemented the Elementary Cellular Automata (CA) rules (Rule 30, 110 and 250, see SI Section 4) in a fully deterministic way, see Rule 30 as an example Fig. 4C and movie SM3. In the deterministic model as there is a one-to-one mapping between stirrer ($DS$) and chemical states ($CS_i^t$), information loops through digital and chemical domains where the chemical state machine mirrors the digital state machine precisely, see Fig. 4A. Next, we introduced the probabilistic computational logic by introducing hybrid automaton rules, see Fig. 4B. Consequently, the new chemical states in the analogue domain emerge probabilistically where CA rule 30 is modified by creating asymmetric actuation on interfacial stirrers, see Fig 4C. and the probabilistic outcomes of the chemical states can be seen in a single cell with the effect of stirrer speed on the emerging peak amplitudes, see Fig. 4D. By introducing new automaton rules that exploit the probabilistic computation mode enabled by the high dimensional space associated with the Chemical State Machine ($C$), larger configuration spaces can be explored.

It is possible to increase the dimensionality of the space by allowing the CAs to be controlled by the chemical system to give Chemical Cellular Automata (CCA); this uses exploits both deterministic and probabilistic logic in the digital and chemical domains respectively. As such, the one-dimensional CCA rules are defined by **($C_{rule}$-$I_{rule}$)** where **$C_{rule}$** reads chemical states and updates cell stirrer digital states and **$I_{rule}$** reads chemical states and updates interfacial stirrer digital states. With two different chemical states (0 and 1), there are $2^3=8$ possible patterns for the central stirrer and $2^2=4$ for each of the interfacial stirrer. The total number of possible rules in 1D-CCA are given by $2^3 2^2 2^2 = 4096$, see SI Section 5 for quantification of input and chemical states and probabilistic model of 1D CCA.

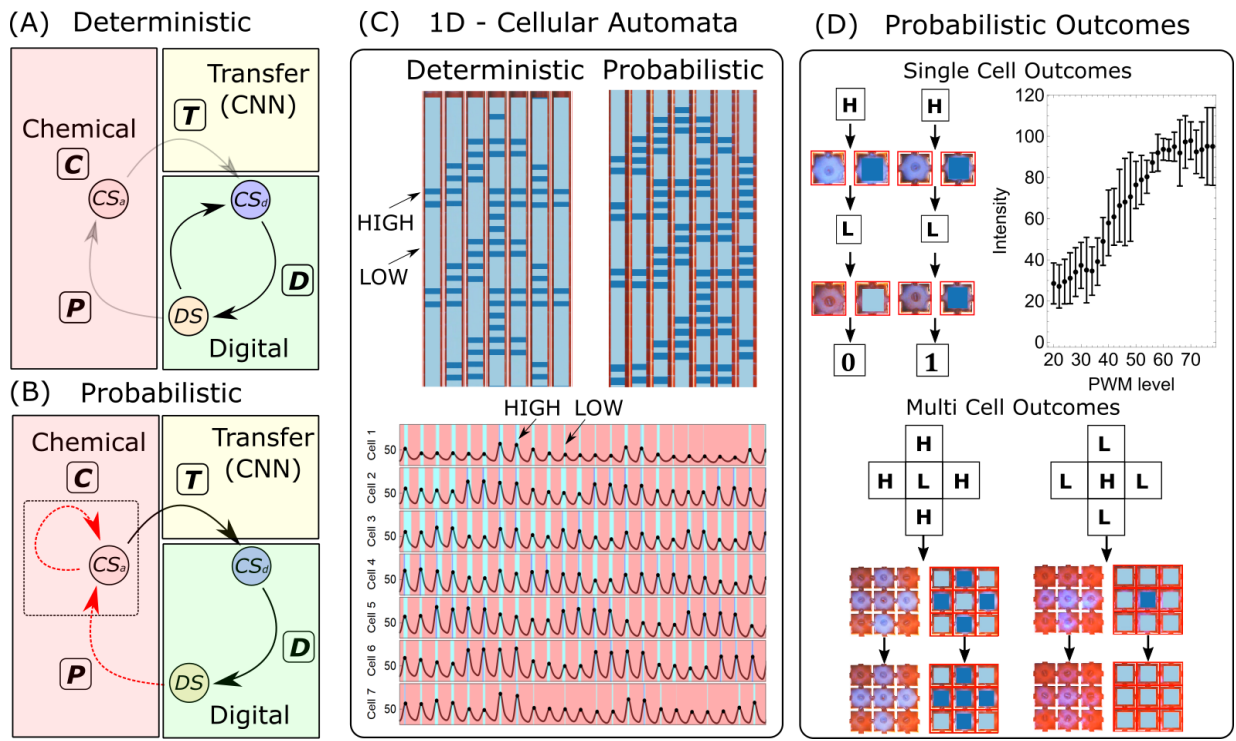

**Fig. 4: Implementation of one-dimensional Chemical Cellular Automata and Configuration Space Quantification.** (A, B) Representation of hybrid electronic-chemical state machine working in deterministic and probabilistic computational modes. (Black arrow: deterministic, red arrow: probabilistic), (C) Top: Implementation of elementary CA (rule 30) in deterministic mode (see movie SM3) and probabilistic modes demonstrating one-to-one and many-to-one mappings (light blue: 0, blue: 1), bottom: observed oscillations with CNN states in the background. (D) Top left and bottom show examples of deviations from one-to-one mapping in single and multiple cells (**H** and **L** represents high and low PWM states). Top right shows the average peak intensity observed at different PWM levels.

**Two-dimensional Chemical Cellular Automata (2D-CCA)**

By extending the CCA into two dimensions it is possible to explore the dynamics and emergence of complex patterns based on local rules defined by the hybrid electronic-chemical logic. In the 2D-CCA, we defined the basic emergent units as Chemical Entities (Chemits), which are comprised of a combination of five nearest-neighbouring cells in von Neumann neighbourhood, see Fig. 5(A&B) for experimental and pictorial representation of the Chemit. The positions and dynamics of Chemits are defined by the combination of digital and analogue representation of chemical states as described previously. We implemented a 2D-CCA digital state machine that takes the chemical states ($CS_i^t = \{0,1\}$) emerging from the probabilistic outcomes of the chemical information processing and implements 4-state PWM logic with states $\{S_0, S_1, S_2, S_3\}$. $S_0$ corresponds to inactive stirrer state, $S_1$ introduces the random fluctuations (weak chemical oscillations on randomly selected cells), $S_2$ creates the Chemit core and $S_3$ introduces the interactions between the Chemit and surroundings. The chemical oscillations created by PWM state $S_1$ in the absence of Chemits only leads to the $CS_i^t = 0$ chemical state. The Chemit core defined by a high PWM value ($S_2$) creates high $CS_i^t = 1$ states. The complete closed-loop probabilistic logic of two-dimensional CCA representing the emergence of Chemits is shown in Fig. 5C.

In the experiments, based on the local rules the Chemits shows propagation, replication and competition analogous to living species (see movie SM4). The emergence of a high chemical state at a specific cell could occur probabilistically due to the interaction of Chemit cells with local neighbours in the analogue chemical domain. These high chemical states at nearest neighbours or next-nearest neighbours in different configurations lead to propagation, replication, and competition events, see Fig. 5B. Propagation and replication events occur when a high chemical state occurs at nearest and next-nearest neighbours. When two Chemits interacts with each other, a competition event occurs, and for one Chemit, it has a 50% survival chance. However, in some cases, multiple strong oscillations causing high chemical states occur at nearest and next-nearest neighbours, which leads to random selection among multiple events, where a propagation, a replication or a competition event among all the neighbours was selected randomly. Fig. 5A shows series of snapshots from the experiments showing propagation and replication dynamics of Chemit in a 7×7 two-dimensional array with periodic boundary conditions. The complete description of the 2D-CCA and pseudo-code is described in the SI Section 6.

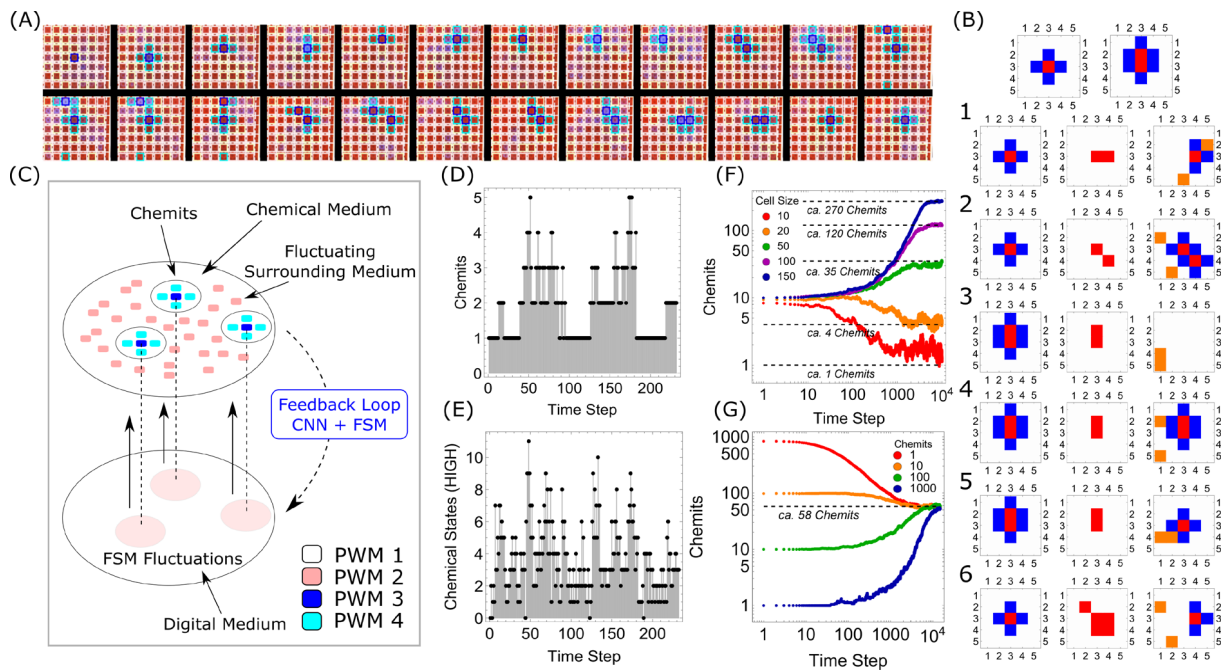

**Fig. 5: Two dimensional chemically instantiated probabilistic automaton.** (A) Snapshots of the two-dimensional experimental platform showing propagation and replication events starting with a single Chemical Entity (see movie SM4). (B) The basic construct of the experimental Chemit and demonstrates propagation (1), replication (2), competition (3-5), and random selection (6) between multiple events in a 5×5 array using the 2D-CCA state machine as a proof-of-principle. (C) Conceptual design of closed-loop chemically instantiated probabilistic automaton. (D) Population dynamics of Chemits on two-dimensional experimental step (7×7 cells) *vs* oscillatory time steps, (E) Number of high chemical states *vs.* time from which Chemits are derived based on PWM logic. (F) Average population dynamics of Chemits with different domain sizes. (G) shows the time evolution of the average number of Chemits with different initial conditions which converge at a steady state depending on the available resource space. All simulations were run 25 times and the mean population at different time steps was estimated.

Over a 7×7 experimental array together with periodic boundary conditions, we observed the emergence of peaks in the population of the Chemits due to sudden localized replication events which later fall due to competition events within the constrained space or resource, see Fig. 5D. The total number of digitally representable high chemical states at a given step leading to Chemits dynamics are shown in Fig. 5E. The population and the propagation dynamics of the Chemits with given control parameters are governed by the initial number of Chemits, available spatial resource, and random fluctuations and it is interesting that the system shows highly complex emergent behaviour as observed in Conway's Game of Life *(25)*, but instantiated in a physical device. To investigate the emergent behaviour over a larger spatial scale, we developed a chemical probabilistic state machine based on the observed phenomenological model to simulate the dynamics of Chemits behaviour up to 150×150 cell array (see SI Section 6 and movie SM5). Similar to the experimental observations, the simulations show the sudden formation of local population clusters due to fast replication and as well as annihilation due to competition events at local clusters of large populations. We observed unstable population dynamics of Chemits on an array with a smaller number of cells, however, with an increase in the number of cells the Chemits population stabilizes at different levels which depends on the available spatial resource, see Fig. 5F. We further investigated the population dynamics by varying the initial population of Chemits over a 100×100 array and observed convergence in the population at the steady state independent of the initial population, see Fig. 5G and SI Section 6 for more characterization. These simulations demonstrate a strong correlation between the global population dynamics and the local probabilistic rules emerging from digital and chemical state machines. The emergent population dynamics and steady-state kinetics of Chemits are analogous to the population behaviours observed in evolutionary biology and can be further extended towards computation operations.

**Solving Combinatorial Optimization Problems using Hybrid Computation Machine**

Building on the dynamic feedback loop between electronic and chemical states, we implemented a hybrid electronic-chemical computing algorithm to solve quadratic combinatorial optimization problems taking advantage of the probabilistic logic to reach the problem solution more efficiently than that of a deterministic logic. In the context of Quantum

Adiabatic Optimization, various combinatorial optimization problems such as partitioning, satisfiability (SAT) and Hamilton cycles can be formulated as energy/cost minimization problems on an Ising lattice (29). Inspired by the Ising or equivalent Quadratic Unconstrained Binary Optimization (QUBO) formulations of these problems, we implemented a hybrid electronic-chemical state machine capable of performing energy minimization using chemical states or PWM states equivalent to Ising spin variables. The generalized Hamiltonian up to a quadratic coupling is given by,

$$H(Q_1 \ldots Q_n) = h^{(0)} + \sum_{i=1}^{n} h_i^{(1)} Q_i + \sum_{1<i<j<n}^{n} h_{ij}^{(2)} Q_i Q_j$$

where, $Q_i$ defines the chemical / PWM state of the cell, $h^{(0)}$ is an offset energy term, $h_i^{(1)}$ defines the self-interaction term of the spin equivalent state $Q_i$ and $h_{ij}^{(2)}$ defines the coupling between states $Q_i$ and $Q_j$. The Ising formulation of the optimization problem can be represented by a connected graph with self-interactions and pairwise couplings for Hamiltonian formulation and mapping to the chemical array (see SI Section 7). The sign of the coupling coefficients describes the positive (ferromagnetic type) and negative (anti-ferromagnetic type) couplings and the magnitude describes the coupling strength. In the ideal case of implementation of a computation algorithm in a physical system, the Ising spin equivalent chemical states should flip according to the interactions defined by local couplings. It should let the problem Hamiltonian reach the global minimum energy configuration and the corresponding chemical and digital states can be interpreted as solutions.

To explore this computation, we have implemented an electronic-chemical hybrid state machine where the two chemical states map directly to Ising spins (−1, +1) and demonstrate the proof-of-principle computation through the information loop without any neighbouring interactions (see SI Section 7 for pseudo codes and implementation details). Based on the emerging chemical states, the configurational energy is estimated from the Hamiltonian *in-silico* and compared with the lowest energy configuration so far and iterated until the minimal energy configuration was achieved. In the extended approach, we utilized a combination of the digital state machine coupled with a probabilistic chemical state machine where the PWM states of the cell stirrers to map directly to the Ising spin variables. The flowchart of the complete computational scheme is shown in Fig. 6A. For the pairwise neighbouring interactions generating probabilistic outcomes, a lookup table of chemical states as shown in Fig.6B was created and the problem was mapped on the platform (see Fig. 6C). As a result, the emergence of the new chemical states not only depends on PWM states of stirrers but also the interactions within chemical states defined by the hybrid state machine. At each step, a comparison was made between ideal states from the lookup table and the emerging chemical states and was utilized for the acceptance of the step towards energy minimization. The probabilistic outcome of the new chemical states arises from the combination of analogue and digital processing. This in turn leads to the lowest energy state due to a higher connectivity in the configuration space. This improvement in efficiency will become more evident with the scaling and complexity of the problem with large local minimum configurations. By distributing the algorithm between the digital and chemical logic, we demonstrate large scale combinatorial problems can be solved efficiently (*30*).

To achieve this advantage, we by map the variables of the problem to the cells such that all the coupling interactions can be introduced between neighbours, see Fig. 6C which shows the mapping of a fully connected four-number partitioning problem (primary spin cell shown in blue) employing multiple instantiations of the same spins (auxiliary cells shown in red) to accomplish pairwise coupling between all the variables of the Hamiltonian. At each step, after flipping the PWM states of the cell randomly, pairwise operations to estimate the energy change were performed in a parallel approach. In the chemical decision-making step, if the emergence of the new chemical state is consistent with the lookup table, the energy change was recorded as such else was recorded with a negative sign.

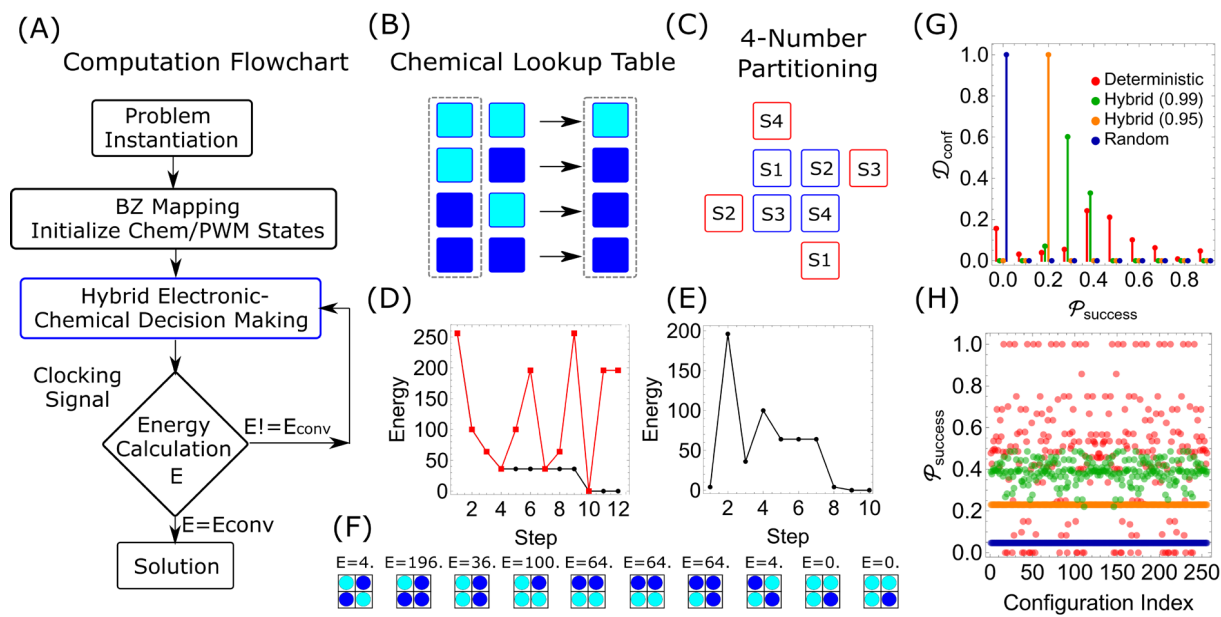

**Fig. 6: Demonstration of chemical computation in the chemical array.** (A) Flow chart describing hybrid electronic-chemical logic for solving quadratic optimization problems using a chemical array. (B) Chemical states lookup table used for chemical decision making based on probabilistic outcomes (light blue: $CS_i^t = 0$ and dark blue $CS_i^t = 1$). (C) Left: Mapping a 4-number partition problem on a chemical array with isolated spins with all couplings defined by neighbouring auxiliary cells. Right: Efficient mapping of 4-number partition problem on the chemical array. (Blue: principal cells, Red: auxiliary cells for spin variables). (D & E) Energy minimization of four-number partitioning solved using a hybrid probabilistic algorithm using two different approaches. (F) Pictorial representation of chemical states from an experiment on energy minimization for a 4-number partition problem (light blue: $CS_i^t = 0$, dark blue: $CS_i^t = 0$), see movie SM6. (G) The distribution of initial configurations over the success probability of solving an 8-number partition problem ($S = \{1, 3, 4, 9, 3, 5, 3, 6\}$) in pure deterministic, random and hybrid approaches. (H) Success probabilities of deterministic (index =1.0), random (index = 0.5) and hybrid approach (index = 0.99/0.95) vs. initial configuration indices for the 8-number partition problem. The deterministic index with value 1.0 corresponds to a pure deterministic algorithm, 0.99/0.95 as our tuneable hybrid approach and 0.5 as a random algorithm.

The overall energy change was estimated via summation over all the pairwise spins, and the flipping was accepted or rejected similarly to the first hybrid logic scheme according to the principle of energy minimisation. Once the energy associated with the current PWM states estimated from the Hamiltonian reaches the global minimum, the solution to the optimization problem is interpreted from the PWM states, see Fig. 6 D and E for the solution to the four-number partitioning problem for the number set $S = \{1, 3, 4, 8\}$ using the two-hybrid algorithms. The spin state configurations of the primary spin cells for the 4-number partitioning problem using the second hybrid algorithm with the corresponding energies are shown in Fig. 6F. Other examples including number partitioning, Boolean Satisfiability and Travelling Salesman Problems are demonstrated in SI Section 7.

To investigate the influence of probabilistic logic from the chemical state machine in the hybrid approach, we quantified the algorithmic performance by defining hybrid logic as a combination of the digital algorithm coupled with analogue chemical processing. The probabilistic decision making occurring in the chemical state machine is tuneable by selecting the PWM levels, see Fig. 4D. We formulated an 8-number partitioning problem on the number set $S = \{1, 3, 4, 9, 3, 5, 3, 6\}$ and estimated the success probability with all possible starting ($2^8$=256) configurations. By calculating the stationary distribution of the possible configurations, we could estimate the probability of finding the configuration with global minimum energy at different values of the deterministic index, see SI Section 7. We observe by the reduction in the deterministic index taking advantage of the probabilistic chemical state machine; the success probability distribution shrinks leading to higher chances for finding the solution independent of the initial configuration (see Fig. 6G and H). The distribution shows for a given 8-number partition problem, the hybrid approach finds a solution with all starting configurations however even though the deterministic algorithm shows a high probability of success, many configurations get trapped in local minima and gave no solution.

We have demonstrated a novel computational architecture that processes the information in an electronically programmable chemical medium with tuneable probabilistic logic utilizing the natural behavior of physicochemical processes (see SI Section 7, Qualification of Chemical Computation). The distribution of information between the digital and analogue chemical state machines is generic and implementable on any digitally programmable physicochemical process. Our architecture can not only implement novel cellular automaton rules but demonstrates and discovers large scale emergent behavior such as Chemits resulting from the local probabilistic interactions between digital and chemical state machines. We also demonstrated two hybrid schemes that solve combinatorial optimization problems by minimizing Ising Hamiltonian formulation utilizing chemical states and nearest-neighbouring couplings. Importantly we showed the distribution of information processing between the digital and the chemical domains, demonstrating that the chemistry is taking an active part in the computation.

**References and Notes**

1. M. M. Waldrop, More than Moore. *Nature*. **530**, 144–147 (2016).
2. Y. Taur, D. A. Buchanan, W. Chen, D. J. Frank, K. E. Ismail, L. O. Shih-Hsien, G. A. Sai-Halasz, R. G. Viswanathan, H. J. C. Wann, S. J. Wind, H. S. Wong, CMOS scaling into the nanometer regime. *Proc. IEEE*. **85**, 486–503 (1997).
3. C. Hu, Future CMOS Scaling and Reliability. *Proc. IEEE*. **81**, 682–689 (1993).
4. F. Arute, *et al.* Quantum supremacy using a programmable superconducting processor. *Nature*. **574**, 505–510 (2019).
5. J. Gorecki, K. Gizynski, J. Guzowski, J. N. Gorecka, P. Garstecki, G. Gruenert, P. Dittrich, Chemical computing with reaction-diffusion processes. *Philos. Trans. R. Soc. A Math. Phys. Eng. Sci.* **373** (2015), doi:10.1098/rsta.2014.0219.
6. T. Rueckes, K. Kim, E. Joselevich, G. Y. Tseng, C. L. Cheung, C. M. Lieber, Carbon nanotube-based nonvolatile random access memory for molecular computing. *Science*. **289**, 94–97 (2000).
7. B. Schrauwen, D. Verstraeten, J. Van Campenhout, An overview of reservoir computing: theory, applications and implementations. *Proc. 15th Eur. Symp. Artif. Neural Networks*, 471–482 (2007).
8. L. M. Adleman, Molecular computation of solutions to combinatorial problems. *Science*. **266**, 1021–1024 (1994).
9. C. H. Bennet, D. P. DiVincenzo, Quantum Information and Computation. *Nature*. **404**, 247–255.
10. Y. Fang, V. V. Yashin, S. P. Levitan, A. C. Balazs, Pattern recognition with materials that compute. *Sci. Adv.* **2**, e1601114 (2016).
11. G. Katsikis, J. S. Cybulski, M. Prakash, Synchronous universal droplet logic and control. *Nat. Phys.* **11**, 588–596 (2015).

12. X. Lin, Y. Rivenson, N. T. Yardimci, M. Veli, Y. Luo, M. Jarrahi, A. Ozcan, All-optical machine learning using diffractive deep neural networks. *Science*. **1008**, eaat8084 (2018).

13. A. Adamatzky, B. D. L. Costello, T. Asai, *Reaction-Diffusion computers* (Elsevier Inc., 2005).

14. J. Torrejon, *et al.* Neuromorphic computing with nanoscale spintronic oscillators. *Nature*. **547**, 428–431 (2017).

15. L. Kuhnert, K. I. Agladze, V. I. Krinsky, Image processing using light-sensitive chemical waves. *Nature*. **337**, 244–247 (1989).

16. J. M. Parrilla-Gutiérrez, S. Tsuda, A. Sharma, G. Cooper, G. Aragon-Camarasa, K. Donkers, L. Cronin, A programmable chemical computer with memory and pattern recognition. *ChemRxiv*. **7712564** (2019), doi:10.26434/chemrxiv.7712564.v1.

17. O. Steinbock, Á. Tóth, K. Showalter, Navigating complex labyrinths: Optimal paths from chemical waves. *Science*. **267**, 868–871 (1995).

18. M. A. Tsompanas, C. Fullarton, A. Adamatzky, Belousov-Zhabotinsky liquid marbles in robot control. *Sensors Actuators, B Chem*. **295**, 194–203 (2019).

19. F. Hadaeghi, H. Jaeger, Computing optimal discrete readout weights in reservoir computing is NP-hard. *Neurocomputing*. **338**, 233–236 (2019).

20. D. Deutsch, *Proc. R. Soc. London A Math. Phys. Eng. Sci.* **400**, 97–117 (1985).

21. N. Tompkins, N. Li, C. Girabawe, M. Heaymann, G. B. Ermentrout, I. R. Epstein, S. Fraden, Testing Turing's theory of morphogenesis in chemical cells. *Proc. Natl. Acad. Sci.* **111**, 4397–4402 (2014).

22. V. Horvath, D. J. Kutner, J. T. Chavis, I. R. Epstein, Pulse-coupled BZ oscillators with unequal coupling strengths. *Phys. Chem. Chem. Phys.* **17**, 4664–4676 (2015).

23. V. Petrov, V. Gáspár, J. Masere, K. Showalter, Controlling chaos in the Belousov - Zhabotinsky reaction. *Nature*. **361**, 240–243 (1993).

24. S. Wolfram, Mathematical Physics Computation Theory of Cellular Automata. *Commun. Math. Phys*. **96**, 15–57 (1984).

25. M. Gardner, The fantastic combinations of John Conway's new solitaire game "life." *Sci. Am.* **223**, 120-123. (1970).

26. R. E. Korf, Artificial Intelligence A complete anytime algorithm for number partitioning. *Artif. Intell.* **106**, 181–203 (1998).

27. P. Hansen, B. Jaumard, Algorithms for the Maximum Satisfability Problem. *Computing*. **44**, 279–303 (1990).

28. D. L. Applegate, R. E. Bixby, V. Chvatal, W. J. Cook, *The Traveling Salesman Problem: A Computational Study* (Princeton University Press, Princeton, NJ, USA, 2007).

29. S. Y. Guo, *et al.* A molecular computing approach to solving optimization problems via programmable microdroplet arrays. *Matter*. **4**, 1107–1124 (2021).

30. D. Pierangeli, G. Marcucci, C. Conti, Large-Scale Photonic Ising Machine by Spatial

Light Modulation. *Physical Review Letters* **122**, 213902 (2019).

**Acknowledgements**
We would like to thank Hessam Mehr and Liam Wilbraham of the University of Glasgow for discussions.

**Funding Sources**
The authors gratefully acknowledge financial support from the EPSRC (Grant Nos EP/H024107/1, EP/I033459/1, EP/J00135X/1, EP/J015156/1, EP/K021966/1, EP/K023004/1, EP/K038885/1, EP/L015668/1, EP/L023652/1), the ERC (project 670467 SMART-POM), and the DARPA molecular informatics project.

**Author Contributions**
L.C. conceived the original idea and together with A.S. designed the project and the research plan. L.C. designed the reactor array and A.S. and M.T.-K.N. designed and built the robotic platform with help from J.M.P.G. J.M.P.G. implemented the computer vision and chemical clocking algorithms, and M.T.-K.N. created the training dataset. M.T.-K.N and A.S. implemented the 1D-CCA. A.S., Y.J. and M.T.-K.N implemented the 2D-CCA. A.S. and Y.J. implemented the chemical computation and M.T.-K.N, Y.J and A.S. performed the experiments. A.S. did the data analysis, and developed models and ran the simulations with help from Y.J. A.S. helped benchmark the system with L.C. Finally, A.S and L.C wrote the paper with help from the rest of the authors.

**Data and Code availability**
Due to the large total size (>500 Gb) of the experimental and theoretical data, the data used in this work are available upon request to the corresponding author at lee.cronin@glasgow.ac.uk. The code used to operate the platform and various implemented simulation models are available of <https://github.com/croningp/BZComputation>.

Supplementary Materials
Materials and Methods
Figs. S1 to S69
Tables S1 to S4
References 31 to 36
Movies SM1 to SM6

**Supplementary Video 1**: SM1 Description of the automated experimental platform.
**Supplementary Video 2**: SM2 Hydrodynamic coupling between nearest neighbouring cells monitor using ink as a tracer.
**Supplementary Video 3**: SM3 Implementation of elementary cellular automata rule 30 using a dynamic feedback loop.
**Supplementary Video 4**: SM4 Experimental implementation of 2D-CCA.
**Supplementary Video 5**: SM5 Simulation of Chemits on 100 ×100 array with different initial conditions.
**Supplementary Video 6**: SM6 Solving four number partitioning problem demonstrating chemical decision making.

Supplementary Information for

**A Probabilistic Chemical Programmable Computer**

Abhishek Sharma, Marcus Tze-Kiat Ng, Juan Manuel Parrilla Gutierrez, Yibin Jiang and Leroy Cronin*

School of Chemistry,

The University of Glasgow, University Avenue, Glasgow G12 8QQ, UK, *Corresponding author email: Lee.Cronin@glasgow.ac.uk

# Table of Contents

# 1 Overview of the automated platform

The overall hybrid electronic-chemical computational platform consists of three main control domains as shown in Fig. S1, namely the (1) chemical domain which consists of stock solutions required for BZ reaction that were pumped using syringe pumps sequentially in the right proportion into the mixing chamber. The mixing chamber contains a magnetic stirrer bar that rotates at 140 RPM constantly to ensure the stock solutions were well mixed. Using another pair of syringe pumps, the reaction mixture in the mixing chamber was then transferred to the 3D printed experimental arena with stirrers in the (2) experimental setup as shown in Fig. S4. In this experimental setup, the rotation of stirrers is controlled by DC motors equipped with Neodymium-based permanent magnets located at the bottom of the arena. Each motor speed and direction can be individually addressed by the supported electronics control. The BZ chemical oscillations occurring in the experimental arena on the response of stirrer actuation were then observed and recorded by a camera.

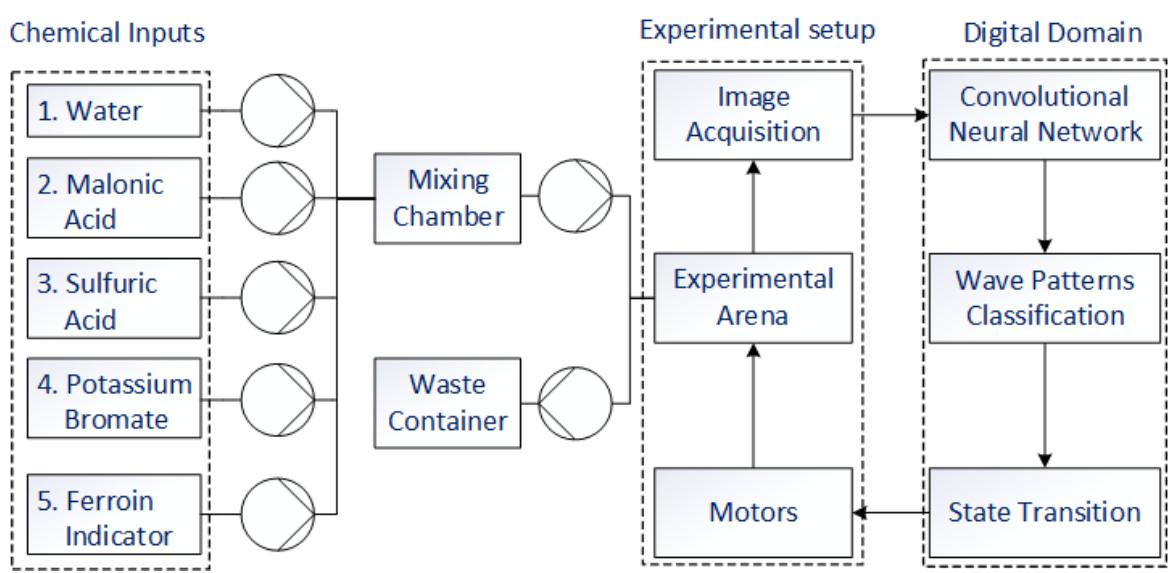

**Fig. S1:** Complete schematic of the experimental setup. The schematic diagram shows three main parts of the automated experimental platform, chemical inputs, experimental setup and digital domain. The chemical inputs create the BZ mixture from the stock solutions using pump control. The experimental setup runs the experiment with motor control and imaging. The digital domain runs real-time data analysis and state machines for hybrid electronic-chemical information processing.

These temporal oscillatory patterns were then passed into the (3) digital domain where further information processing occurs. The oscillatory patterns were classified into three different states using a convolutional neural network (CNN). There are three different classification states, RED, LIGHT BLUE and BLUE. These classified states were used for creating a common chemical clock over all the cells as well as the basic programmable chemical states (CS) for computation. The chemical clocking logic is used over all the experiments as a sync signal for a single feedback loop step. These patterns in the given time frame were then converted into the observed chemical states, high state: CS=1 and low state CS=0 by using a Finite State Machine (FSM). This FSM reads the temporal CNN states over a given time and return the digital CA state based on the observed oscillatory behaviour and resets. The hybrid electronic-chemical computational logic is then implemented on these chemical states using various problem-dependent state machines which are discussed further in the later sections. The hybrid electronic-chemical computational logic utilizes these chemical states which then dynamically controls the stirrers speed and completes the feedback control loop over the complete experimental period. This feedback loop together with the chemical clocking logic were used to create novel one-dimensional and two-dimensional Chemical Cellular Automata (CCA) and performing useful computation by solving combinatorial optimization problems. Once an experiment is finished, the remaining solution was drained into the waste container using a pair of syringe pumps and the experimental arena underwent a series of rinsing and cleaning cycles to get ready for the next experiment. This is the overall description of the complete experimental protocol and full details of individual steps and experimental protocols is discussed in the following sections. The video of the platform working can be found in Supplementary Video 1.

## 1.1 Materials

Sulphuric acid was sourced from Fisher Scientific, ≥95 % analytical reagent grade. Malonic acid was sourced from Sigma Aldrich, Reagent Plus 99%. Ferrous sulphate heptahydrate was sourced from Sigma Aldrich, ≥ 98%. 1,10-phenanthroline was sourced from Sigma Aldrich, ≥ 99%. Potassium bromate was sourced from VWR and Sigma Aldrich > 99%. The 3D-printed parts were designed using a cloud-based CAD software OnShape and printed with Stratasys Connex500 3D printer using the VeroWhitePlus, for the batch system that holds the chemical mixtures and FullCure 720 material for the rest of the platforms, detailed descriptions of the CAD designs are shown in Section 1.2.4. The TriContinent C3000 syringe pumps were used for fluid handling over the complete platform. The supporting structures for the experimental box enclosure were purchased from Ooznest Ltd and acrylic sheets were purchased from

www.plasticsheets.com (Wanna Ltd.). Table S1 gives the details of the various components used in the experimental arena.

**Table S1:** Details of the various parts used in the complete experimental setup

| Part | Description |
|---|---|
| Stirrers | Two different types of stirrers for cells and interface<br>1. VWR micro. 7 x 2 mm<br>2. VWR micro. 5 x 2 mm |
| DC motors (6V) | 1. 200 RPM Mini Metal Gear Motor with Gearwheel Model: N20 3 mm Shaft Diameter<br>2. RS PRO, 6 V dc, 73 gcm, Brushed DC Geared Motor, Output Speed 1490 rpm |
| Magnets for motor shaft | First4Magnets 2 x 2 mm Neodymium Magnet – 0.15 kg Pull |
| Arduino | Arduino Uno REV3 SMD |
| PWM servo driver shield | Adafruit 16-Channel 12-bit PWM/Servo Shield – I2C interface |
| Power Supply Unit | Programmable Two-Channel Power Supply unit (RS Components) |
| Computers | 1. Ubuntu 18.04 LTS, Intel Core i7 8700 3.20 GHz, 32 Gb RAM<br>2. Arch Linux AMD Ryzen 7 2700, 32 Gb RAM<br><br>Both are capable to run Python 3.7.X, OpenCV 3.X, and TensorFlow for image processing |
| Camera | Logitech C 920 HD PRO WEBCAM |
| Syringe Pumps | TriContinent C3000 Syringe Pumps |
| Tubing and fluidic connections | 1. Flangeless fittings Cole Palmer Ltd.<br>2. PTFE tubing (1/8- and 1/16-inch diameter) |
| Experimental Arena Housing | 5 mm Translucent Acrylic Sheets and support structure using v-slot linear rails (Ooznest Ltd.) |
| Arena/ stir bar array | 3D printed |
| Motor array | 3D printed |
| Arena support structures | 3D printed |
| Motor shaft magnet holder | 3D printed |

## 1.2 Platform description

The complete description of the experimental setup where the platform was placed in a light-controlled box made up of translucent acrylic sheets with a matt finish is shown in Fig. S2. This ensures an even light distribution in the platform as well as minimises the reflection of the solution when in contact with direct background light. The light levels were kept consistent over all the experiments to avoid problems in classifying the three discrete states from the analogue signal of the chemical oscillator. The coordinates of the experimental platforms and the height of the cameras were fixed to ensure consistent and reliable image acquisition in every experiment. Two different types of stirrers are placed in the experimental platform and input/output flows are selected such that no stirrer moves from its position when BZ reaction mixtures flow in or experimental waste goes out.

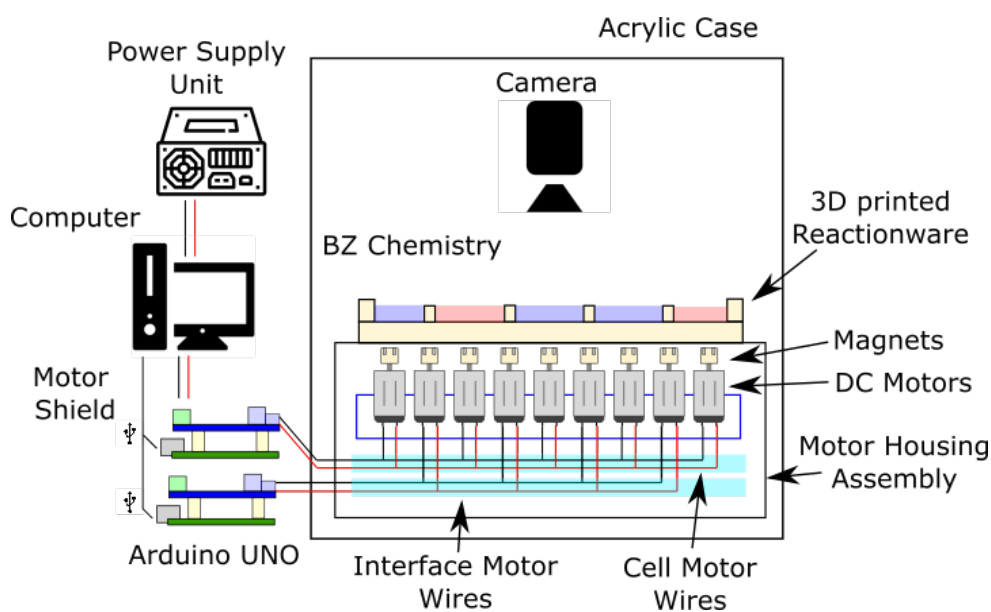

**Fig. S2:** Pictorial representation of the experimental setup. The experimental set-up which comprises of the BZ platform (3D printed reactor assembly, addressable motors and electronic control assembly), a camera that does the image acquisition. The data recorded from the camera was then passed to the digital domain of the platform, depending upon the state machine implemented, new stirrers' operation was delivered to the DC-motors via microcontroller prototype units (Arduino UNO).

The exploded view of the one- and two- dimensional 3D printed reactors is shown in Fig. S3. The actual implemented of the complete experimental setup is shown in Fig. S4. The complete details are described in the Section 1.2.4, OnShape (CAD) links and STL files are also available for download from <https://github.com/croningp/BZComputation>. The experimental

S6

hardware, electronic components and control software in complete detail will be discussed in the subsection below.

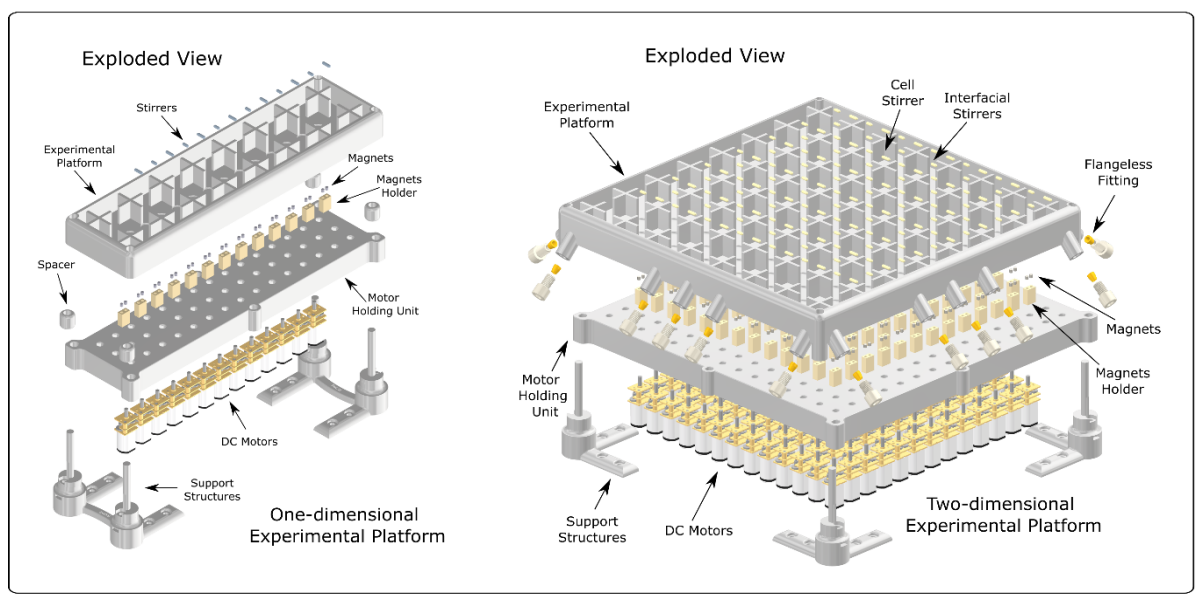

**Fig. S3:** Exploded views of one- and two- dimensional setup. Figure shows exploded view of one- (left) and two- (right) dimensional experimental setup as designed by using CAD software OnShape. Full description of the setup with CAD drawings is shown in detail in Section 1.2.4.

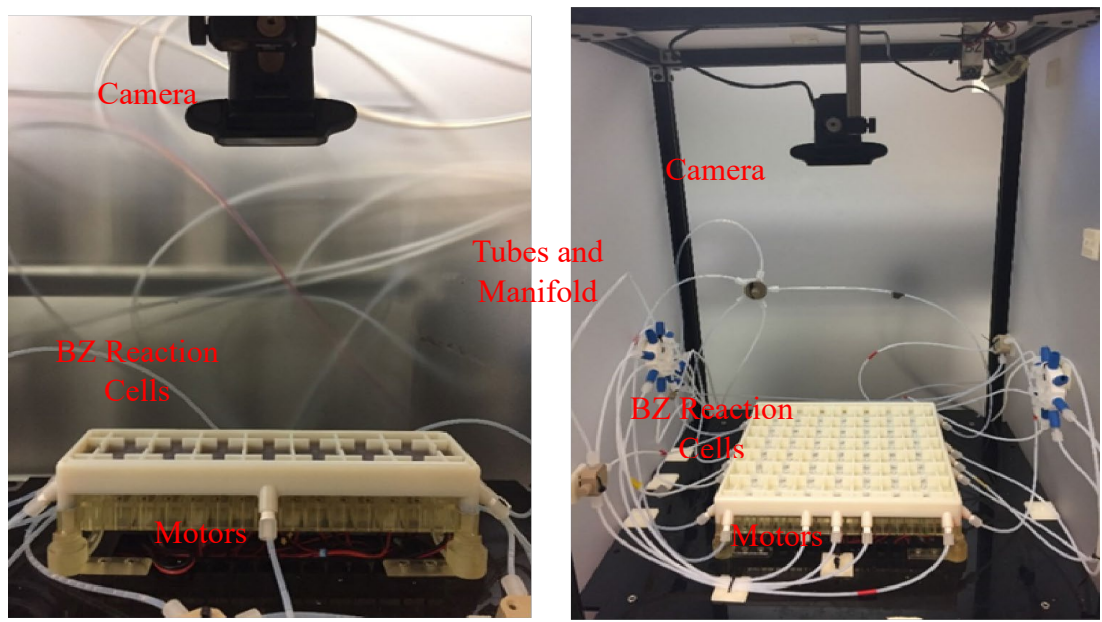

**Fig. S4:** One- and two-dimensional experimental setup. Real experimental set-up of the 1D platform (left) and 2D platform (right) where the PTFE tubes act as inlets and outlets for the various reagents and wastes. The cameras were mounted above the platform in a fixed height to have a clear and complete image of the 3D printed platform. The motor housing setup and wire connectivity go beneath the experimental setup.

## 1.2.1 DC motors for stirrer control

Two types of DC geared motors were used with different RPMs for cell and interfacial stirrers for different speed requirements. Cell stirrers used to initiate BZ chemical oscillations requires high-speed motors and interfacial stirrers require lower speed for coupling localized between nearest neighbouring cells. For one-dimensional setup, 7 cell motors and 6 interfacial motors and for two-dimensional setup, 49 cell motors and 84 interfacial stirrer motors were used.

The description of these motors are as follows,

1) **Cell Motors**: The active cells motors used were purchased from RS Components under the description of "RS PRO, 6 V dc, 73 gcm, Brushed DC Geared Motor, Output Speed 1490 rpm"
2) **Interface Motors**: The interfacial cells motors used were bought from "AliExpress" under the description of "DC 6V 200 RPM Mini Metal Gear Motor with Gearwheel Model: N20 3mm Shaft Diameter"

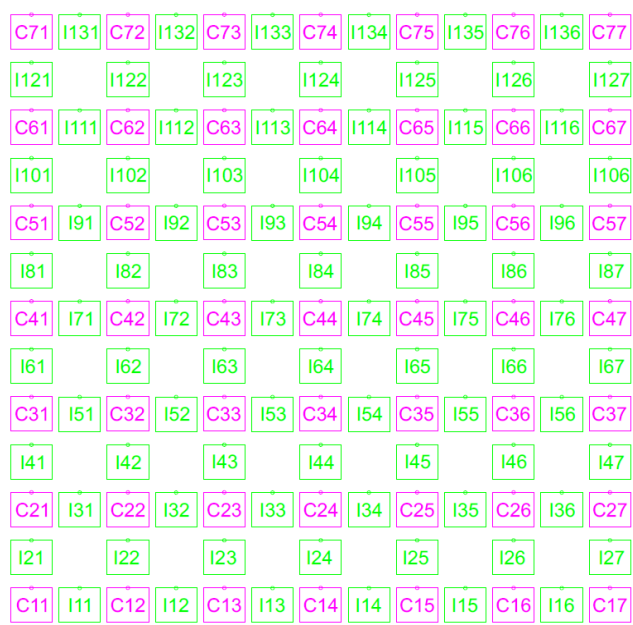

**Fig. S5:** Motor Numbering Scheme for Cell and Interfacial motors. The figure shows a motor numbering scheme for cell (magenta) and interfacial motors (green) which allows direct command of the neighbouring interfacial and cell motors. These motor IDs were stored as a variable in the firmware on the microcontroller board for fast access to the motor.

The motors were placed in a periodic array aligned with 3D printed reactor cells. For the two-dimensional setup, the placement scheme together with motor Ids for both cell (magenta) and interface (green) motors is shown in Fig. S5. The same scheme was used in the control software to address individual motors.

### 1.2.2 Electronics

The DC motor speed (RPM) was controlled by applying a modulated voltage between 0-6V using a Pulse Width Modulation (PWM) signals from the microcontroller prototype board Arduino UNO. To control both motor speed and direction, Adafruit Motor Shields (Ver. 2) which were stacked to control multiple DC motors for each Arduino UNO. The exact number of Arduino Uno, Adafruit Motor Shields and the total DC Motors controlled by the one- and two- dimensional set-up can be found in Table S2. The motors were directly connected to the Motor Shield using pin-screw terminals with a wire connection scheme to separate cell motors and interfacial motors. Each shield can control the speed and direction of four motors and uses I2C communication protocol to communicate with the microcontroller unit. Each motor shield consists of 5 address-select pins (via soldering) to provide an address for communication to the specific shield. By stacking multiple shields for the overall platform, each motor can be addressed by a combination of shield address and motor ID (1-4) on the shield. To power all the motors, programmable 2-channel external power supply unit (RS Components Ltd.) capable of supplying enough current to power all 133 DC motors were used. To simplify the nomenclature of the motors and addressing, four Arduino UNO boards, two each for cell stirrer and interfacial stirrer motors were used. These four microcontroller units allowed us to individually address 133 motors for the two-dimensional platform. Table S2 shows the number of motors, shields and prototype boards used for one- and two-dimensional setup. Each motor address is given by the combination of three addresses: {Arduino UNO ID, Motor Shield Address, Motor Number}. The TriContinent C3000 pump control uses daisy-chaining to connect with a physical address pin (0-15) on each pump. An in-house designed PCB for daisy-chaining pumps data and power pins were employed for the TriContinent C3000 pumps. A 24 V power supply from RS Components Ltd. were used to power the pumps.

**Table S2.** Electronic components required in the 1D and the 2D platform.

| Platform | No. of Arduino Uno | No. of proto shields per Arduino Uno | Total DC Motors controlled |
|---|---|---|---|
| **1D** | 1 x Active Cells + Interfaces | 2 x Active Cells<br>2x Interfaces | 7 x Active Cells<br>6 x Interfaces |
| **2D** | 2 x Active Cells<br>2x Interfaces | 13 x Active Cells<br>21 x Interfaces | 49 x Active Cells<br>84 x Interfaces |

### 1.2.3 Control software

The software layer that is responsible for controlling the platform can be divided into two parts, namely (i) The firmware that runs on all the Arduino UNO boards to activate all the individual motors with I2C communication and (ii) A Python script that communicates between the Arduino and Python via the serial interface as well as an in-house developed library to control TriContinent C3000 series pumps via a python script. The firmware for Arduino UNOs was written using opensource Arduino IDE and an available library from Adafruit Industries Ltd. were used to communicate with the stacked motor shields. A high-level interface python script was written to communicate with the Arduino via serial interface (using pyserial). This ensures a more intuitive way to program the experimental system by a researcher.

Adafruit 16-Channel 12-bit PWM/Servo Shield – 12C interface is employed to generate the PWM signals that would actuate the motors. By using the corresponding Arduino library: https://github.com/adafruit/Adafruit-PWM-Servo-Driver-Library, any users can send different commands for PWM signals. The function is as followed "setPWM (pin, direction, speed)", where pin refers to a unique motor and direction refers to the direction of the motor i.e. 0 is clock-wise and 1 is anti-clockwise and speed generate a PWM signal resulting in specific RPM.

## 1.2.4 3D-parts and hardware

Detailed computer-aided software (CAD) images with detailed descriptions of the specifications of the platforms can be found in this section. More information can be found in <https://github.com/croningp/BZComputation>. The main 3D printed parts are as follows:

- Experimental arena: Located at the top of the platform, where the BZ-solution is held, programmed, and imaged.
- Motors holder: Located at the bottom of the platform, where the DC-motors are fitted and held at the coordinates that mapped directly to the experimental arena.
- Columns and stand: Elevate and fix the platforms in a static position.
- Magnets holder: Act as an adaptor that mounts magnets and DC-motors together.

Note that the 3D designs were first created with OnShape and exported to Autodesk® Inventor® Student Edition to generate the figures. The various components used in the experimental setup are shown in Fig. S6– S15.

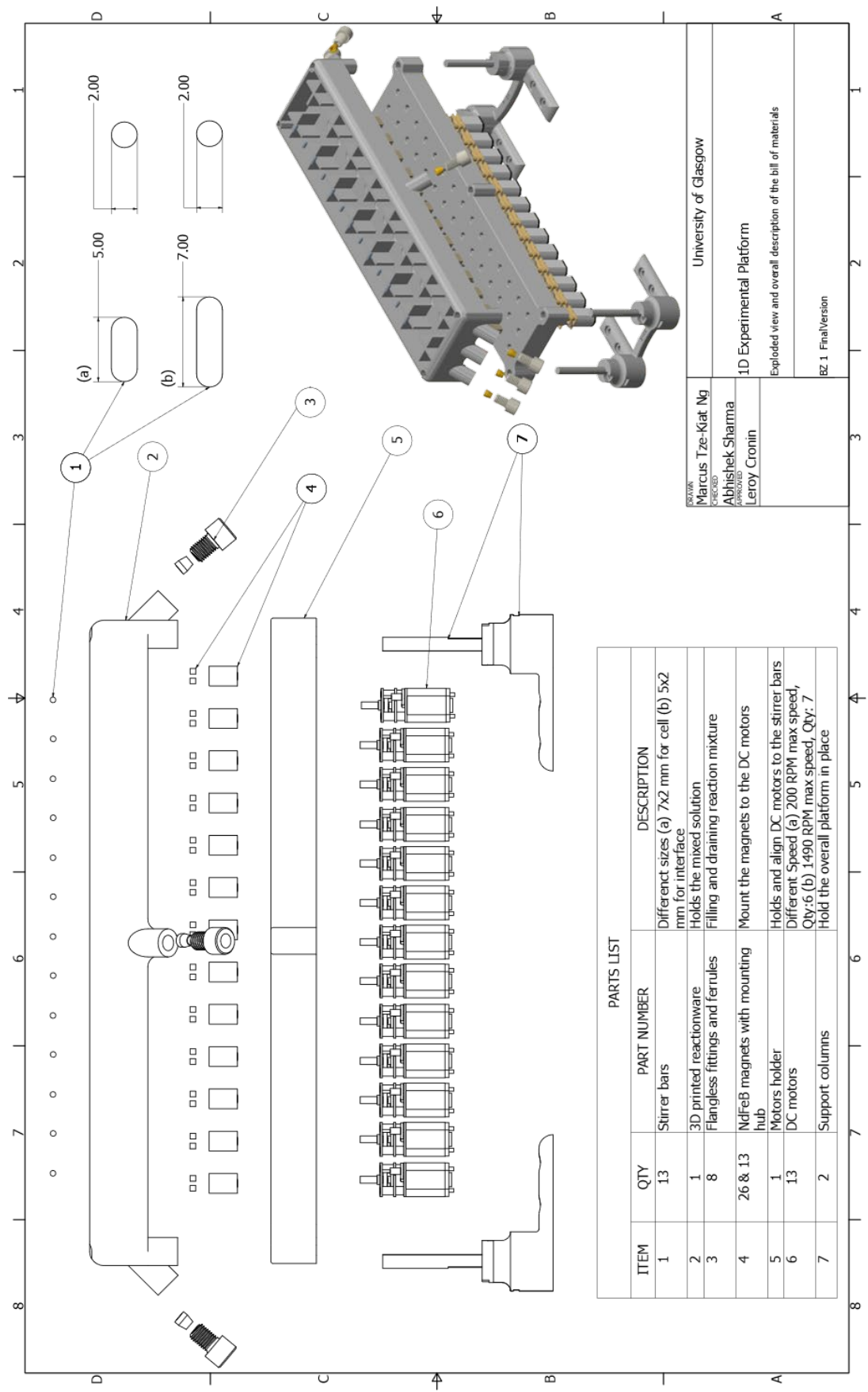

**Fig. S6:** Overall set-up of the one-dimensional platform with a bill of materials.

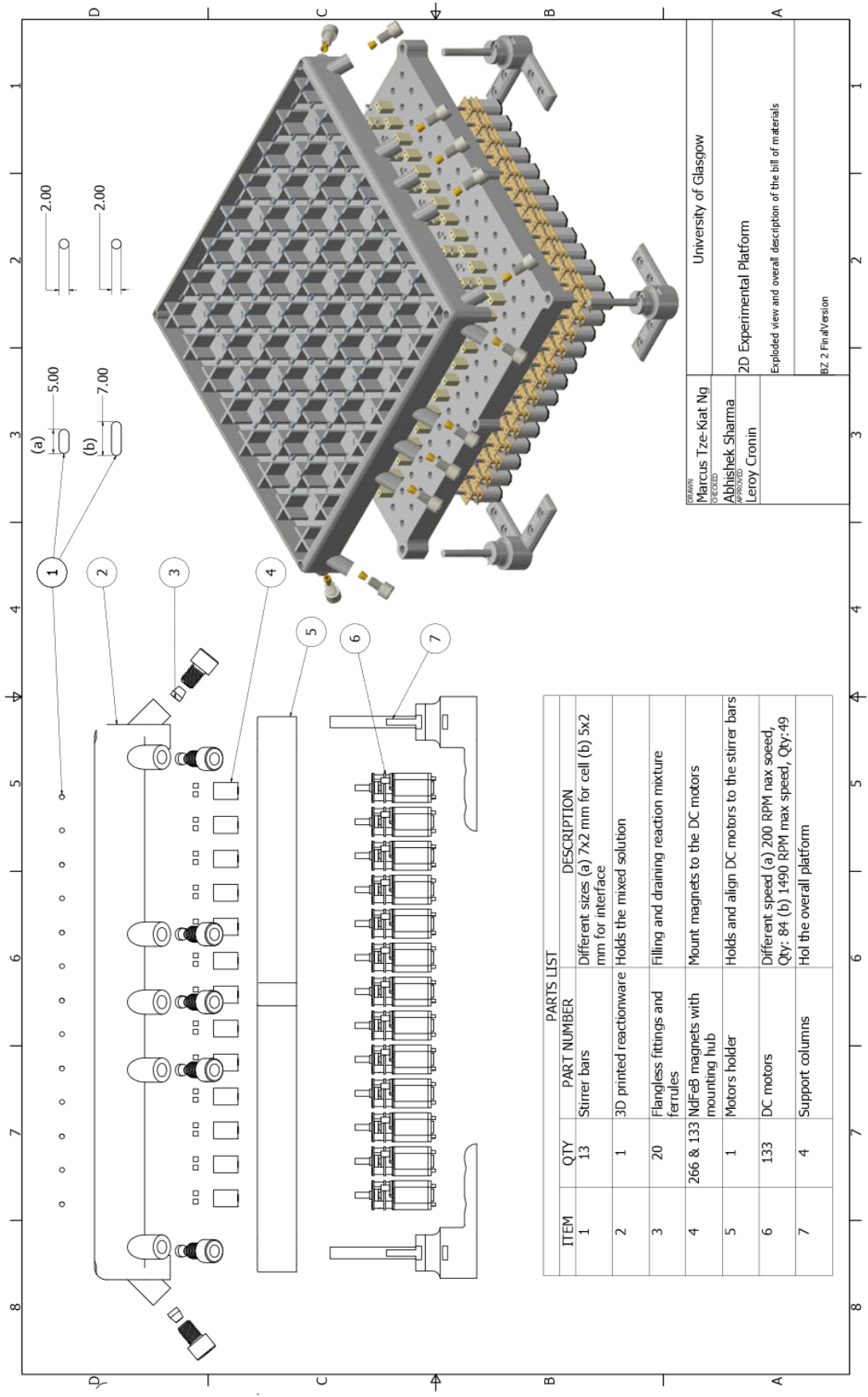

**Fig. S7:** Overall set-up of the two-dimensional platform with a bill of materials.

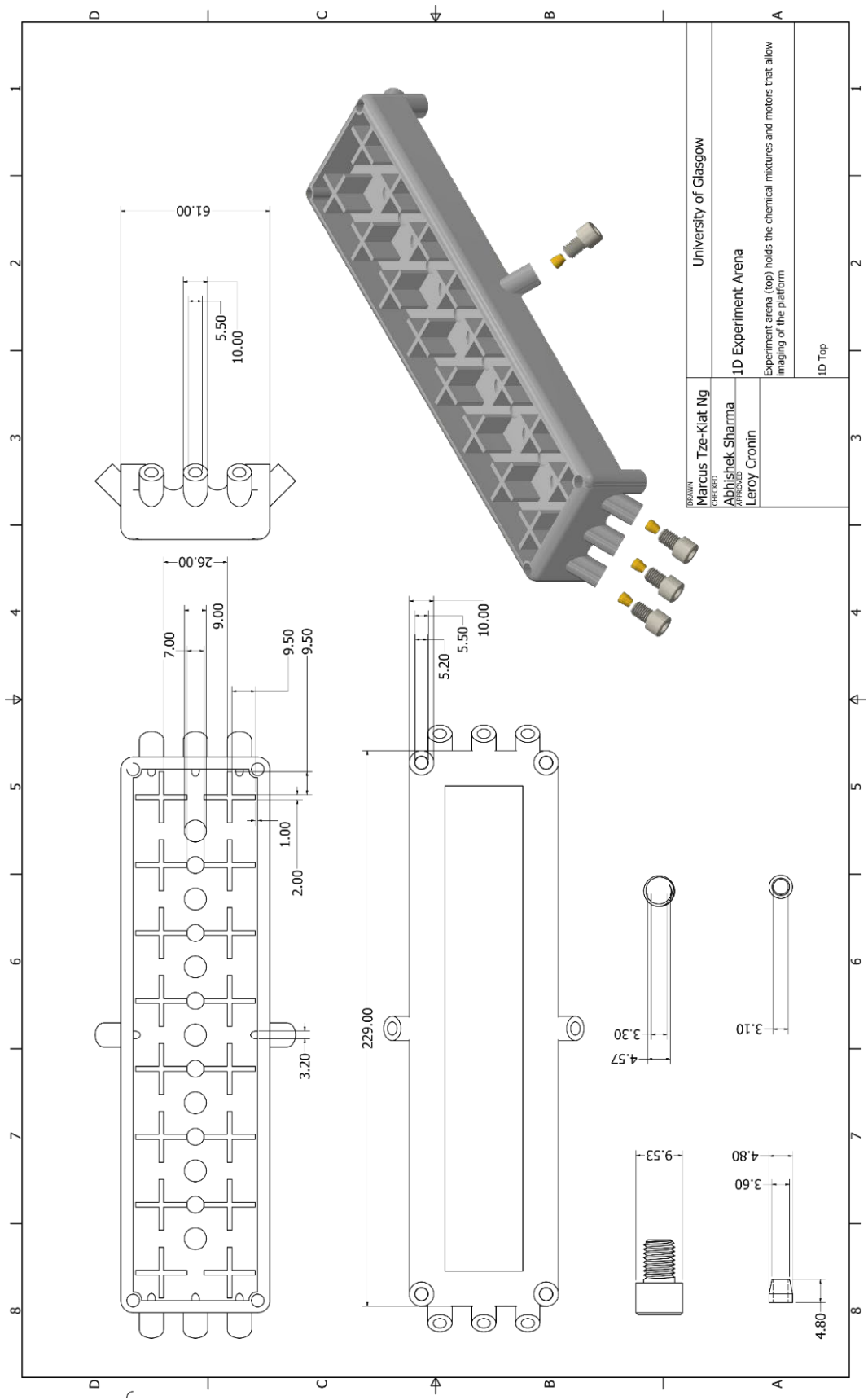

**Fig. S8:** One dimensional experimental arena with an interconnected network of reactors.

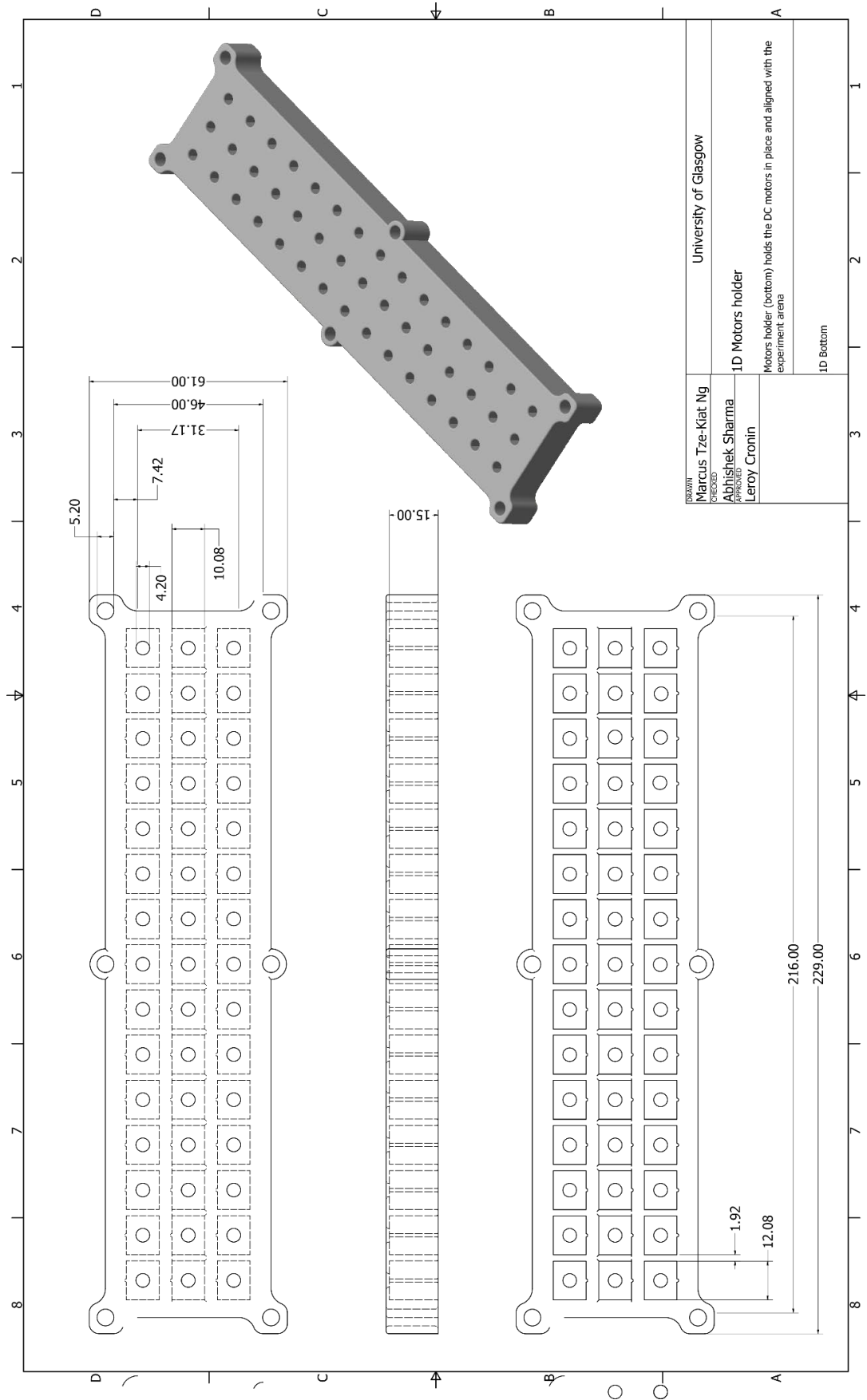

**Fig. S9:** One-dimensional setup for holding motors to map with reactor cells.

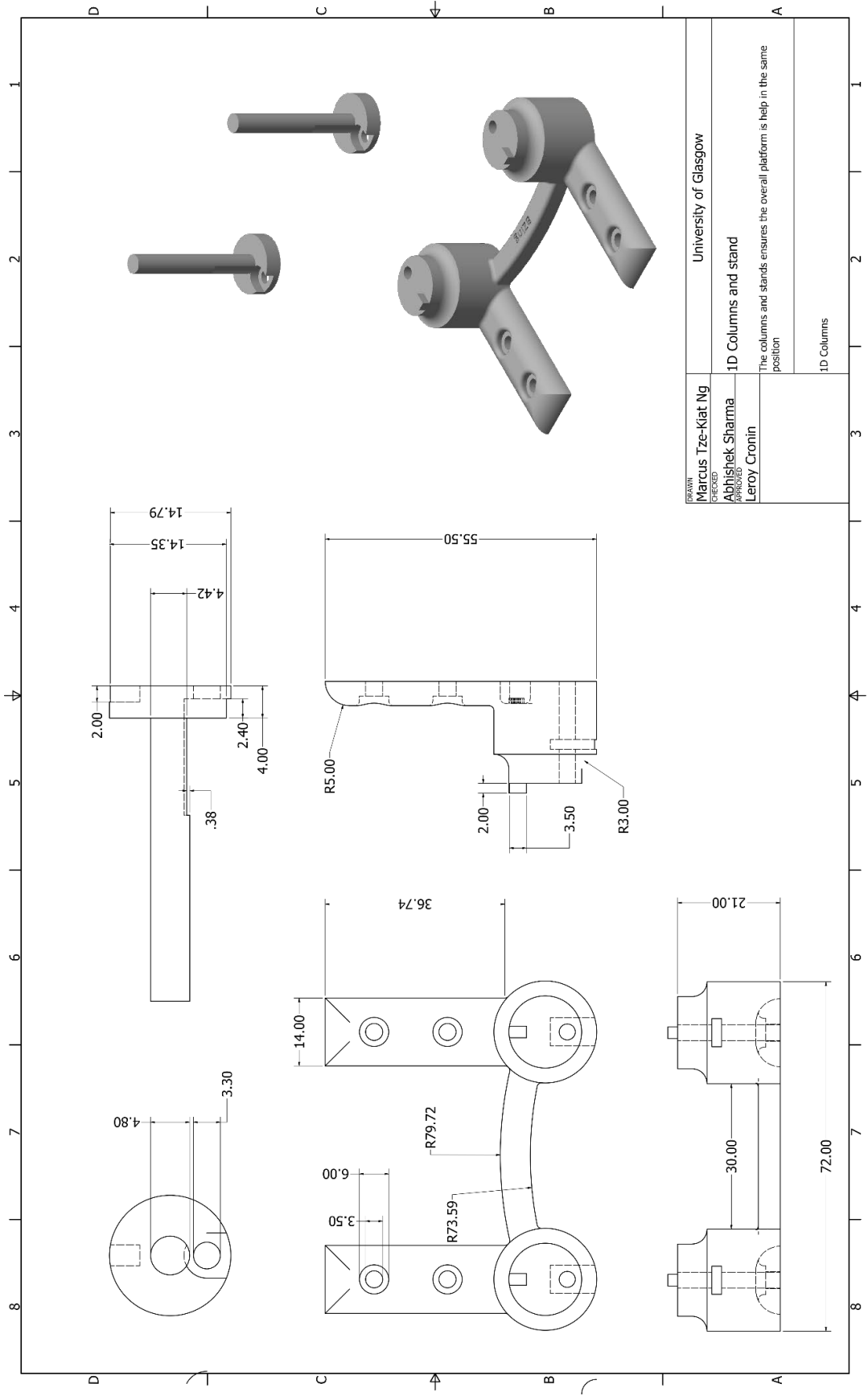

**Fig. S10:** Supporting column structure to hold the complete 1D experimental setup.

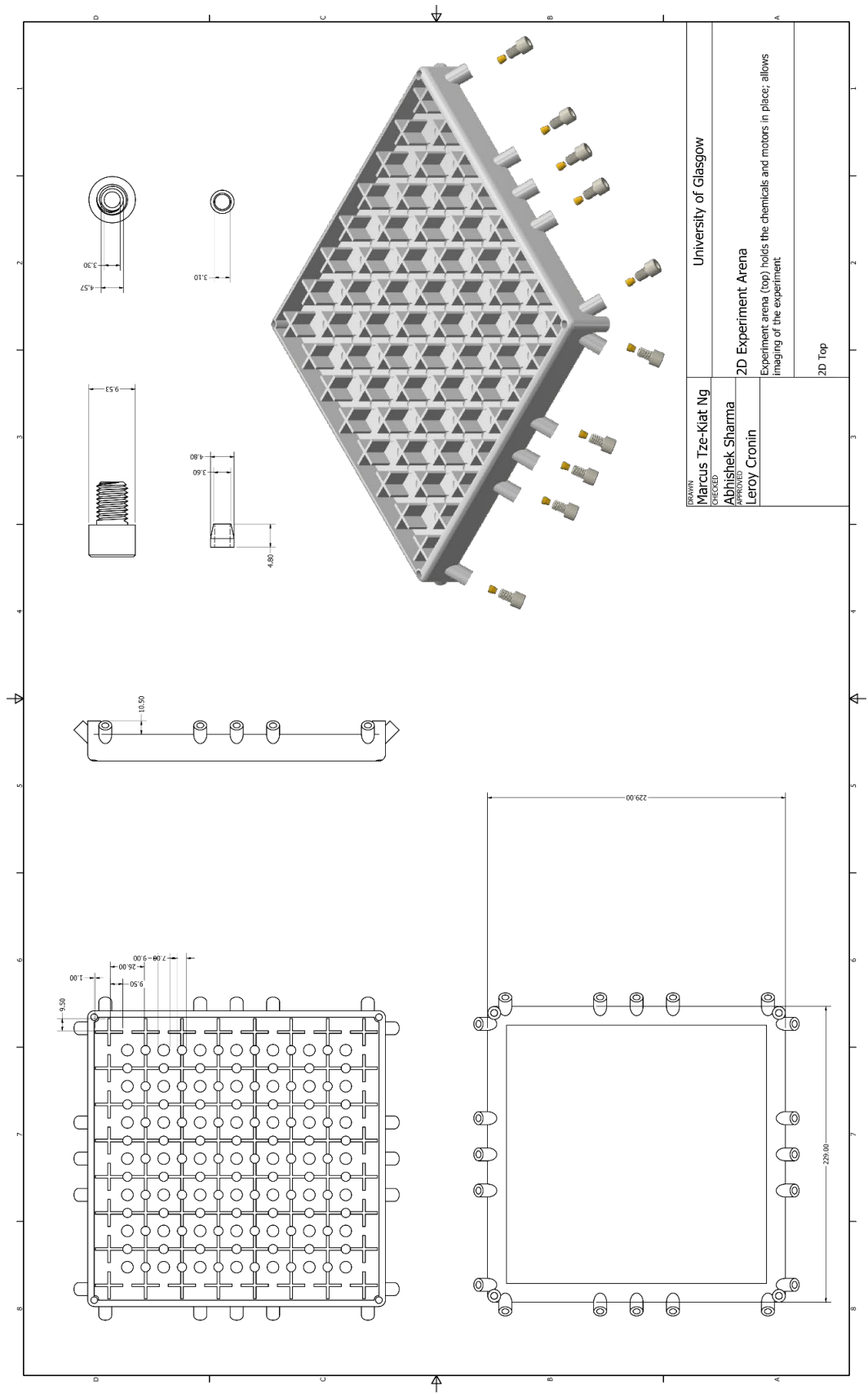

**Fig. S11:** Two-dimensional experimental arena with an interconnected network of reactor cells.

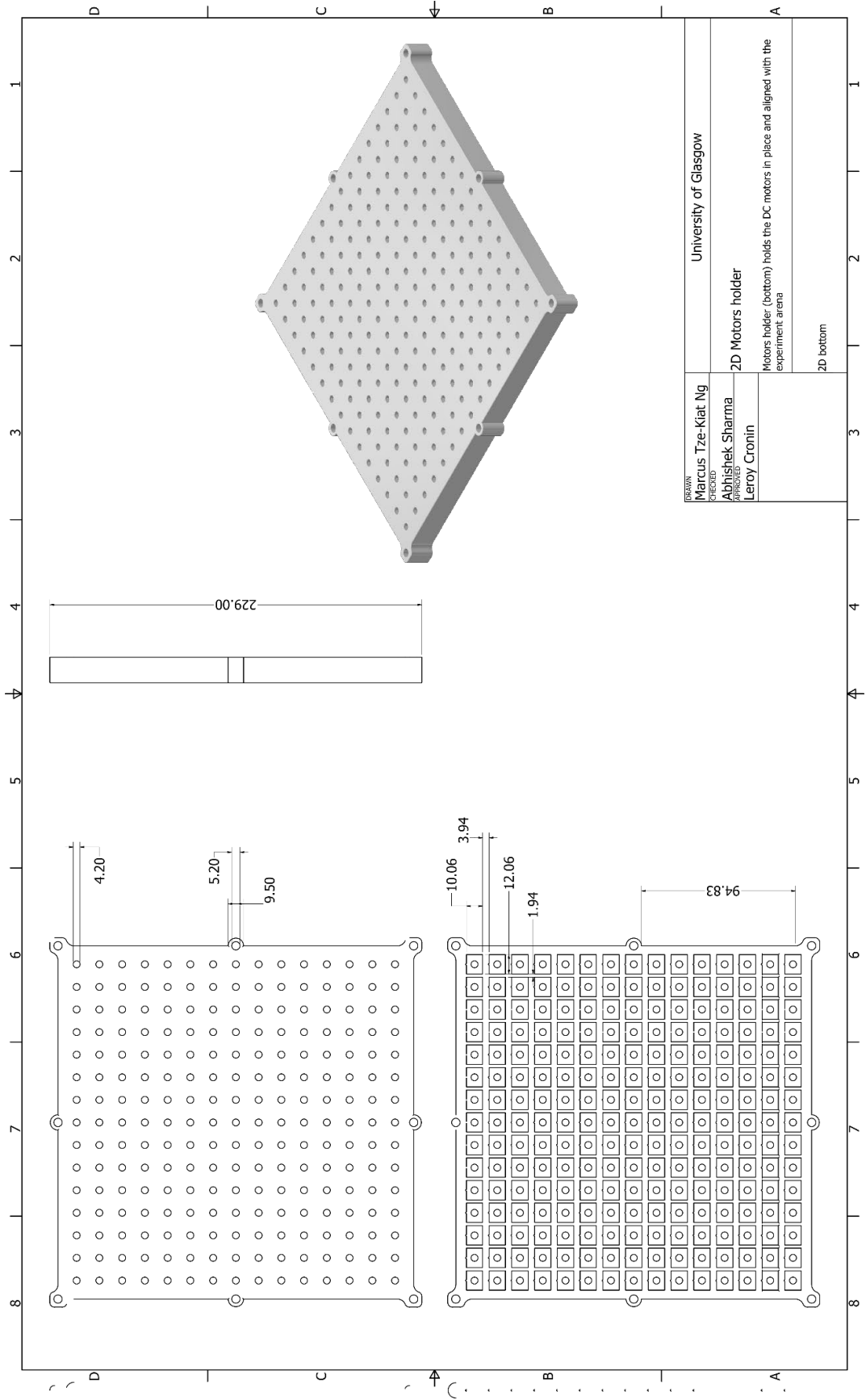

**Fig. S12:** Two-dimensional setup for holding motors to map with reactor cells.

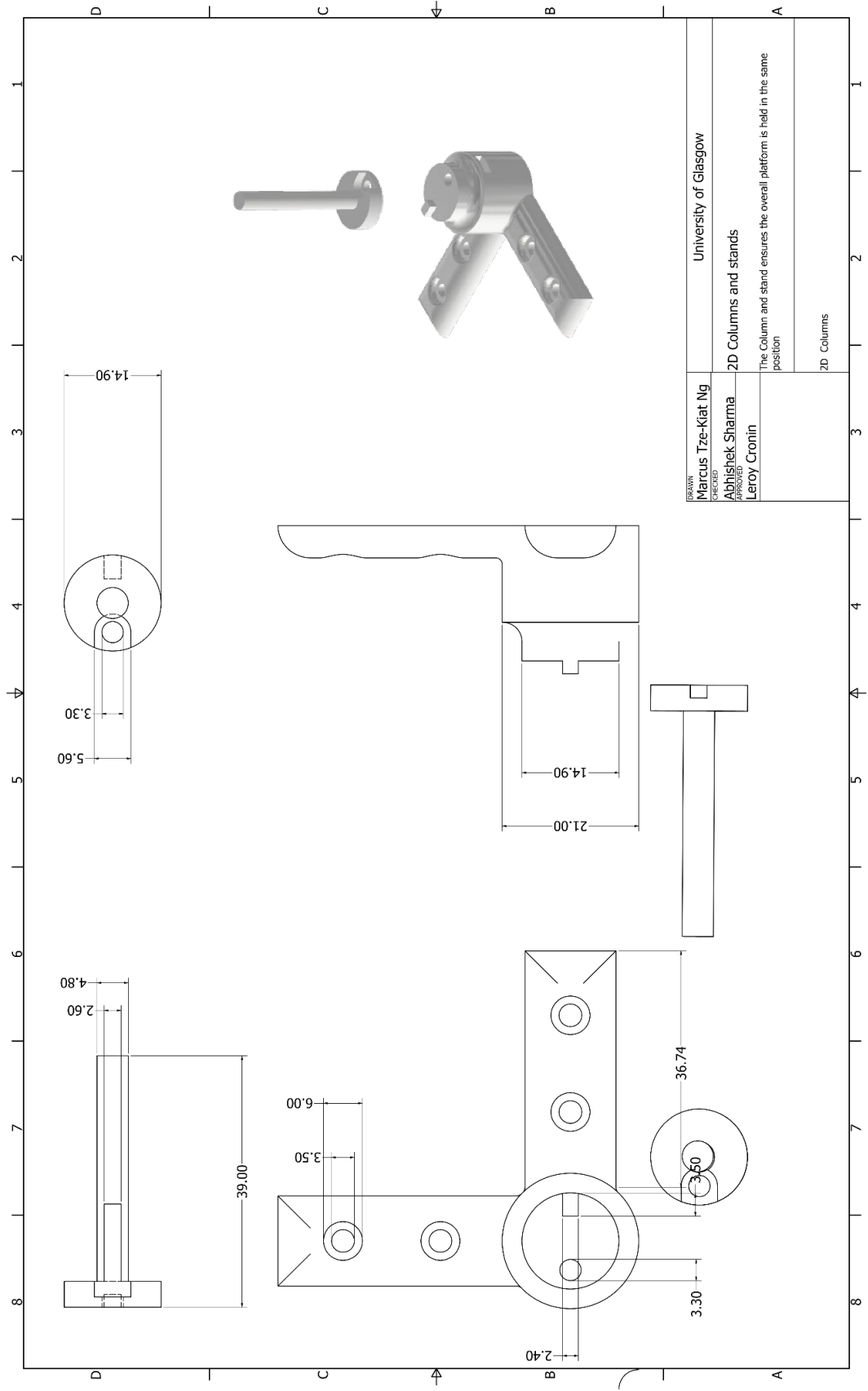

**Fig. S13:** Supporting column structure to hold the complete 2d experimental setup.

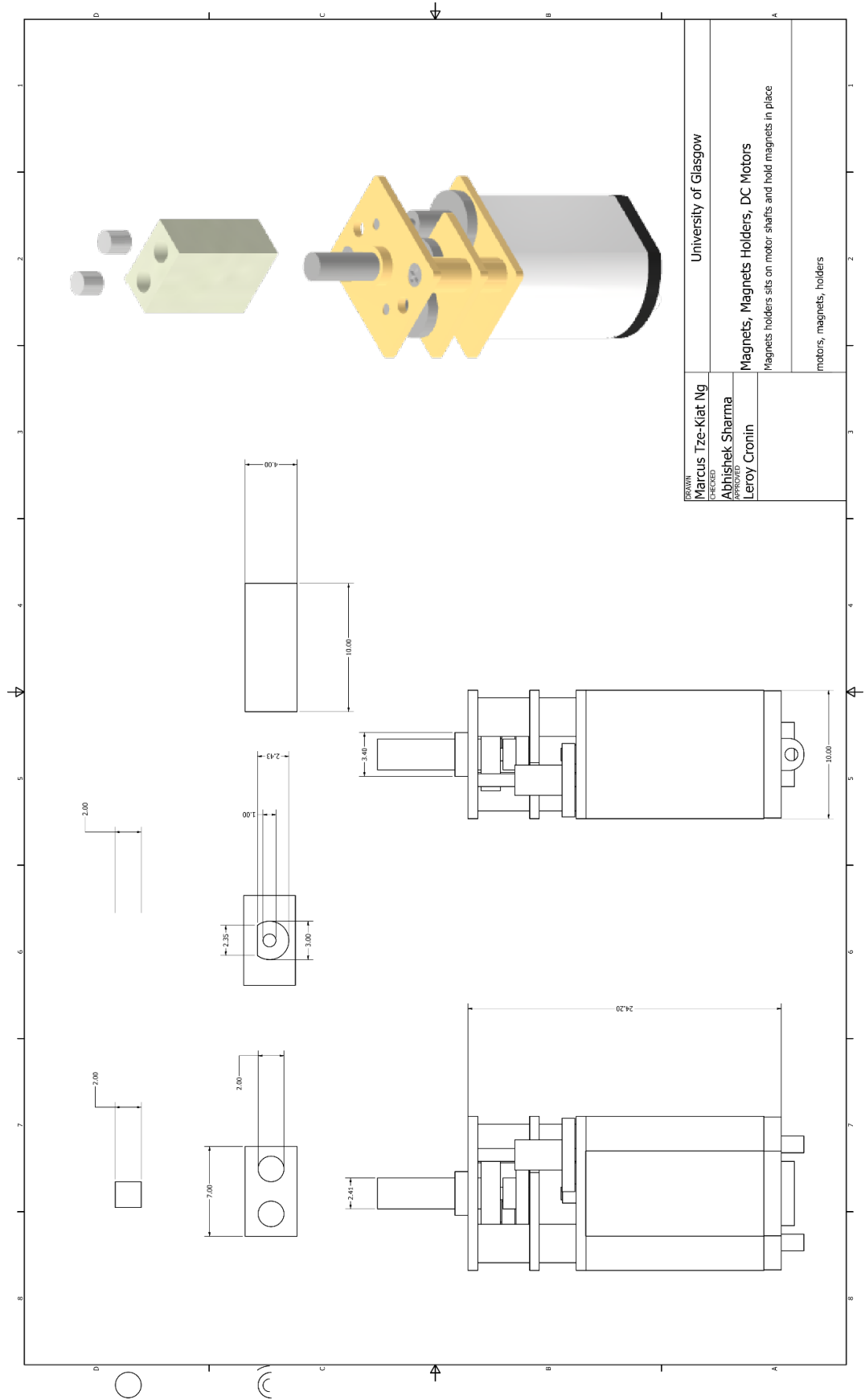

**Fig. S14:** Magnets, magnets holders and DC motors used in 1D and 2D platforms.

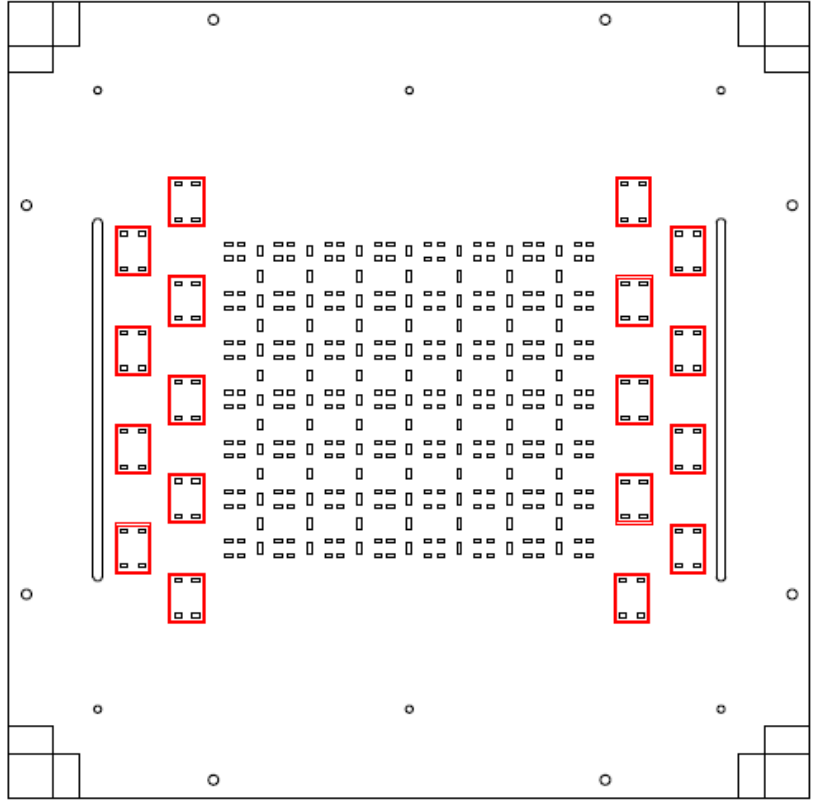

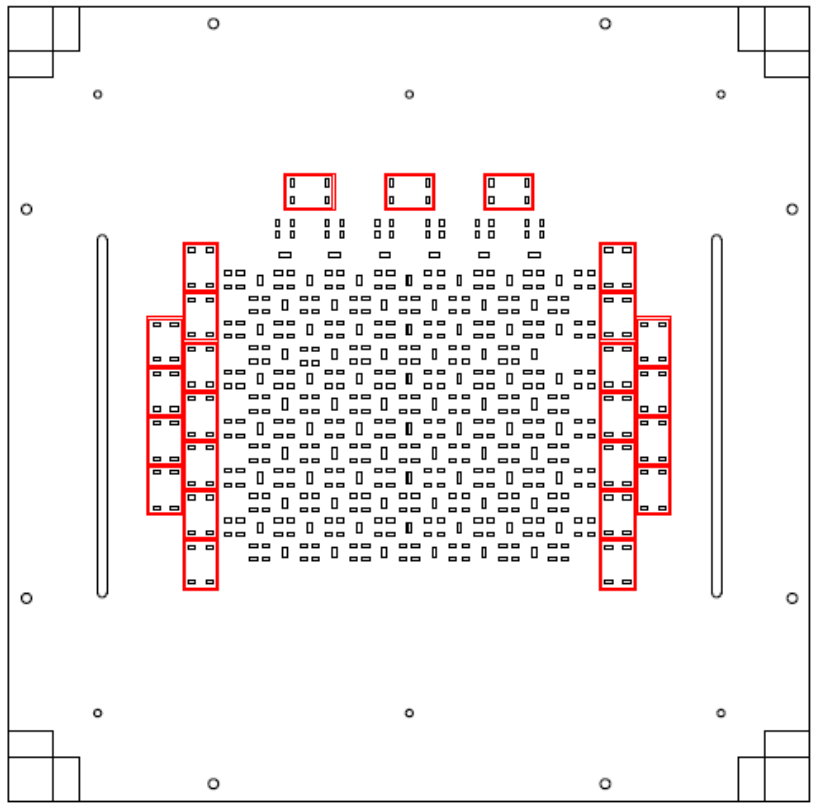

**Fig. S15:** Laser cuts for guiding wires from DC motors to respective motor shield (cell and interfacial motors).

## 1.3 Chemical oscillator

Amongst the oscillatory reactions, the Belousov-Zhabotinsky (BZ) reaction was chosen due to its robustness and easily accessible reagents. For instance, the BZ reaction can last up to an hour or more with consistent oscillations even in a closed system. This allows the system to be programmed and observed without the necessity of constant replenishment of reactants. Admitedly, continuous-stirred tank reactor (CSTR) system would allow lengthier experiment and a more controlled $O_2$ uptake may provide a more consistent period and aplitude of oscillation. Reports have shown the oxygen inhibition effect *(31)* which may change the behaviour of the system ever so slightly with the decarboxylation of malonic acid which cause $CO_2$ bubbles. However, we have opted to a batch reactor system due to its simplicity and we have implemented chemical clocking logic which would compensate any phase-drift and change in amplitude of the oscillation that may occur in the experiment which is discussed in Section 4. An upside of the closed system with 3D compartments, as opposed to CSTR can be attributed to the fact that the state of localized chemical oscillations and hysteresis effects can be preserved and could be utilized for implementing logic and computation. This localized memory-like effect provides the key towards the implementation of closed-loop dynamic logic and computation. The iron-catalysed variant (ferroin) of the BZ reaction was used in all the experiments. Ferroin was used as the indicator and a single-electron redox catalyst, which results in colour changes from red when the iron is in (+2) state and blue when the iron is in (+3) state as shown in the equations below.

$$Fe^{2+}(phen)_3 \leftrightarrow Fe^{3+}(phen)_3 + e^- \qquad (1)$$

$$Red\ (Reduced) \leftrightarrow Blue\ (Oxidised) \qquad (2)$$

The oscillation of the BZ reaction can be easily stimulated mechanically *(32)* and photochemically *(33)*. Mechanical stimuli via magnetic stirring were employed in this system due to its simplicity towards large-scale implementation and its capability to create a well-controlled local coupling between the nearest neighbouring cells.

The stock solutions for the automated platforms were prepared as followed:

- Ferroin: 0.1 M of the solution was prepared by dissolving 2.78 g of ferrous sulphate heptahydrate and 5.40 g of 1,10-phenanthroline in 10 mL of deionised water. The solution was then further diluted to 0.001 M for the experiment.

- Sulfuric Acid: 1.0 M of the solution was prepared by diluting 56 ml of concentrated $H_2SO_4$ to 1 L of deionised water.

- Potassium Bromate: 0.5 M of $KBrO_3$ solution was prepared by dissolving 83.5 g of $KBrO_3$ in 1 L of 1 M $H_2SO_4$.

- Malonic Acid: 1.0 M of the solution was prepared by dissolving 104 g of $CH_2(COOH)_2$ in 1 L of deionised water.

- Deionised water: PURELAB® Option-S/R 7/15 was used as the source of all water used in the experiments and preparation of stock solutions.

## 1.4 Experimental protocols

The complete schematic diagram of the single experiment protocol is shown in Fig. S16. A single experiment takes the following steps.

The following reagents were added sequentially via Tricontinent C-Series syringe pumps into the mixing chamber during the sequential chemical inputs stage:

1) 18 mL of water

2) 18.0 mL of 1.0 M malonic acid

3) 12.5 mL of 1.0 M sulfuric acid

4) 19.0 mL of 0.5 M potassium bromate in 1 M sulfuric acid

5) 2.5 mL of 0.001 M ferroin indicator

When the ferroin indicator was added into the reaction mixture, immediate colour change from red to blue was observed whilst the mixture was stirred. The reaction mixture was then pumped into the respective platforms; note that the volume of reagents mentioned above was used in the 1D platform and they were scaled up by four times for the 2D platform.

After the mixture was pumped into the platform, the mixtures went through a series of pre-experiment treatment using magnetic stir bars that were controlled by DC-motors as followed:

1) All the stirrers were activated at 50 PWM (max PWM: 255, 8-bit) for a minute

2) Deactivation of all stirrers and the solution was rested for four minutes

3) Activation of interfacial stirrers and pulsing of cell stirrers for five minutes

4) The initialisation of the experiment

The pre-treatment before each of the experiments as shown in the list above ensures the bulk oscillations of the BZ reaction appearing due to the filling step break down as well as the emergence of the synchronous periodicity of every cell before the start of the experiment. The initialisation of the experiment was based on the type of experiment that was carried out.

Once the experiment was finished, a series of draining and rinsing sequences were carried out on the platform to ensure most of the reagents from the current experiment goes to waste container. Upon cleaning, the platform either stops or performs a new experiment depending on the condition that was set before starting the experiment.

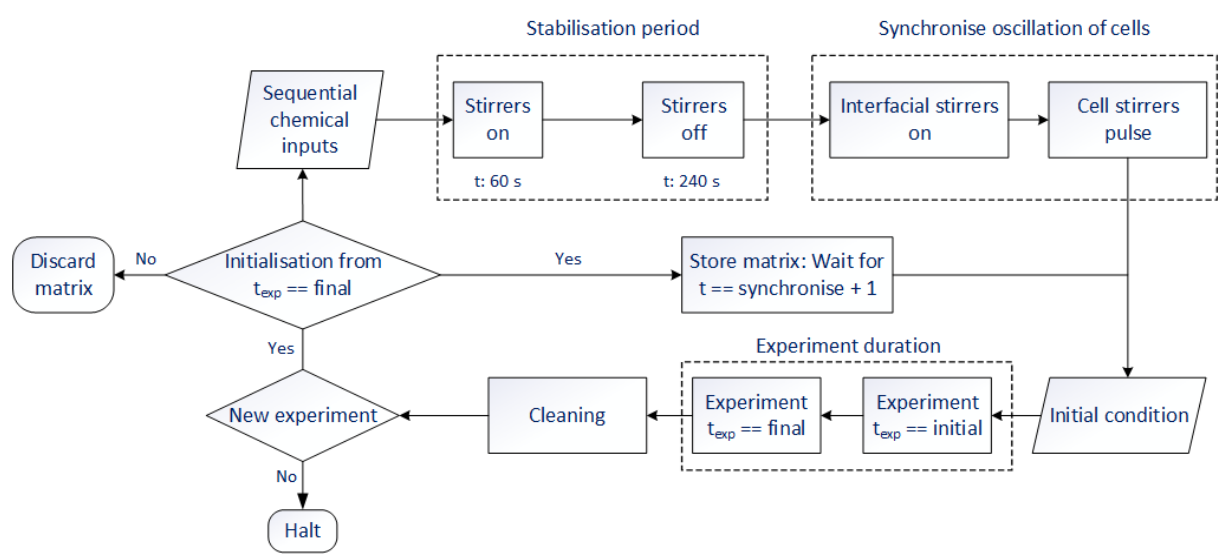

**Fig. S16:** The workflow of a single experiment. The schematic figure shows experimental protocols carried out in every experiment which consists of the stabilisation period, a synchronised oscillation in all the cells, actual experiment, cleaning, and a pre-defined logic to run the following experiments.

## 1.5 Camera and configurations

The 1D and the 2D platforms used Logitech C920 HD PRO Webcam. The webcam was situated 20.5 cm above the 1D platform arena and 33.4 cm for the 2D platform arena. These distances were chosen carefully to fill the complete field-of-view to get the best resolution. The video was configured to 1280 × 720 pixels and 10 frames per second (FPS) for the 1D platform while the 2D platform was configured to 800 × 600 pixels and 15 FPS. The camera was configured using the opensource GUVCView software with camera configuration details as given in Table S3. It is important to note that the selected parameters were chosen based on the

S24

light levels in the experimental enclosure such that oscillatory coloured patterns are distinct and clear for image processing. XVID compression was used in both of the platforms. During the experiment, the camera stream was fed into a running Python (3.7.1) OpenCV (3.4.4) script.

**Table S3.** Camera configuration of the 1D platform (left) and 2D platform (right): The configuration mode for all the camera settings were set to manual and fixed accordingly to ensure the lighting and colour of the oscillation produced in each experiment were consistent.

| Platform / Settings | 1D | 2D |
|---|---|---|
| Brightness | 145 | 155 |
| Contrast | 222 | 222 |
| Saturation | 200 | 255 |
| Gain | 140 | 85 |
| Power Line Frequency | 50 Hz | 50 Hz |
| White Balance Temperature | 2900 | 2900 |
| Sharpness | 128 | 128 |
| Backlight Compensation | 0 | 0 |
| Exposure Mode | Manual | Manual |
| Exposure | 300 | 400 |
| Pan | 0 | 0 |
| Tilt | 0 | 0 |
| Focus | 10 | 10 |
| Zoom | 120 | 125 |
| LED 1 Mode | Auto | Off |
| LED 1 Frequency | 0 | 0 |
| Frame Rate | 10 fps | 15 fps |
| Resolution | 1280 x 720 | 800 x 600 |
| Camera Output | RGB3 – RGB3 | RGB3 – RGB3 |

## 2 Control experiments

Multiple control experiments were performed to find the optimal parameters for controlling the stirrers to initiate strong or weak oscillations and localized nearest neighbour couplings between the cells necessary for the implementation of programmable hybrid electronic-chemical logic. This also includes the selection of appropriate dimension of individual reactor cells, stirrers' positions and wall dimensions to create ideal hydrodynamic coupling between the nearest neighbouring cells. Depending on the requirements, experiments were performed in one- and two- dimensional platforms. The details can be found in the following sections, 2.1 to 2.3.

### 2.1 Basic oscillation test and recording

As an initial step to estimate the time scale and stability of chemical oscillations in the closed system, basic actuation test by constantly stirring the cell stirrers up to an hour was carried out and the oscillations over all the cells in the one-dimensional platform was recorded. Constant oscillations over all the cells with a minor decrease in measured intensity up to one hour was observed. This provides the evidence for running each experiment up to an hour without changing chemical reagents. Fig. S17(A-C) shows stable oscillations with the same frequency and intensity over three neighbouring central cells of the one-dimensional platform which comprises of seven cells in total.

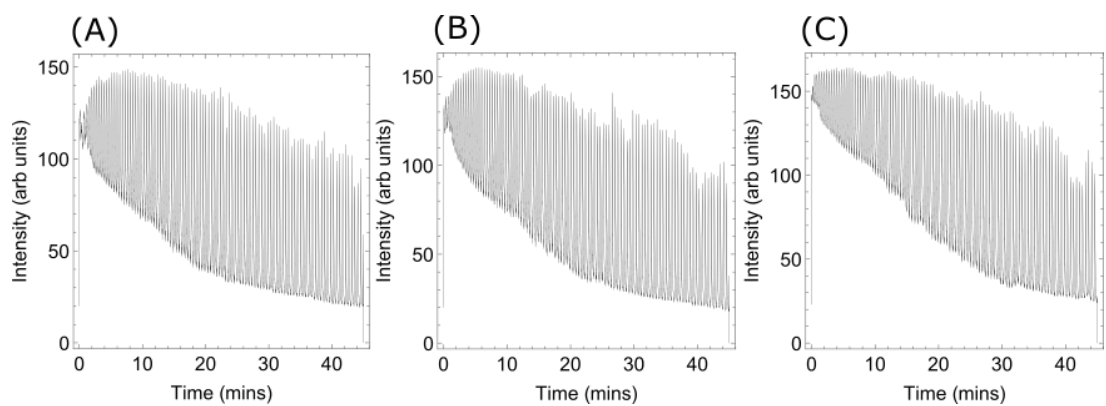

**Fig. S17:** Basic oscillations with constant stirring in a one-dimensional experimental setup. The figure shows BZ oscillations recorded over an hour on three neighbouring cells by actuating the individual cell stirrers at the same PWM level.

To prove that the effect of the cell stirrers is localized to the specific cell in the absence of the interfacial stirrers, experiments on the one-dimensional platform was carried out by actuating every alternate cell (1, 3, 5 & 7) up to 5 seconds and record the oscillations. The recorded oscillation data over all the seven cells is shown in Fig. S18, where activated cells (1, 3, 5, and 7) shows strong oscillations due to stirrer actuation as well as dampening effect when the stirrers were inactivated.

The other cells 2, 4, 6 shows almost negligible oscillations due to extremely weak interactions in the absence of hydrodynamic coupling when the interfacial stirrers were inactivated. This proves that actuation of only cell stirrers creates only local interactions and hence, can be used to program localized chemical states in a controlled manner.

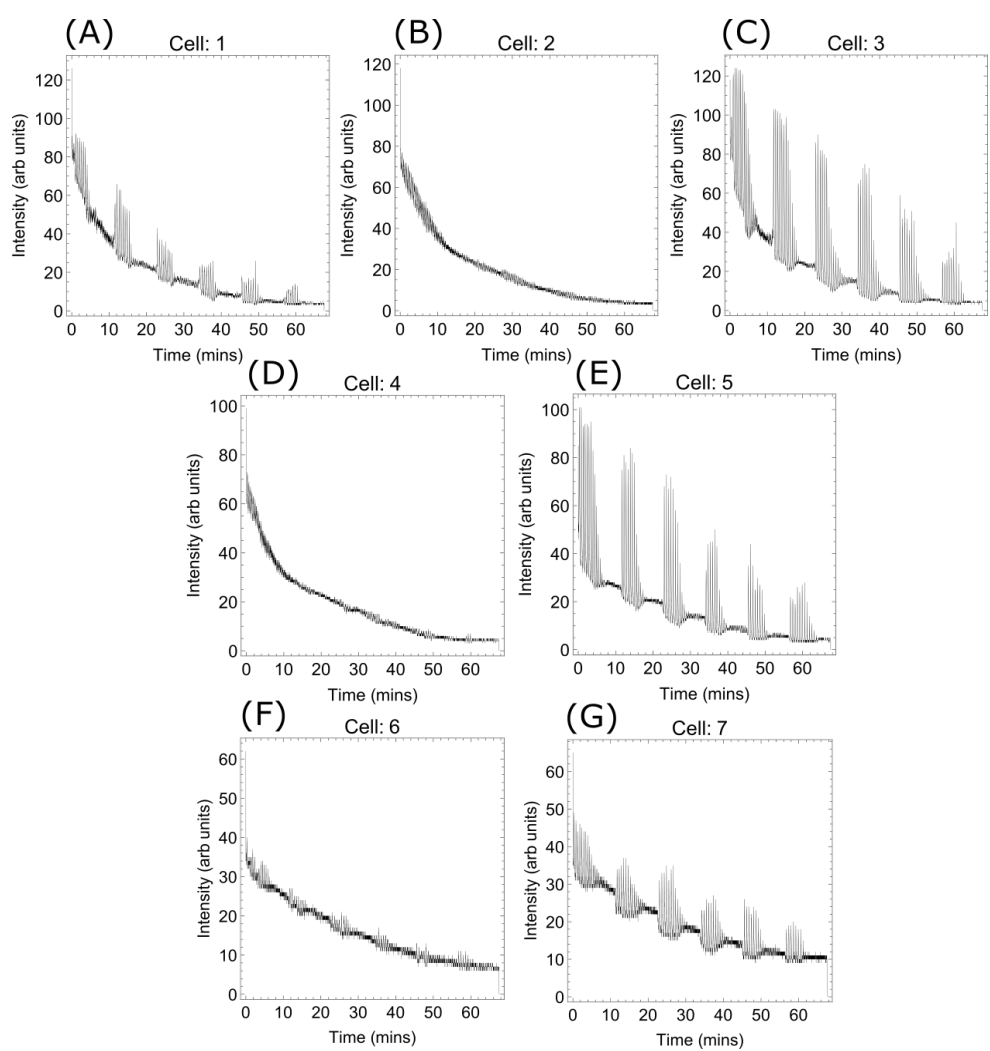

**Fig. S18:** Experiments to demonstrate actuation of cell stirrers is localized to the specific cell. The figure shows oscillations recorded over all the seven cells of the one-dimensional platform with cell 1, cell 3, cell 5 and cell 7 were activated with periodic actuation from cell stirrers.

## 2.2 Hydrodynamic tests with markers to test neighbouring cell coupling

To investigate the coupling between the neighbouring cells by activating interfacial stirrers, we performed hydrodynamic tests by using ink as a tracer to monitor fluid flow by tracking the colour change between the cells. The interfacial stirrer was placed symmetrically between the two cells, such that the interfacial stirrer couples the two vortices of the neighbouring cells where the cell stirrers were active. Utilizing this concept, the experimental setup was designed considering three important criteria,

The cells architecture was built based on three main criteria:

1. Vortex of the fluid must be localised within a single cell when the cell stirrer is turned on and the interfacial stirrer is inactive. The vortex in the cell must be contained.
2. On activating the interfacial stirrer when the two neighbouring cell stirrers are also active, the interfacial stirrer should be able to couple the two cell vortices causing neighbours to interact with one another.
3. The fluid flow at the interface should create a symmetric bidirectional flow between the two cells.

Fig. S19 shows a pictorial representation of the two neighbouring cells whose vortices do not couple with each other when interfacial stirrer is inactive and hence no coupling and symmetric bidirectional coupling occurs when interfacial stirrer is active. All the active cell stirrers rotate in the same direction and interfacial stirrers rotate in the opposite direction to the active stirrers.

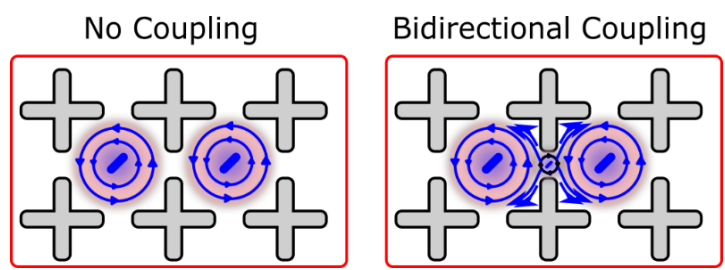

**Fig. S19:** Hydrodynamic coupling between two neighbouring cells with inactive and active interfacial cells.

In principle, it is possible to run Computational Fluid Dynamics (CFD) simulations to simulate the fluid flow over the network of cells with interacting vortices. However, CFD simulations with moving objects are usually computationally intensive. Hence, experiments using ink as a marker to monitor fluid flow were carried out. Diluted "Sumi" ink was used as a fluid marker to monitor the fluid flow.

As the first example, two neighbouring cell stirrers were activated at the same PWM level or RPM. It was then followed by the activation of the shared interfacial stirrer between the two cells. A drop of ink using a fine syringe needle was placed on the interface and the flow of the ink was recorded by a camera. Due to symmetric flow of fluid between the neighbouring cells, the ink flows symmetrically between two neighbouring cells and as soon as the ink boundary travelled close to the centre of the cell stirrer, it coupled strongly with the inner vortex in the cell and this led to homogeneous mixing in both cells, see snapshots in Fig. S20. So, by activating the interfacial stirrer, the two neighbouring cells with the same PWM values can be coupled with each other symmetrically which could be used for programming the couplings between the cells for computation experiments. Video of the described phenomena can be found in Supplementary Video 2.

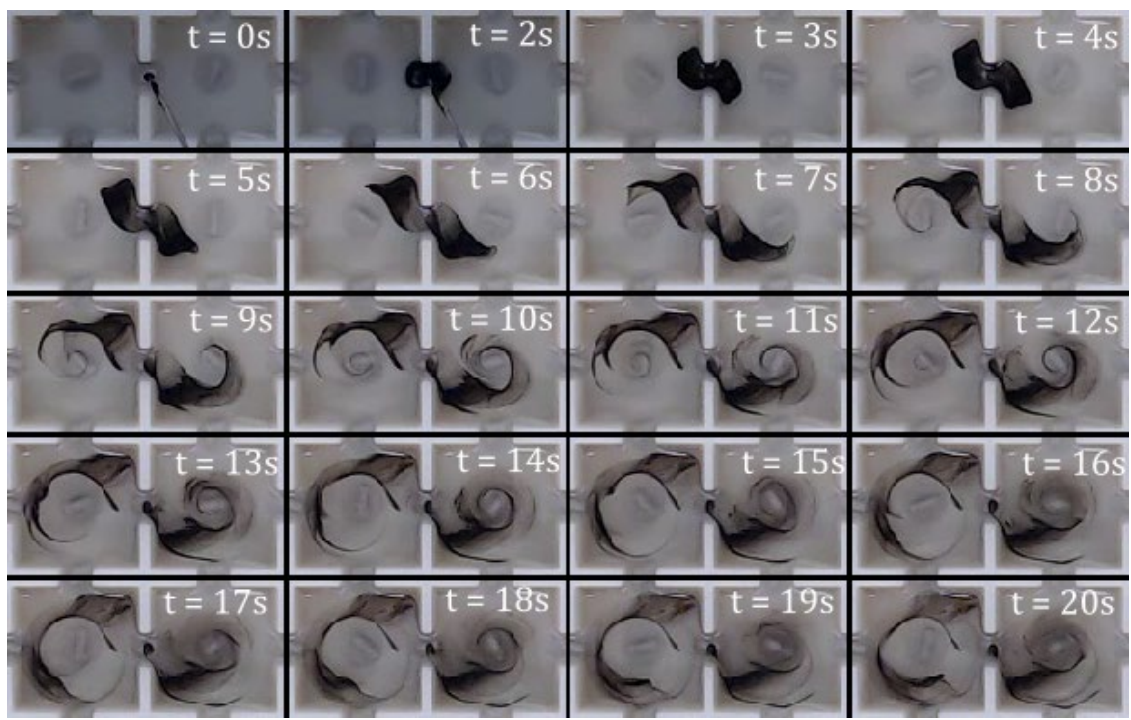

**Fig. S20:** Hydrodynamic tests towards symmetric coupling. Shows snapshots at different times demonstrating symmetric coupling between the two cells.

As an additional test, we performed a similar experiment with two active cell stirrers but with different PWM levels and activating the interfacial stirrer between them. Due to the difference in the speed of the cell stirrers, the cell-cell coupling becomes asymmetric as demonstrated in

Fig. S21. The PWM of the right cell is stronger as compared to the left cell. At the start, the coupling appears to be symmetric, but as soon as the ink boundary reaches strong vortex of the right cell, it starts mixing well throughout the cell, which was not observed in the left cell at the same time. The asymmetric coupling between the cells is much harder to quantify at different speed levels as compared to the symmetric coupling due to non-linear coupling between the interactions fluid flow within the cells and at the interface. Thus, symmetric coupling was used as the basis for computation experiments.

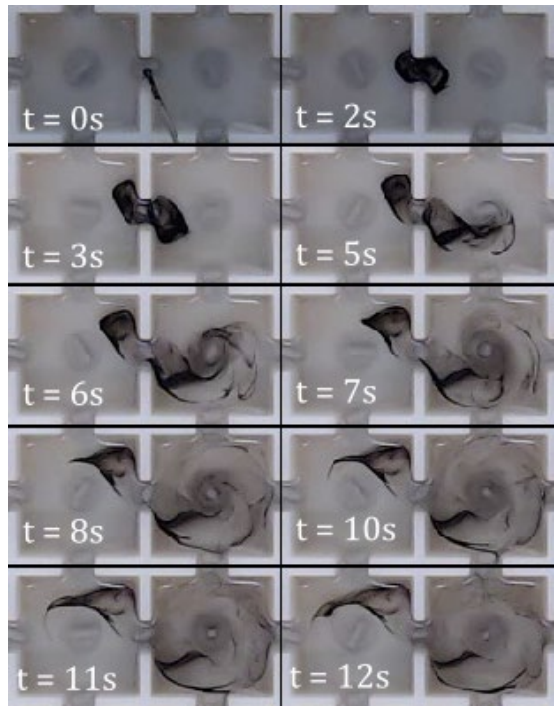

**Fig. S21:** Hydrodynamic tests towards asymmetric coupling. Shows snapshots at a different time demonstrating symmetric coupling between the two cells.

Additionally, hydrodynamic coupling tests in the two-dimensional platform were performed to demonstrate coupling between neighbouring cells occurs only when interfacial stirrers between them are active. To demonstrate this effect, a 3 × 3 grid of cells with only one set of cell stirrers and interfacial stirrer between them was activated. In this case, instead of dropping ink at the interface, a drop of ink was placed in the central cell and then activate the cell stirrer. Subsequently, we activate the interfacial stirrer between the central and the left cell. A series of snapshots in Fig. S22 shows mass transfer between the two active cells and as a result the concentration of the ink in the central cell decreases and increase in concentration on the left cell. No mixing was observed in the rest of the cells where interfacial stirrers were inactive, note that the apparent leakage observed on the right, top and bottom of the cell is due to the experimenter whilst injecting the ink.

These hydrodynamic tests gave a good indication about developing electronic actuation operations which could be used for creating a hybrid electronic-chemical logic. However, understanding the effect of stirring on the oscillation of BZ reaction together with the mass transfer is complex phenomena. In the later sections, further tests were carried out to investigate BZ oscillations and their coupling between cells. Based on these hydrodynamic tests, the optimized parameters for the cells and the experimental arena are shown in Table S4.

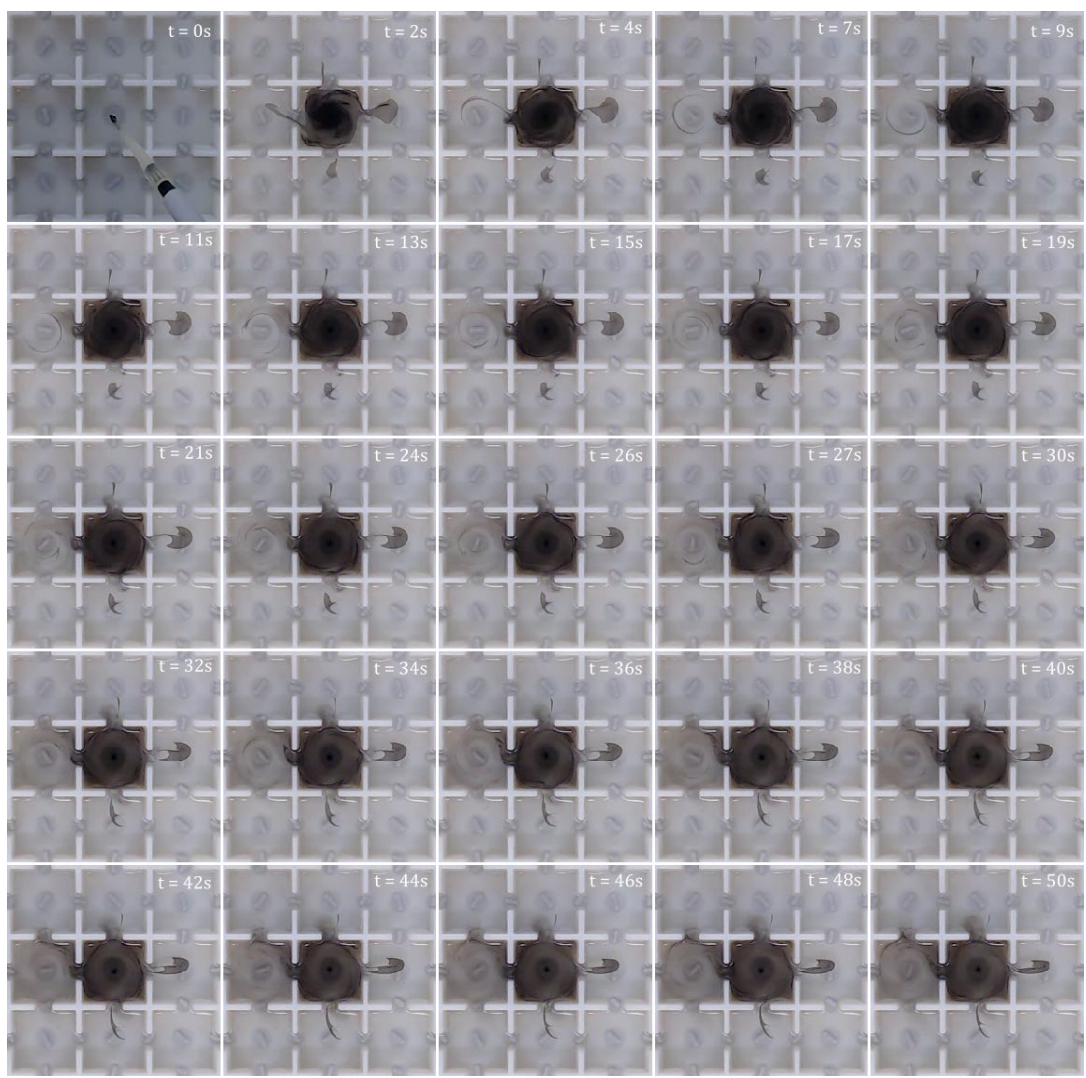

**Fig. S22:** Hydrodynamic tests towards programmable coupling. (A-Y) shows snapshots of the two-dimensional platform at different times showing the coupling between the centre and left cell by activating interfacial stirrer between them. No mass transfer was observed between other neighbours.

## 2.3 Chemical and hydrodynamic tests to investigate the effect of depth

In order the investigate the effect of the depth of fluid layer as compared to the size of the stirrer on the homogeneity of the chemical oscillations, we performed and characterized experiments on 3D printed single cell reactor with higher transparency (VeroClear). The 3D printed single cell reactor was designed with dimensions exactly similar to the single cell from the reactor array as described in Table S4 (26.0 x 26.0 x 13.5 mm). The oscillations were observed from the side by mounting a webcam horizontally. With the mechanical actuation i.e. stirring in this case, the emerging chemical oscillations appeared to be homogeneous throughout the majority of the cell in both horizontal and vertical directions. The temporal snapshots of the chemical oscillations with a time interval of 10 seconds are shown in Fig. S23.

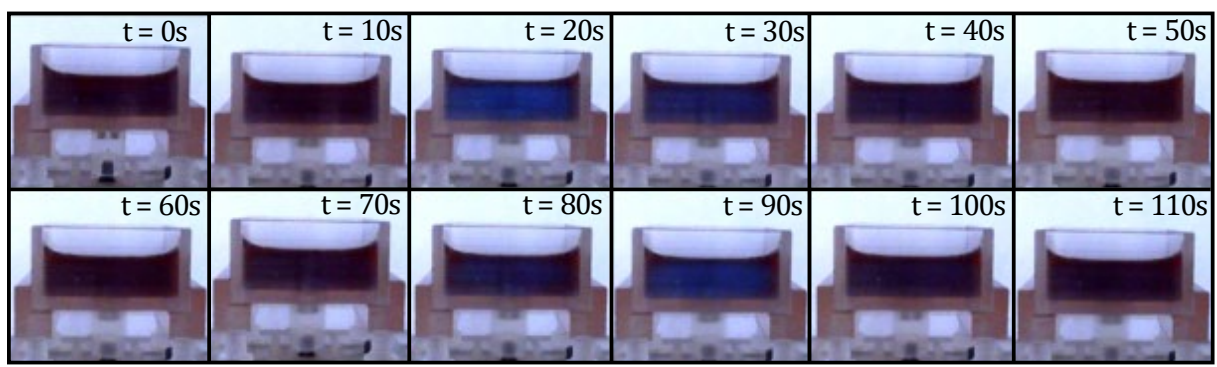

**Fig. S23:** Snapshot of the cross section of the BZ oscillation at different time points.

Inhomogeneities in colours at the edges and boundaries are expected due to lack of fluid flow which is constrained by the no-slip boundary conditions of the fluid-solid interface. However, during the image acquisition and its analysis using CNN, the observed colour at the edges/boundaries were not considered. This confirms that the fluid profile of the vortex created by actuation of the stirrer is homogeneous along the vertical axis over the complete depth across the observational domain.

A further experimental investigation was performed to visualize the temporal pattern of the emerging fluid vortex on the stirrer actuation. Initially, the cell was filled with water with ink (Sumi) as a fluid flow marker, was carefully injected at the bottom of the well. Upon activation of the stirrer, the ink disperses quickly and creates a steady state profile along the complete cross section. The temporal snapshots of the profile create by ink dispersion along the cell is shown in Fig. S24.

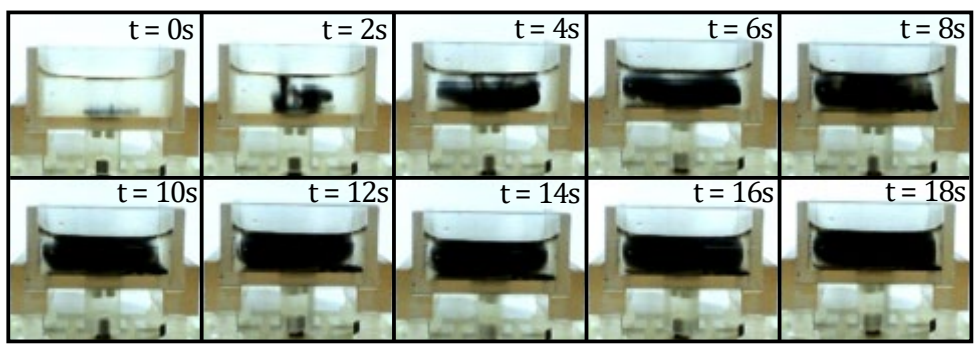

**Fig. S24:** Snapshot of the cross section of the vortex induced by cell stirrer at different time point, tracked by patterns of the ink (Sumi).

The temporal snapshots as shown in Fig. S24 demonstrates the homogeneous mixing of the ink in approximately 12-14 seconds. The no fluid flow zone at the edges of the cells can be seen clearly at times 12-18 seconds. This no flow zone at the edges helps to avoid the interactions between next-nearest neighbours. The interactions between the nearest neighbours occurs due to the coupling of vortices due to interfacial stirrers and the no fluid flow zone minimizes the coupling between diagonally placed cells. Fig. S25 (A, B) pictorially shows the permitted pair-wise interactions and forbidden interactions leading to complete fluid flow loops which is the basis of the design criteria.

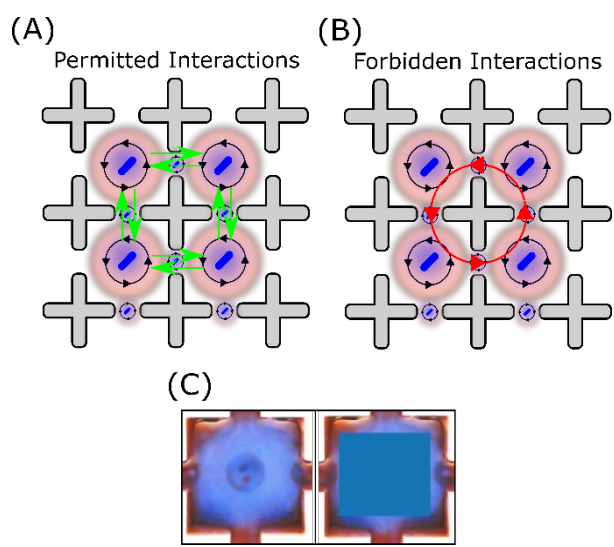

**Fig. S25:** Interaction between neighbouring cells via hydrodynamic coupling. (A) Permitted bidirectional interactions between the neighbouring cells. (B) Forbidden interactions involving next nearest neighbours due to loops in fluid flow. (C) Left: Chemical oscillation of a single cell imaged from the top showing no oscillation in no flow zones at the edges, Right: the region of interest detected by the CNN (indicated by the blue square showing that CNN detected Blue colour).

Due to the presence of no flow zone at the edges, no oscillations occur at the edges which leads could lead to inhomogeneity in the measurement. However, the region of interest detected by the CNN is smaller than the dimension of the cell such that the measurement does not suffers from the spatial inhomogeneity (see Fig. S25(C)). The details of the data preparation can be found in Supplementary Information Section 3.1.

**Table S4**. Parameters of the optimised cell design.

| Cell dimension | 26 x 26 mm |
| --- | --- |
| Cell wall thickness | 2 mm |
| Cell stirrer's rotation | Anti-clockwise |
| Interfacial stirrer's rotation | Clockwise |
| Size of the magnetic stir bar for the cells | 7 x 2 mm |
| Size of the magnetic stir bar for the interfaces | 5 x 2 mm |

## 2.4 Effect of stirring rate on chemical oscillations

In this section, we investigate the effect of stirring rate i.e. PWM levels on the observed amplitude of the chemical oscillations. We used one-dimensional BZ platform and scan the PWM levels in the range [20, 80] with the interval of 2 unit in forward direction and waiting for 40 seconds at each PWM value. Fig. S26 shows the observed chemical oscillations scanning in the forward direction. We observed continuous increase in the amplitude of oscillations with increase the in stirring rate. At the lower PWM levels, the rate of increase in amplitude is slow and weaker oscillation persists in the PWM range 20-30. In the range 30-60, the oscillation amplitude increases at higher rate which eventually saturates in the range 60-80. This transition in amplitudes from the lower to higher PWM levels allows us to select from a range PWM levels to define programmable states based on temporal oscillation patterns as discussed in detail in the next section.

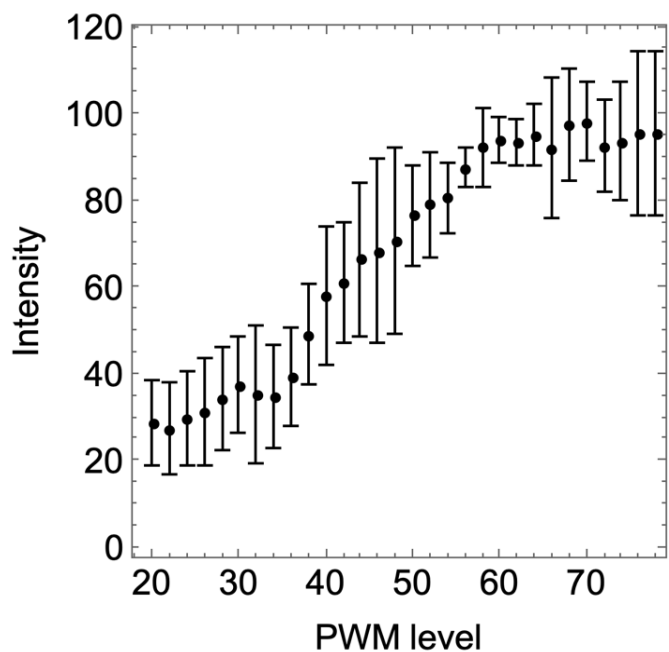

**Fig. S26:** Peak intensity of the oscillation vs stirring speed defined by PWM levels. Figure shows peak mean and deviation of measured peak amplitude of oscillations based on two different experiments vs. applied PWM level on single cell.

## 2.5 Phenomenological behaviour of the hybrid electronic-chemical system

Besides the optimization of the cell dimensions and enhancing control of the cell-to-cell coupling, the oscillator chemistry plays a significant role in the observed spatiotemporal behaviour of the overall experimental platform. In this section, we will investigate the dynamics of chemical oscillations within a single cell as well as coupling dynamics between the nearest neighbouring cells due to the actuation of the cell and the interfacial stirrers. The different behaviours observed were later used to define hybrid electronic-chemical logic for dynamic closed-loop experiments for implementing cellular automata and computation. The various phenomenological behaviours as described are as followed:

1. Single cell
    a. Chemical oscillations only appear when the cell stirrer is active which is equivalent to a forced oscillator.
    b. Once the active cell stirrer is deactivated, the forced oscillator turns into a damped oscillator and after a few cycles, the chemical oscillations disappear.

2. Coupled neighbouring cells

a. Neighbouring cells do not interact with each other when the interfacial stirrer is inactive.
   b. On activating interfacial stirrer, weakly coupled neighbouring cells come in phase independent of their initial phase differences.
   c. Interactions between two or more cells in one and two-dimensional geometry are confined to nearest neighbours only.

## 2.5.1 Chemical oscillator as a forced and damped oscillator

BZ chemical reaction oscillating under the actuation of a cell stirrer and falling back to the ground state once the stirring actuation is turned off can be described as a combination of a forced and a damped oscillator. As soon as the cell stirrer is active, the system reaches an active oscillating state in around 1-2 oscillations. However, it takes at least three oscillations to come back to the ground state once the cell stirrer is deactivated. This damped oscillatory behaviour can be used as short-term localized chemical memory for developing key components of the hybrid electronic-chemical logic such as clocking signal, self-interactions etc. To estimate the time scale of the dampening effect, we performed experiments on the one-dimensional experimental setup by activating the cells for a finite amount of time and followed by deactivating them. The oscillations over the complete setup were recorded over the whole experimental time.

Fig. S27 shows the oscillations of three different cells and the position of the peaks with time. In all three cells, it only took time scale of one oscillation *ca.* 30-40 seconds to reach the forced oscillation state from the ground state, and around 3-5 oscillations to arrive ground state once the cell stirrer was deactivated.

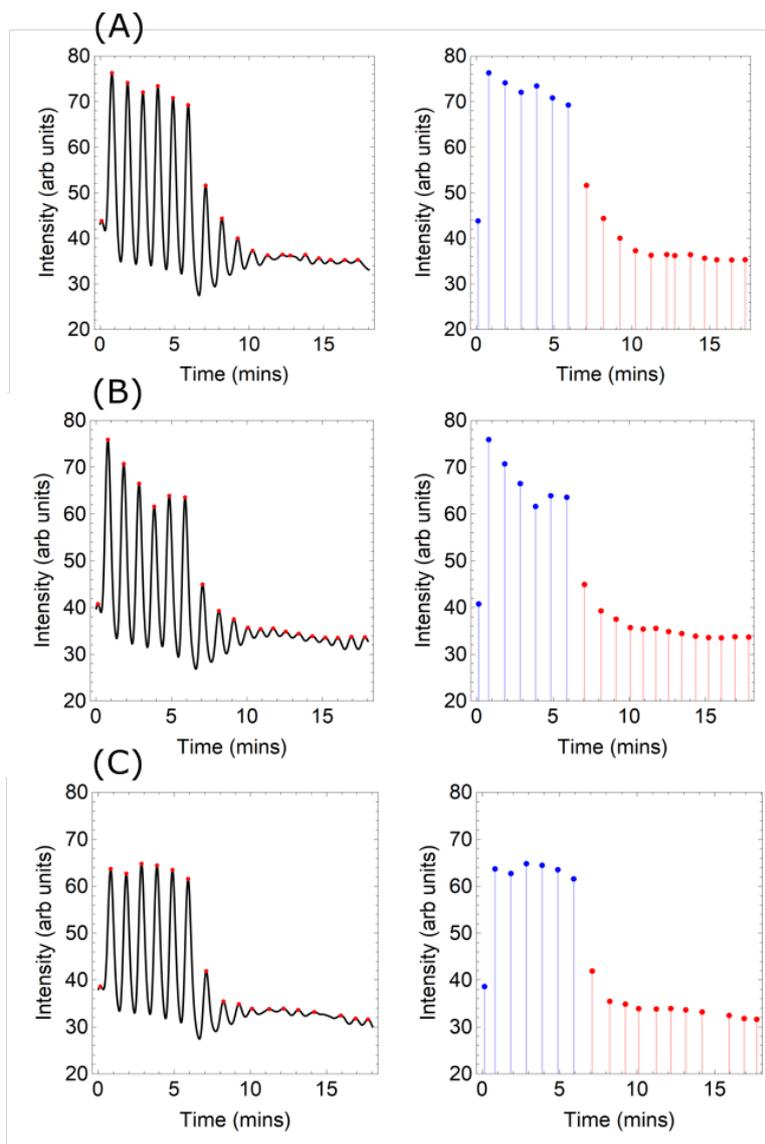

**Fig. S27:** BZ Oscillator as a forced-damped oscillation. (A-C) The left column shows the actual oscillations detected from the camera and the right column shows the peak intensity for all individual peaks detected when the cell stirrer was active (blue) and turned off (red).

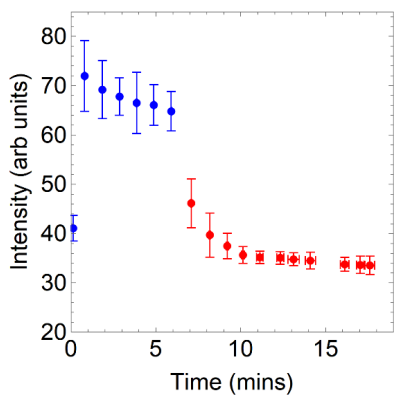

**Fig. S28:** Average Intensity of oscillations of the cells showing forced-damped oscillator. The figure shows oscillation peaks when cell stirrer is turned on (blue) and dampening oscillation when cell stirrer is turned off (red). The damped oscillation characteristic time scale is *ca.* 1.5

mins, which acts as a short-term memory for the feedback control experiments including cellular automata and computation.

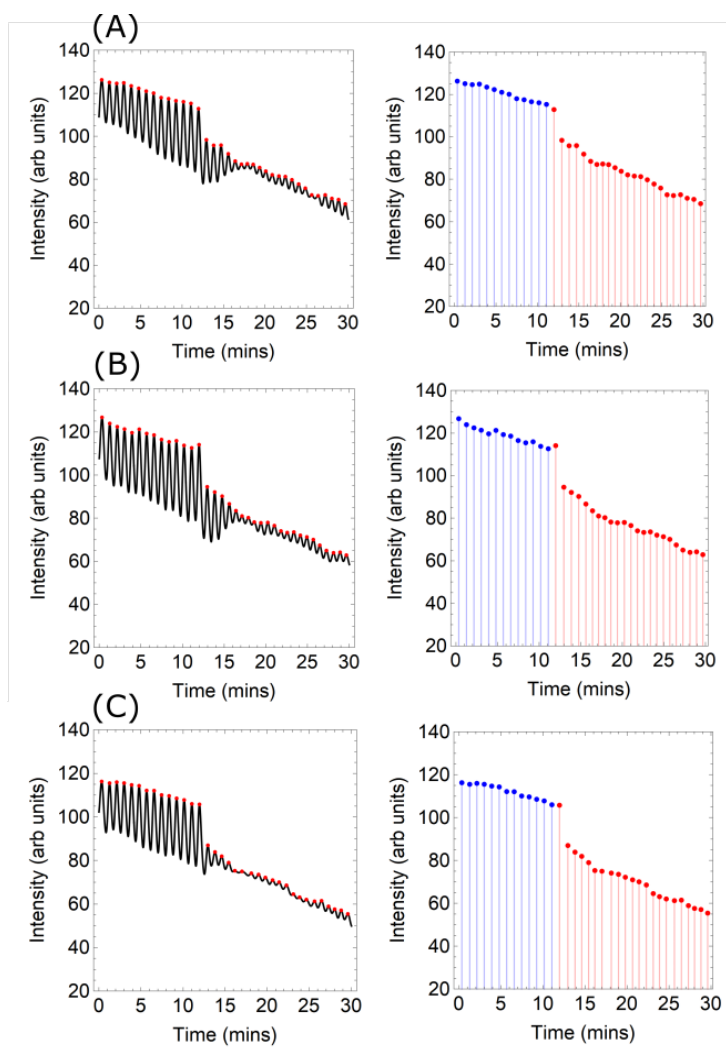

**Fig. S29:** BZ Oscillator as a forced-damped oscillation for a longer time scale up to 30mins. (A-C) The left column shows the actual oscillations detected from the camera per and right column shows the peak intensity for all individual peaks detected when the cell stirrer was active (blue) and turned off (red).

From the experimental data presented in Fig. S27, the average oscillation intensity and the time of the peak position together with the standard deviations as shown in Fig. S28. The position of the peak of all the oscillations seems well-defined and the characteristic timescale for damping can be estimated. Additional experiment for a longer time scale and another set of results for three cells in one-dimensional setup is shown in Fig. S29 and Fig. S30.

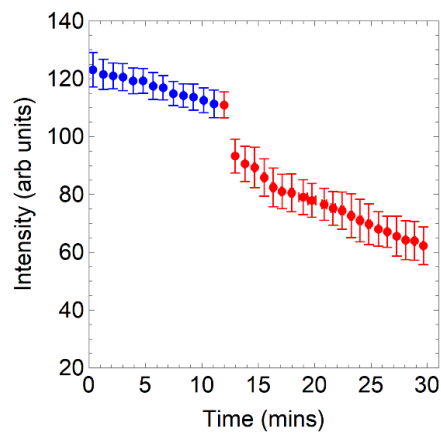

**Fig. S30:** Average Intensity of oscillations of the cells showing forced-damped oscillator. The figure shows oscillation peaks when cell stirrer is turned on (blue) and dampening oscillation when cell stirrer is turned off (red). The damped oscillation characteristic time scale is *ca.* 1.5 mins, which acts as a short-term memory for the feedback control experiments including cellular automata and computation.

## 2.5.2 Interactions between nearest neighbour cell oscillations

A key feature necessary for the experimental platform towards efficient computation is to develop a "decision-making" logic based on the observed chemical oscillations. As the chemical oscillations within different cells could emerge at different times, it is important to create a clocking logic which acts as a sync signal, analogous to the one used in electronic devices, to update the temporal oscillatory states occurring in all the cells. This decision-making logic could be massively simplified if all the oscillations occurring in the cells stays in the same phase. This could be possible by creating a weak interaction amongst all the cells by utilizing combinations of cell and interfacial stirrers. To investigate the effect of interfacial stirrers between the two oscillating nearest neighbouring cells with cell stirrers activated, we monitored the phase difference between the oscillations. We started by actuating two cells stirrers at different initial times so that we can observe an initial phase difference between the emerging chemical oscillations. Then, the interfacial stirrer was activated between them followed by monitoring the phase difference between the two neighbouring cells. Fig. S31 shows the observed BZ oscillations and their peak positions recorded over the time scale of 35 mins. When the interfacial stirrer was turned on, minor change in the intensity of the BZ oscillations was observed. This is due to stronger hydrodynamic coupling between two cells leading to mass transfer. The phase difference between the chemical oscillations occurring in the two cells can be estimated by taking the difference between the nearest peak position times between two cells. Fig. S32 shows observed phase difference estimated at the peak positions between the two weakly coupled cells.

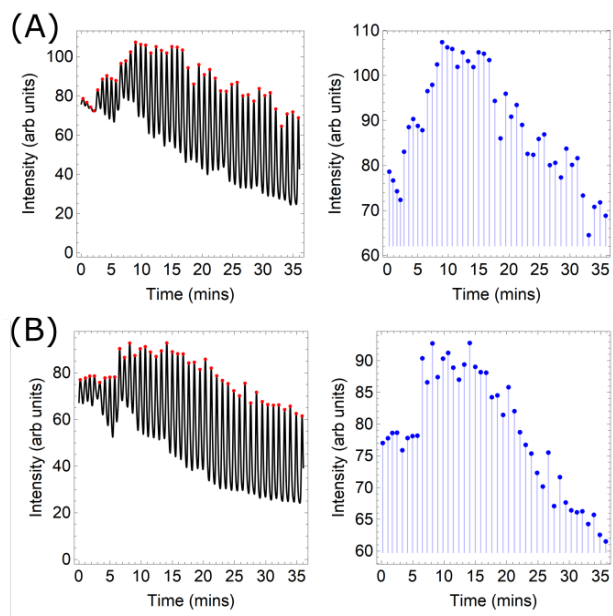

**Fig. S31:** Chemical oscillations and peak positions of two coupled neighbouring cells. Left (A, B) shows BZ oscillations recorded over two cells once the interfacial stirrer is active and Right (A, B) shows the corresponding peak positions vs. time.

A consistent decrease in the phase difference between the two cells was observed and within the time scale of 7-8 oscillation cycles, the two cells arrived the same phase as an outcome of the weak coupling. The temporal oscillatory state of each cell is due to the time-dependent localized concentration of the reagents. Due to symmetric weak coupling, mass transfer leads to the exchange of various ionic species which eventually lead to a similar state of concentration in both cells. Once the two coupled cells arrived at the same phase, as long as the weak coupling is active, the phases of the two cells stay consistent with minor deviation over the full experimental period as evident by the Fig. S32.

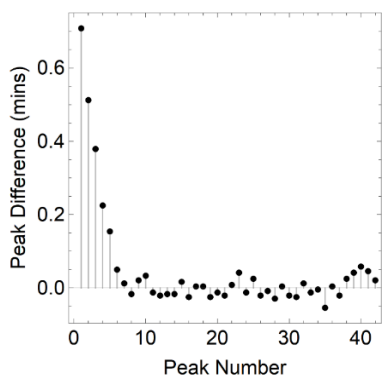

**Fig. S32:** The phase difference between two weakly coupled cells. The figure shows the phase difference between the two neighbouring cells when the interfacial stirrer was active.

The investigation was further extended by coupling three adjacent cells of the one-dimensional platform cells to demonstrate that the weakly coupling scheme could be extended over a long-range by utilizing localized neighbouring interactions. Analogous to the two-cell case, we started activating three cell stirrers with initial time shifts so that they are out-of-phase before the two interfacial stirrers get activated.

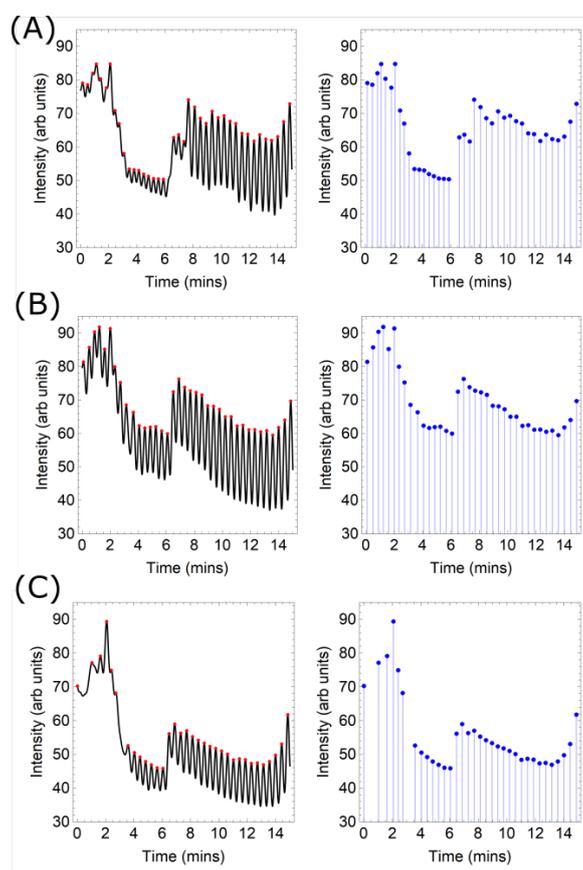

**Fig. S33:** Chemical oscillations and peak positions of two coupled neighbouring cells. Left (A-C) shows BZ oscillations recorded over three adjacent cells once the interfacial stirrers were active and Right (A-C) shows the corresponding peak positions vs. time.

Once the two interfacial stirrers were active, the oscillations over all the three cells were recorded. Fig. S33 shows chemical oscillations and the peak positions vs. time of all the three cells. The phase differences between the central cell (cell 2) and two other two neighbouring cells (cell 1 and 3) were calculated as shown in Fig. S34. Similar to the two-cell case, all the three cells came into the same phase in around ten oscillations. This proves that in a similar way all the cells in an experiment can be bought to the same phase due to the weak coupling between neighbouring cells by activating all the interfacial stirrers. This global coupling created by weak interfacial stirring action was later used for clocking logic for hybrid decision making which is described in detail in later sections on Chemical Cellular Automata and computation.

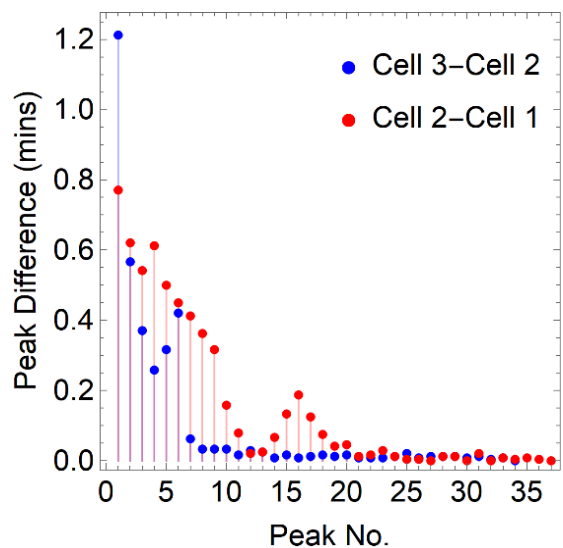

**Fig. S34:** Pairwise peak differences between the three neighbouring cells. The figure shows the calculated phase differences at the peak positions with the central cell (2) as the reference.

## 3 Image processing

The BZ reaction is a chemical oscillator that oscillates in the analogue domain which shows a continuous transition between the red and the blue colour. Visually, it is easy to identify at least three distinct states, red, light blue and blue. As discussed in Section 1.5, a camera was placed on the top of the 3D printed reactor array to visually analyse the evolution of the oscillatory states. The video and corresponding frames captured from the camera during the experiment were processed by different image recognition algorithms which could classify the analogue signals into discretized signals with three distinct states. The temporal patterns of these discretized states were then interpreted as chemical states to define further operations throughout the experiment in a closed-loop approach. This approach of translating the analogue signal into a digital signal creates an information link between the chemical oscillations (analogue domain) and chemical states (digital equivalent) such that any Finite State Machine (FSM) can be implemented into the analogue domain via chemical states.

## 3.1 Data preparation

To characterise and recognise the colour of the oscillation waves, a "supervised learning" strategy (X features → y label) was followed. Thus, it was required to create a dataset that would relate images representing individual cells from the BZ medium in our platform (X), to three possible categories that related to their colour: "red", "light blue" and " blue" (y). To build this dataset, the following steps were executed (see Fig. S35):

1. Multiple experiments with different stirring sequences, speed and positions were performed on the experimental setup. Each experiment was recorded from a camera placed on top of the 3d printed reactor array (see Fig. S4) and saved as a video in MKV and AVI format.
2. A researcher would invoke the "generate_dataset.py" Python script to individually load each one of the videos. This script would play the video, overlaying on top of the displayed platform a grid of squares which aligned with the platform cells. This script would wait for user input.
3. A researcher would press the "space" key to either pause or play the video. When the video was paused, the researcher would then click on some of the squares/cells to label them in one of the three categories based on his/her judgement. To label a cell as "red" a right-click was needed. "Light blue" needed a middle click, while " blue" needed a left-click.
4. Once a square was clicked, a Portable Network Graphics (PNG) file of the cell was saved. Each of these PNG files was saved with a unique file name which included five random characters followed by the RGB value of the average global colour of the previous 3000 frames. The PNG file was automatically placed on a folder named as "Red", "Light blue" or " Blue" depending on the mouse button (left, right or middle) that the user used to click on it.

Following this procedure, two databases were created. Database 1 was comprised of colour tagged images from the one-dimensional experimental setup and it contained >13,000 images. Database 2 was comprised of tagged images from the two-dimensional experimental setup and it contained >7,000 images.

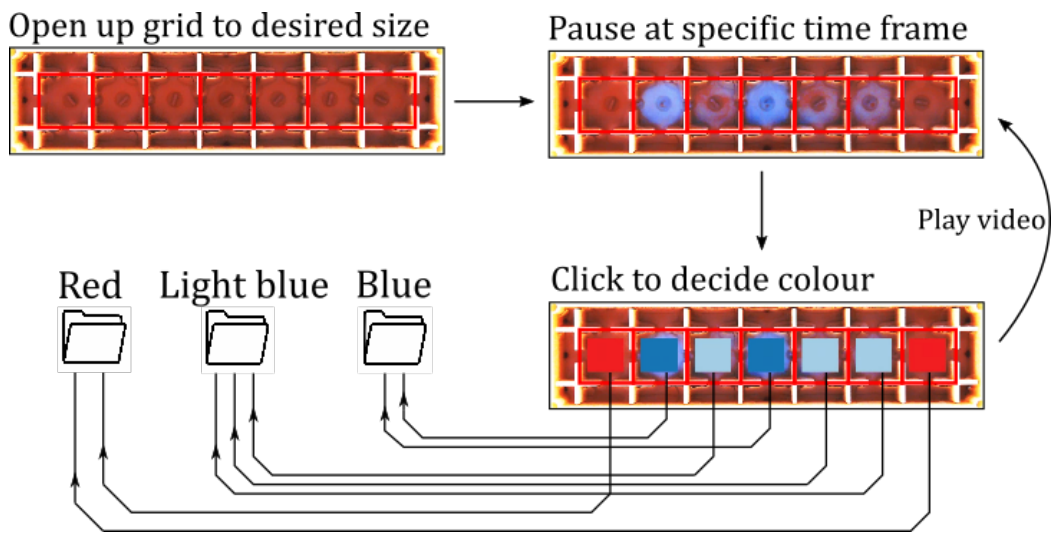

**Fig. S35:** Flow diagram describing steps to create the data set for image classification.

The dataset was generated using user input following the workflow as shown in Fig. S35. A researcher would load a video containing an experiment. The researcher would then stop the video at different times and click on the cells labelling them as "Red", "Light blue", or "Blue". Once a cell was clicked, an RGB image was extracted from its contents, saved as Portable Network Graphics (PNG), and stored in a folder named following one of the three labels. All the data gathered for the 1D and 2D platforms underwent the same workflow.

Initially, a simple linear discriminator was tested to classify the differences between the three different classes of colour. However, during a BZ experiment, the concentration of $KBrO_3$ decreases over time in a non-linear way, and this dramatically changes the colour of the solution towards a more reddish shade. The 'Red' and 'Light Blue' discrete states emerging at the beginning of an experiment may look like the 'Light Blue' and 'Blue' states emerging at the middle and the later stage of the experiment, as illustrated in Fig. S36, which makes them indistinguishable. Several classifiers such as Support Vector Machine (SVM), K-Nearest Neighbours (KNN), Multilayer Perceptron (MLP) and Random Forests, were tested, but none of them was able to distinguish the difference between light blue and blue reliably. To overcome this issue, Deep Learning, and in particular, Convolutional Neural Networks (CNN) was employed as this architecture has shown it can detect objects with higher accuracy than humans. Therefore, we expected that CNN would be able to classify correctly the three labels (red, light blue, and blue) throughout a full experiment.

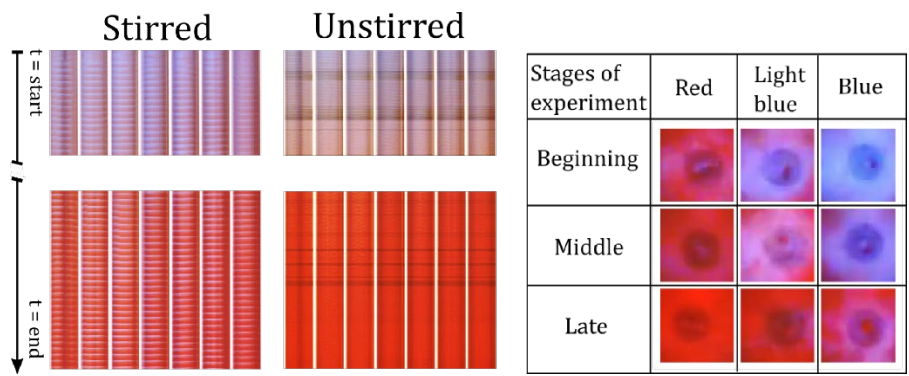

**Fig. S36:** Visual representation of the transformation of the colours at different stages of an experiment.

Before training the CNN, further data pre-processing was needed. As shown in Fig. S35, the PNG files generated using the user input were placed in folders named as their respective labels, namely Red, Light Blue and Blue. Using this set-up, a script would execute the following instructions to prepare the data to be inputted to the CNN:

1. A Python list was created where each element was a tuple. This tuple contained the filename of each PNG file and its label. The label was extracted based on the name of the folder where the PNG file was stored. The label was coded as an integer number, being 0 for Blue, 1 for Light Blue and 2 for Red colour.
2. For each element of this list, the first step was to extract the average background colour, which was hardcoded in the filename of the PNG file, as described on step 4 in the previous list.
3. For each element of the list described the in the first step, each PNG file was loaded into a 3D Numpy array using OpenCV, being the 3 dimensions: height, width and channels (RGB). All the elements of this array were divided by 255 to keep their values between 0 and 1.
4. The Numpy array (3D) was then split into three different arrays representing each of the RGB channels. Green channel was discarded.
5. From both blue and red channels, the average background colour from step 2 was subtracted. These two channels were then merged into a 3D array, representing width, height and channel (red or blue).
6. This 3D array was resized to be 50 by 50 (and by 2 for the blue and red channels) using cubic interpolation.

7. All the 3D arrays generated this way, with their associated label, were randomly split into the train and test dataset. The split ratio was 95% for training and 5% for testing.
8. Both train and test datasets were shuffled using Numpy's "shuffle" functionality.

## 3.2 Convolutional neural network (CNN)

The input to the CNN consisted of (50,50,2) Numpy arrays. The CNN was trained using batches of size 100. The output of the CNN was a single integer value representing the class of the input array. These integer values could be 0, 1 or 2, being 0 for Blue, 1 for Light Blue and 2 for Red. The architecture of the CNN used to classify the experimental output in three different discrete states is shown in Fig. S35.

Tensorflow 1.X was used to define the architecture shown on Fig. S37. Within the Tensorflow library, "Conv2D" was used as the convolutional layers. The stride was set to 1 in all of them, and the padding was set to "same". Tensorflow "nn_max_pool" was used as the max-pool layers. When specified, the dropout was 30% (using Tensorflow layers dropout). Unless specified, the activation function is the rectified linear unit (ReLU).

The model of the CNN is as follows:

- Two Conv2D with a kernel size of 9 and 64 filters
- Max-pool with kernel size [1, 2, 2, 1], strides [1, 2, 2, 1].
- A dropout layer.
- Two Conv2D with kernel size 3 and 128 filters
- Max-pool with kernel size [1, 2, 2, 1], strides [1, 2, 2, 1].
- A dropout layer.
- Dense layer with 64 neurons.
- A dropout layer.
- Dense later with 3 neurons
- Softmax layer

The network was trained using Adam's optimizer. Its parameters were left to the default values. The loss function was "tensorflow.reduce_mean" paired with "minimize" from Adam's optimizer. Finally, it was trained over 1000 epochs.

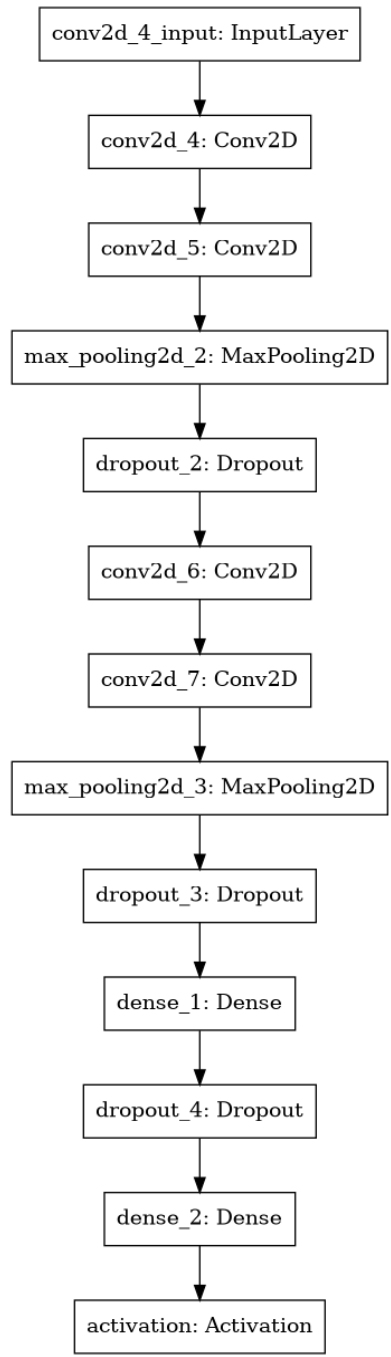

**Fig. S37:** The architecture of the CNN used to train the images generated by the user.

## 3.3 State classification

Using the CNN, our system could visually detect three distinct colour levels of BZ oscillations in real-time: Red, Light Blue, and Blue. In order to perform logic operations, the next step was to transform this visual information into binary data, similar to the digital states 0 and 1 used in almost every electronic device. We call the three distinct colour states from the analogue oscillatory signal using the CNN as CNN states. The binary states emerging from the temporal patterns of CNN states are called as Chemical States (CS). The Chemical States defined here are the digital representation of the analogue Chemical State which exists as oscillations in the chemical domain. It is important to note that, our Chemical States do not need to be binary, it could be generalized to n distinct chemical states. However, in this work, we developed our hybrid electronic-chemical logic based on two distinct chemical states (low state: CS=0 and high state: CS=1). These chemical states when defined as binary states, also represented as "0" and "1" states in this section for simplicity.

The step from the three levels of visual oscillations to the digital states of CS=0 and CS=1 was performed using a Finite State Machine (FSM), also referred as a recognition Finite State Machine (rFSM). It is important to distinguish the recognition FSM from the digital Finite State Machine which is used for digital processing in the Chemical Cellular Automata and Computation in later sections. Given the colour of a cell and the colour of the same cell in the next iteration, the FSM would define the chemical state of that cell as CS=0 or CS=1, see Fig. S38. We further introduce an accumulator, in which the CS from several clocking cycles are recorded and accumulated. This increases the robustness and minimises the error when observing the state of the cell.

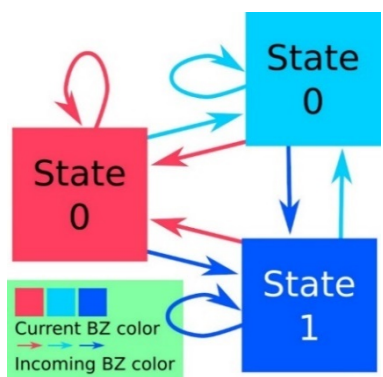

**Fig. S38:** Finite State Machine detailing the transition rules between BZ states. The figure shows the basic implementation of the FSM to binarized chemical states where state "0" represents the chemical state CS=0 and "1" represents the chemical state CS=1.

Every experiment always started with all the cells at the red BZ visual state. This is the default BZ colour when the cells are unstirred. At this colour, the binarized chemical state of the cell is defined as "0". Once an experiment starts and the chemical medium is stirred, it will initially oscillate between the red and the light blue colours, and while the BZ medium is following this pattern, the state of each cell would be "0". When a cell is stirred at a high speed, that cell would oscillate into blue, and only at this point the binary state of that cell would be "1", and as soon as this cell oscillated back to red or light blue, it returned to "0", see Fig. S39-Top. Fig. S39-Bottom shows images from the one-dimensional experiment platform of (a) image acquisition from the camera and (b) detected by the CNN in real-time, followed by (c) transformation of the detected signals to two distinct states of 1s and 0s using the described FSM.

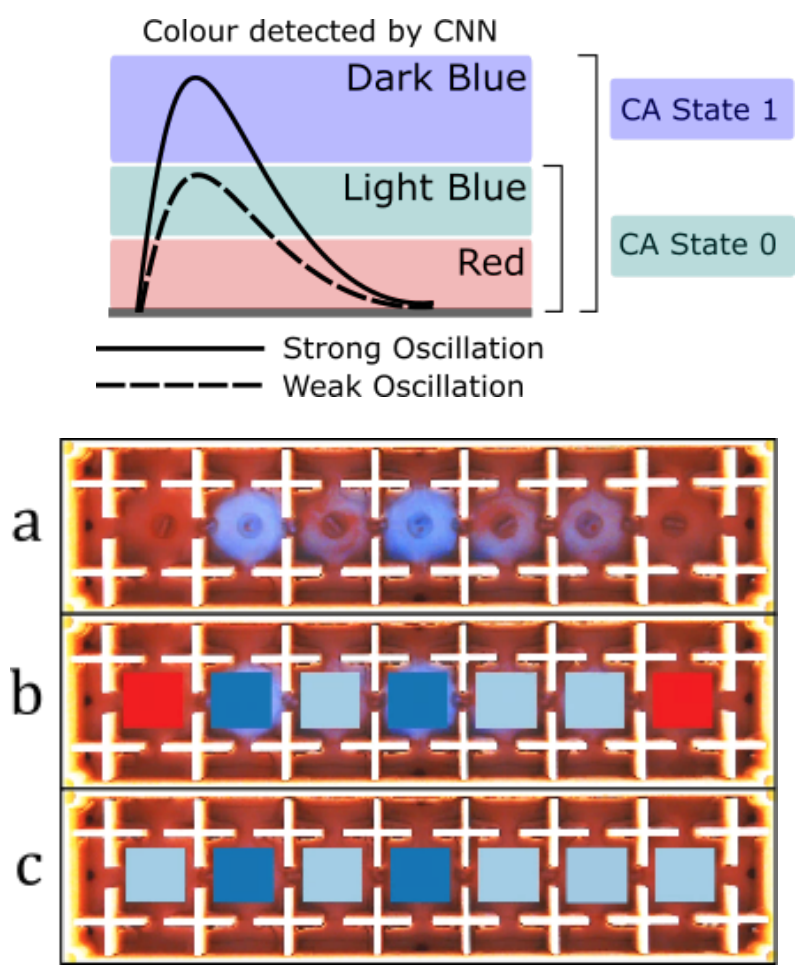

**Fig. S39:** Discretization of continuous oscillation using CNN. <u>Top</u>: If a given cell detects strong oscillation (indicated by blue detected by the CNN), the state of the cell becomes "1". The cell that was detected as weak oscillation or no oscillation at all was classified as the state "0". <u>Bottom:</u> (a) Unprocessed video from the camera which was followed by (b) colour recognition from the CNN and (c) the digitisation of oscillation to state (0: light blue, 1: blue in figure C).

Initially, the FSM described on Fig. S38 was implemented using the "Transitions" Python library. This worked well when used with the one-dimensional experimental platform, as it only had to update 7 cells, but in the 2D platform, when it was updated to 49 cells, this library became the bottleneck of the whole data pipeline and therefore it was not possible to run the system at 20 FPS. Thus, it was decided to simplify the FSM, by implementing an in-house FSM by using Numpy. Initially, the state of a cell would only swap to 1 if it follows the pattern "red – light blue – blue", but the condition was relaxed in this case, and a cell could also swap to 1 if it also followed a pattern "red – blue". Even though a cell can only go to blue after being light blue, sometimes this transition happened very fast and the CNN was not able to detect it, therefore the straight transition from red to blue was also enabled to represent a state "1" for a cell.

### 3.4 Mapping between stirrer PWM levels and the Chemical States

In the previous section, we have demonstrated our BZ chemical system with programmable stirrer control act as a damped oscillator which shows strong hysteresis behaviour when the oscillations are classified as the Chemical States based on a convolutional neural network and a recognition state machine. If we assume a one-to-one mapping between PWM states and the emerging Chemical States, then at each step we should have chemical state 0 for low PWM and high chemical state 1 for high PWM. This is possible only in the absence of the hysteresis effect and weak or no interactions between the nearest neighbouring cells. To investigate both hysteresis effect and local interactions, we perform simple test experiments on a single cell and nearest neighbouring cells to demonstrate deviations from one-to-one mapping.

**Probabilistic Mapping in Isolated Cells due to the Hysteresis Effect:**

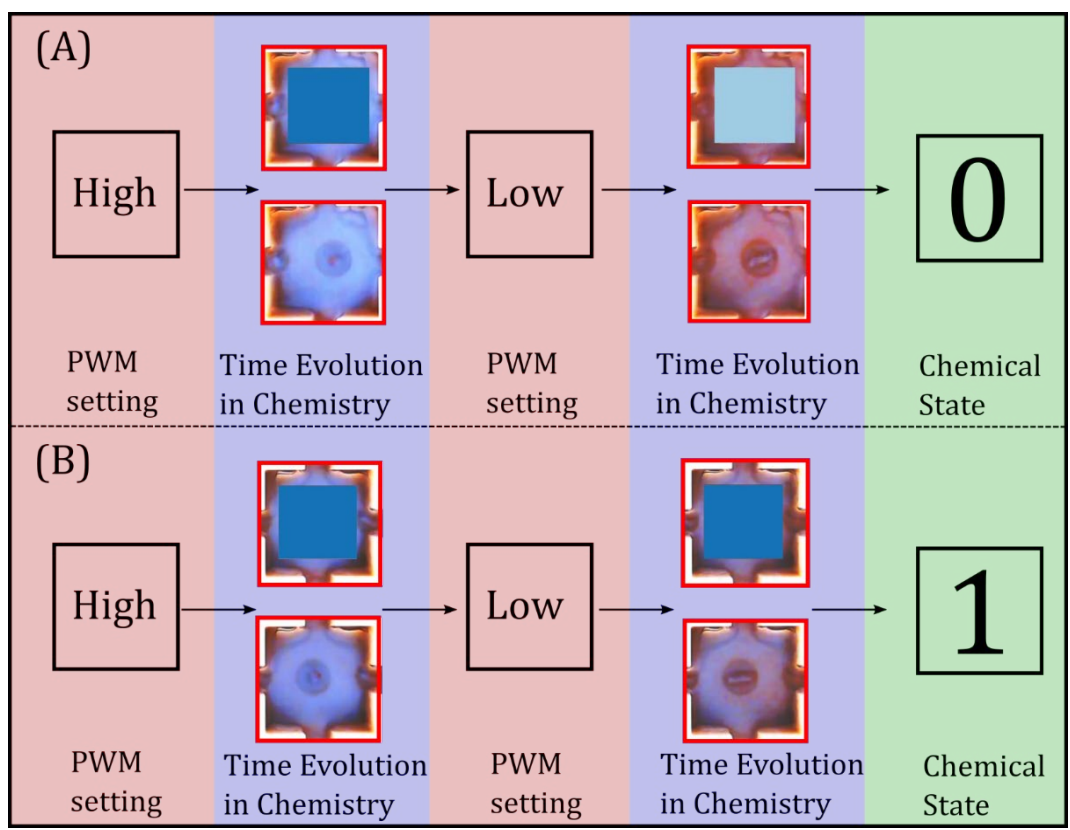

**Fig. S40:** Deviation from one-to-one mapping of single cell. In both (A) and (B), the cell was first stirred in high PWM (50) producing high chemical state by strong oscillation and when the same cell switches to low PWM (30), it can decay to chemical state 0 (a) or retain chemical state 1 (b), which is recognised by the CNN.

To study the hysteresis effect, we investigated the changes in the chemical states from an isolated cell. In a single cell experiment, we activated the cell stirrer at a high PWM for 2 mins so that the chemical state 1 appears. Then we switched to low PWM. If it is indeed the case of one-to-one mapping, we should expect the emergence of chemical state 0. We can observe the system decay to chemical state 0 or retains the chemical state 1, which occurs due to the damping effect from the stronger oscillation. In Fig. S40, we showed that the chemical state produced can either fade away or retain its original chemical state 1 after changing the PWM from high to low. The probabilistic mapping is more pronounced when local interactions with the nearest neighbours are introduced in the system.

**Probabilistic Mapping in Interacting Cells due to the Local Interactions:**

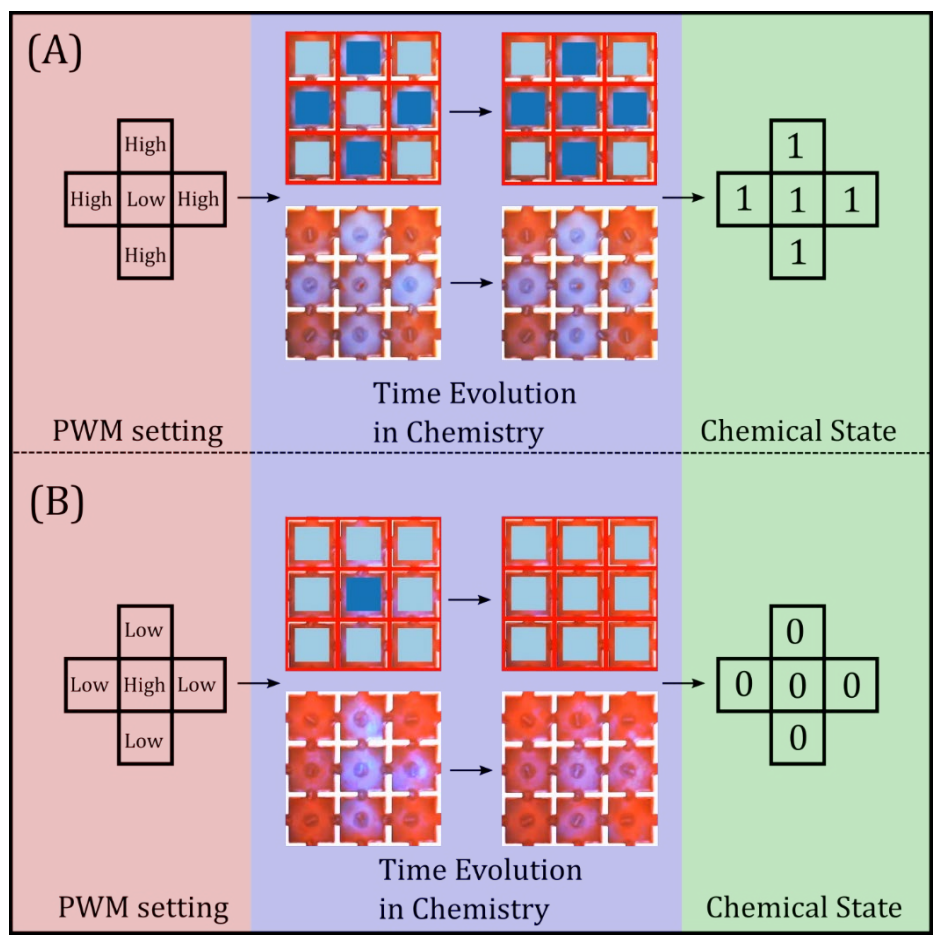

**Fig. S41:** Deviation from one-to-one mapping of single cell with nearest neighbours. (A) Shows the cell was first stirred in low PWM (30) producing low chemical state by weak oscillation, and due to the effect from the nearby high PWM (50), the chemical state switched to high. (B) The opposite case of (A) which shows the decay of the high chemical state to low due to the nearby low PWM interactions.

To study the local interactions between cells, we further investigate the change of the chemical states when the nearest neighbouring cells are interacting with the central cells. The chemical state of the central cell is not retained and is dominated by the PWM states of the neighbouring cells instead, as shown in Fig. S41. In both cases, we activated the central cell stirrer at a high(low) PWM and the nearby cells at low (high) PWMs, and obverse the transformation of chemical state from high (low) to low (high).

# 4 Dynamic feedback

So far, the description on the BZ medium oscillates between different colours following the stirring patterns executed by DC motors were mentioned. A camera records the chemical oscillations based on colours, and the CNN classifies it between three different colour levels: Red, Light Blue and Blue. Finally, the state recognition FSM (rFSM, see Section 3.3) defines the chemical state of a cell as either CS=0 ("0") or CS=1 ("1") depending on the colour history of the cell. This step translates the emerging chemical oscillations (chemical domain) into equivalent chemical states on which any state machine can be implemented (digital domain). By using these current set of chemical states of all the cells, any state machine can be implemented which takes these chemical states as inputs and outputs the RPM states of the motors called as PWM states. These PWM states will update the motor speed, thus generating new oscillatory patterns. This step defines the single closed-loop operation in the context of dynamic feedback. We also defined this process as an information loop, where analogue information from the chemical oscillations accurately enters the digital domain on which various state machines could be implemented. These state machines then dynamically transfer the information back into the chemical domain. Using this approach, complex information processing is possible by distributing information within the chemical and electronic domains. Both chemistry and electronics provide different roles in the framework of information processing, which can be seen in the implementation of chemical cellular automata (CCA) and chemical entity (Chemit) in the later sections.

Here, we describe two different models of the dynamic feedback scheme which were later used in the experiments on cellular automaton and computation. Fig. S42 shows two different models of updating the chemical states in the closed-loop. In both of the models, the electronic state machine connects the chemical oscillations using via mechanical actuation of stirrers. The chemical oscillations were then recorded and processed using CNN and updates the chemical states. However, in the first case, the state machine (shown in red) as the chemical state which is defined as a cellular automata state is updated directly in-silico. This is only possible if there is a direct one-to-one mapping between the chemical state and electronic state (PWM levels of stirrers in our case) and all the individual chemical oscillations directly translate back into the chemical states. In this case, there are no interactions between the analogue chemical states. The exact information from the digital states loops through the analogue domain without providing any additional benefit towards computation using hybrid electronic-chemical logic. We call this implementation as a "display screen" due to direct mapping to the digital domain.

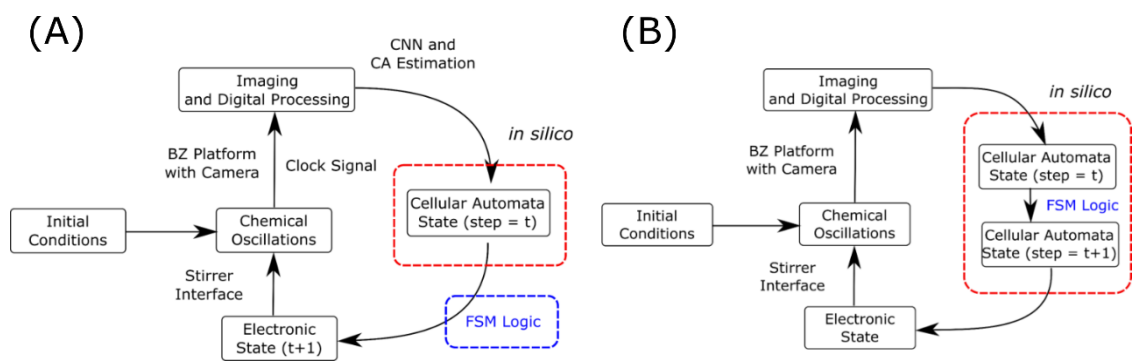

**Fig. S42:** Two models of the dynamic-feedback loop. (A) Shows that the FSM logic is embedded *in silico* such that the new state can be predicted within the digital domain and (B) on the other hand shows that the FSM logic is based on the chemical system.

In the second case, the state machine shown in red reads the chemical states and applies state machine logic to the PWM states of the stirrers, which is shown as an electronic state. In this scenario, it is not possible to predict the evolution of chemical oscillations in the analogue domain and the emerging chemical states. So, there is no direct one-to-one mapping between the chemical states to the digital states. This dynamic loop model describes the true picture where the information processing operations can be split into chemical and digital domains. In the chemical domain, analogue information processing occurs due to interactions between the cells with different oscillatory states, and digital information processing occurs using state machines utilizing the chemical states of the information loop. We have implemented both states machines, the first model to demonstrate successful implementation information loops over the complete experimental time scale by implementing elementary Cellular Automata (ECA). The second model was used to further develop novel one-dimensional chemical cellular automata rules (1D-CCA), two-dimensional chemical cellular automata (2D-CCA) implementation, and computation experiments. In the next step, we introduce the concept of chemical clocking logic which is critical towards implementing logic to develop hybrid electronic-chemical decision-making machine.

## 4.1 Synchronization between the computing cells using a Chemical Clock

As the BZ chemical medium starts oscillating in our experimental platform, we observed a minor phase shift between the different cells due to hysteresis effects of the oscillating reaction and possibly a little stochasticity associated to the stirring actuation signals. Thus, there is a finite phase difference between the oscillations across the whole platform. To get rid of this initial phase difference and not creating strong oscillations which could distort the actual signals, we used the strategy discussed in Section 2.5.2. From our observations as discussed previously, if interfacial stirrers are active, the neighbouring cells come into the same phase and oscillate together. To fix the phase drifting problem, we activate the interfacial stirrers so that all the cells can interact with nearest neighbours and global phase synchronicity can be achieved. Instead of activating the cell stirrers all the time, we pulse them between ON and OFF states. This creates a weak but detectable in-phase oscillation over all the cells. Instead of using no oscillation state as the low state, we use the weak oscillatory state as chemical state CS=0.

Fig. S43(A-G) shows BZ oscillations over all the seven cells of the one-dimensional array with peak positions highlighted with interfacial stirrers active and pulsing cell stirrers. Fig. S43 H shows the time difference of the oscillations occurring at the seven cells of the 1D platform throughout a full experiment when compared against the central cell. We observed an increase in the phase difference as the experiment progresses, up to 10 seconds at the end of the experiment. However, as the difference between the consecutive peaks is *ca.* 40 seconds, the phase drift of 10 seconds does not pose any problem to implement any programmable logic into the system.

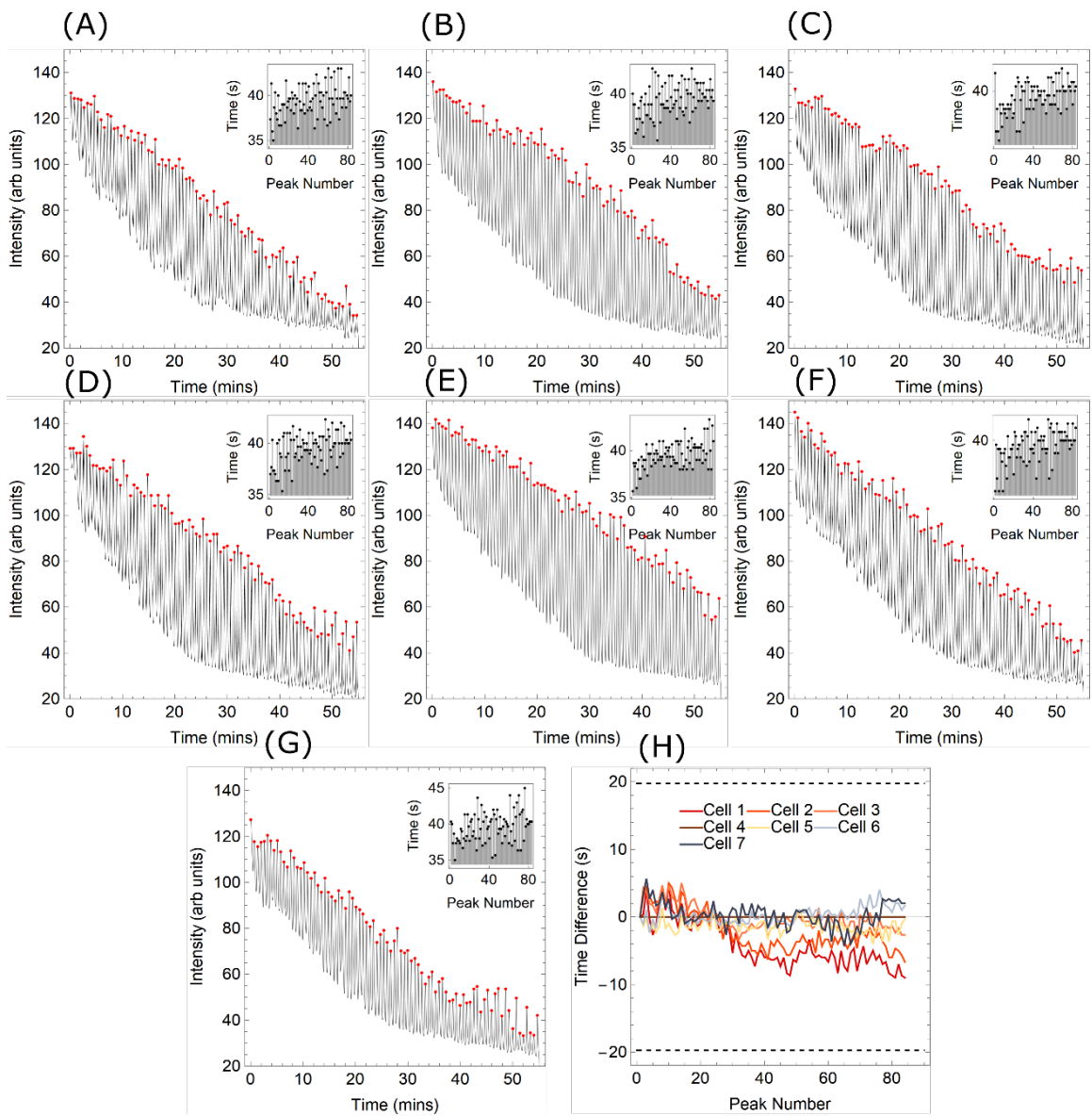

**Fig. S43:** Time difference between peaks amongst all the cell as the function of peak numbers. (A-G) shows oscillations observed in seven cells of the one-dimensional BZ platform. (H) shows the phase difference between all the cells with respect to the central cell.

To perform computations, we define windows of time where all the BZ cells oscillate are needed, despite the drift of oscillations over time. Initially, it was decided to set a fixed quantity of time for each window, based on the average periodicity of the oscillations, but as the temporal phase between oscillations drifted through time, this solution was not suitable. Therefore, it was decided to dynamically alter the windows of computation based on the chemical oscillations. This was implemented similar to the clock element used in digital computers. In a digital computer, when the clock signal goes from 0 to 1 it is often called "tick" and when it goes from 1 to 0 it is often called "tock". Usually, computation only happens during the "tick" phase – that is, when the clock is set to 1. In digital computers, the clock signal is perfectly periodical with a square shape – a 1:1 ratio between tick and tock – while in our case, as it can be seen on Fig. S44(A, C), the portion of time when the BZ medium did not oscillate was much bigger than the portion of time it oscillated.

Following this idea, it was decided to define a "chemical clock" that would track the BZ oscillations. This clock would be 0 when no oscillations happened, and it would be 1 while the BZ medium oscillated. Because each cell could oscillate individually, there were two types of chemical clocks defined: local clocks which represented each of the individual clocks, and a global clock signal, which represented how the BZ medium oscillated as a whole, and it was calculated based on the local clocks.

Each local chemical clock could have three states:

1. **None** – which meant the medium was not oscillating, i.e. the CNN classified the medium as red colour.
2. **Tick** – which meant the BZ medium went from red to blue.
3. **Tock** – which meant the BZ medium went from blue to red.

There were three states defined for the global chemical clock:

1. **None** – which meant the medium was not oscillating: all the local clocks were in the "none" state.
2. **Tick** – which meant at least one cell oscillated: at least one local clock went into "tick" state.
3. **Tock** – which meant all the cells had oscillated: all the local clocks were into "tock" state.

Once a global Tock signal was executed, all the calculations related to the computations using the BZ state would be performed, and the global clock would reset and go back to "None", and it would also set to "None" all the local clocks, see Fig. S45 for schematic diagram. The transitions between these three states were as follows:

1. **To Tick** – Once the first local clock "tick" happened (red to blue), the global chemical clock signal went from none to tick.
2. **To Tock** – Once all the local clocks were in "tock", the global clock signal went from tick to tock.

3. **To none** – Once the computations related to the BZ states were performed, the clock signal went from tock to none.

In the case of the 1D platform, this workflow worked perfectly well, because it was only required that 7 cells would fully go through an oscillation. In the case of the 2D platforms these conditions had to be loosened, this is because the 49 cells may not oscillate in a similar period. Therefore, in the case of the 2D platform the To Tock step was as follows:

2. **To Tock** - Once at least 15 cells have oscillated – thus their local clock was on the tock state - and all of the 49 cells were back in red colour, the clock signal went from tick to tock.

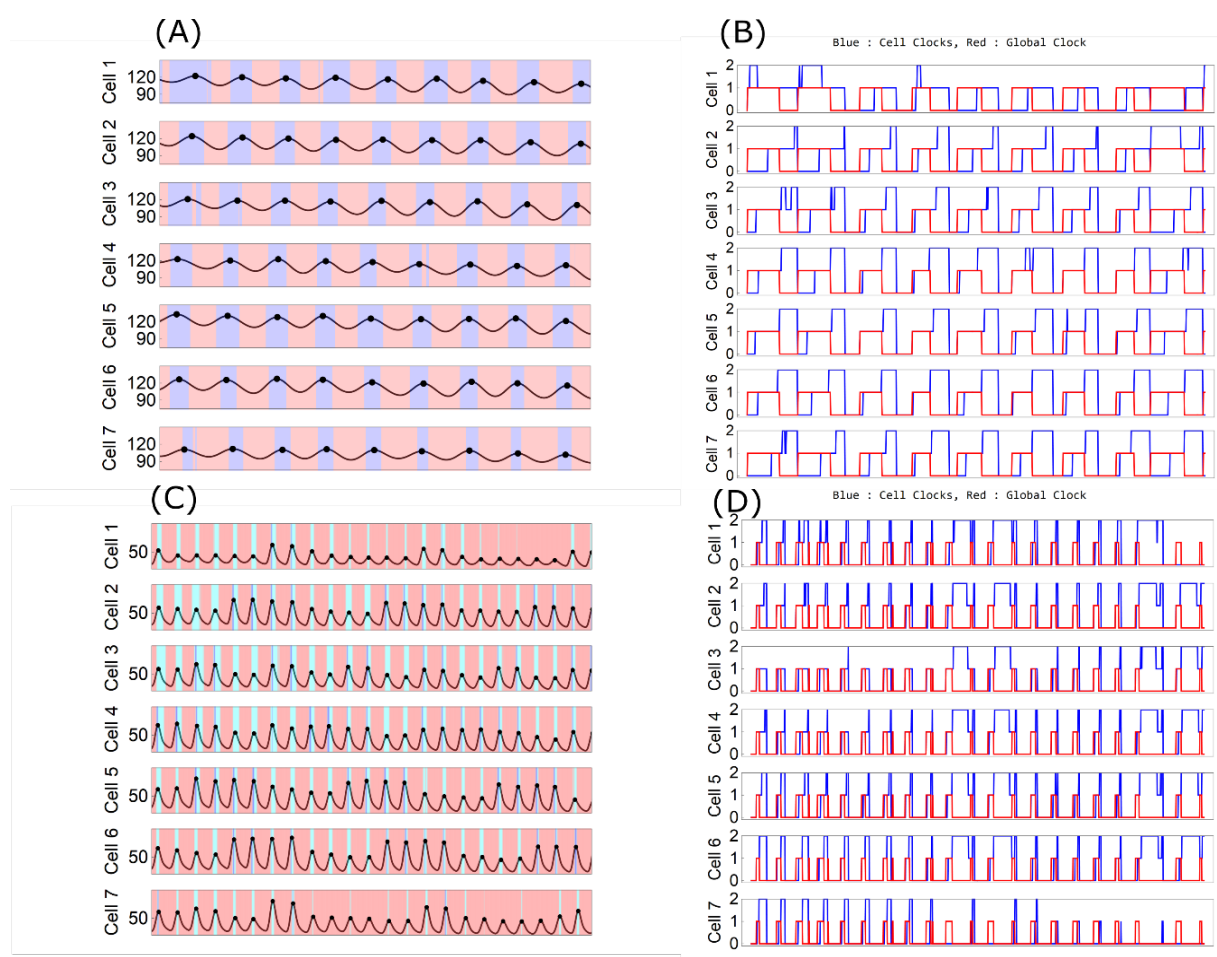

**Fig. S44:** Chemical Clocking logic with one and two chemical states in the one-dimensional experimental platform. (A) Chemical oscillations of all the seven cells in one-dimensional setup testing basic clocking logic without chemical states and corresponding CNN states and (B) Individual cell clocks and the global clock operations from CNN states as shown in (A). (C) Chemical oscillations of all the seven cells in one-dimensional setup demonstrating clocking logic on elementary cellular automaton and corresponding CNN states. (D) Individual cell clocks and the global clock operations from CNN states as shown in (C).

To perform a computation based on the states described by the BZ medium, it was required that the chemical clock signal fully oscillated twice before a decision was taken. This was done to improve the reliability of the chemical state as acquired by the camera and minimise any error that could arise from the phase-drift of the oscillations.

A simplified analysis was performed on clocking mechanism after experiments were complete for representing local clock for individual cells and global clocking scheme for basic clocking test and tests performed with dynamic loops to implement elementary cellular automata (see later sections). The chemical clocking logic was observed as

1. The initial state is when all cells are red.
2. Once a cell oscillates (it only needs one cell), the cell goes to **TICK**, and global clock also goes to **TICK**.
3. Once the cell stops oscillating (going from blue to red), the cell goes to **TOCK** and global clock tests if global **TOCK** is possible.
4. Global **TOCK** is only possible if all the seven cells are red and at least two of them are in **TOCK**.
5. After global **TOCK** the clock resets.

The condition for every decision to be taken place is that the global clock is required to 'tock' twice, this further improves the reliability of the chemical state acquired by the camera.

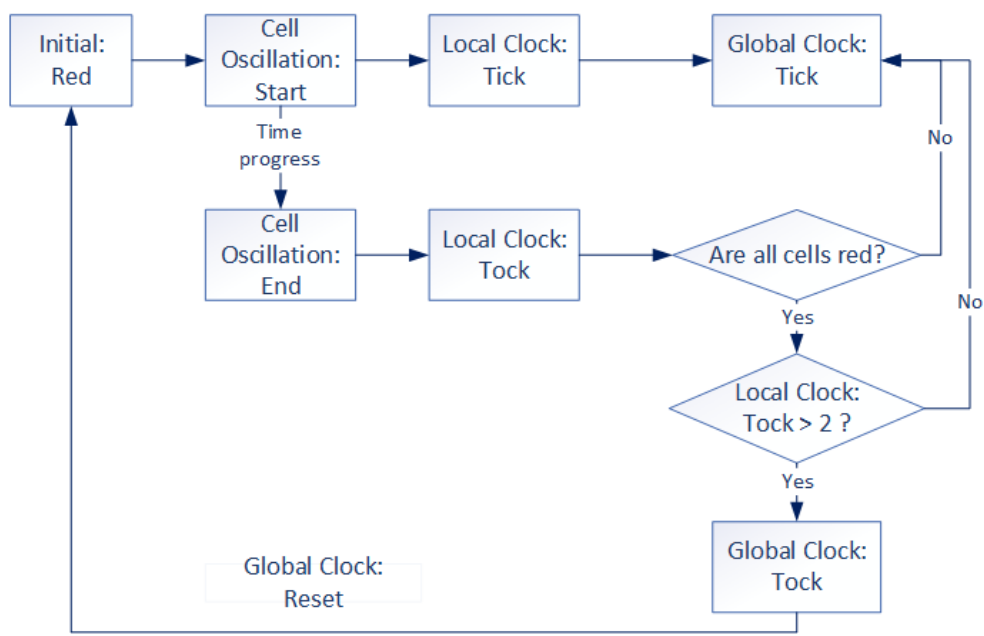

**Fig. S45:** Flow diagram of the logic for the chemical clock in an experiment.

The simplified pseudo-code to analyse the experimental data for representing the clocking logic is shown as,

cell_Lists 0: None, 1: Tick, 2: Tock
global_States 0: Tick, 1: Tock
-------------------------------------------------------------------------------------
Updating cell_Lists:

**if** (cell_Lists[[cell]] == 0 && cnn_State[[cell]] == 0):
    cell_Lists[[cell]] = 2 (cell Tock)

**else if** (cell_Lists[[cell]] == 1 && cnn_State[[cell]] == 0):
    cell_Ticks[[cell]] = 2 (cell Tock)

**else if** (cell_Lists[[cell]] == 2 && cnn_State[[cell]] == 0):
    cell_Ticks[[cell]] = 2 (cell Tock)

**else if** (cell_Lists[[cell]] == 0 && cnn_State[[cell]] == 1):
    cell_Lists[[cell]] = 1 (cell Tick)

**else if** (cell_Lists[[cell]] == 1 && cnn_State[[cell]] == 1):
    cell_Lists[[cell]] = 1 (cell Tick)

**else if** (cell_Lists[[cell]] == 2 && cnn_State[[cell]] == 1):
    cell_Lists[[cell]] = 1 (cell Tick)
-------------------------------------------------------------------------------------
Updating global_states

**if** (global_state == 0 && any cell_List[[cell]] == 1)
    global_state = 1 (Global Tick)

**else if** (global_state == 1 && all cell_List[[cell]] == 2)
    global_state = 0 (Global Tock)
    all cell_Lists[[cell]] = 0 (reset cell_Lists)

**else if** (global_state == 1 && at least n(cell_List[[cell]] == 2) ≥ 2)
    global_state = 0 (global Tock)
    all cell_Lists[[cell]] = 0 (reset cell_Lists)
-------------------------------------------------------------------------------------

## 4.2 Implementation of elementary Cellular Automata

To demonstrate the successful implementation of the dynamic feedback loop of the information transfer back and forth in chemical and digital domains, experimentally, we implemented three elementary cellular automata rules, Rule 30, 110 and 250 in the one-dimensional experimental setup. In this case, we have a direct one-to-one mapping between the PWM state of the cell stirrer and emerging chemical state. In brief, all the interfacial stirrers were turned on to allow coupling between the cells and their nearest neighbouring cells. The cell stirrers can either be CS=0 achieved by constant pulsing (activate stirrer at 16 PWM for 5 seconds and deactivate of stirrer for 15 seconds) and CS=1 achieved by strong stirring operations (activate stirrer at 50 PWM). We implemented these rules to demonstrate successful information propagation from digital logic to analogue oscillatory chemical oscillations which read back into the electronic state via optical imaging and discretization using Convolutional Neural Networks (CNN) as discussed in previous sections. The basic strategy was shown in Fig. S42(A). A detailed description of the BZ oscillations and the CNN states emerging in the elementary CA rule 30 in the one-dimensional experimental setup is shown in Fig. S46(experimental video can be found in Supplementary Video 3). The intensity versus time plot of cell 2 indicated clearly for the lower intensity oscillations, chemical state CS=0 achieved by basic pulsing operation of the cell stirrers and the higher intensity oscillation, chemical state CS=1 strong stirring operations. As seen in the figure, the period of an oscillation varied from 30 seconds to 40 seconds.

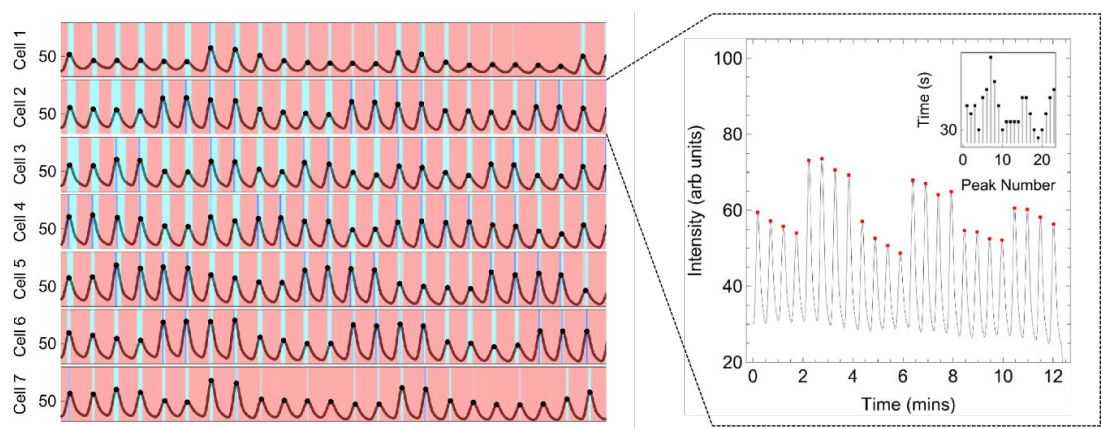

**Fig. S46:** Chemical oscillations and the state detected by the CNN of ECA rule 30. The left figure shows chemical oscillations emerging in the one-dimensional experimental platform, with three discrete CNN states (shown in red, light blue and dark blue) in the background for all seven cells in the one-dimensional platform. The right figure shows the detailed zoomed-in image of the chemical oscillations of the second cell, where lower peaks correspond to chemical state CS=0 and higher peak corresponds to CS=1 state.

The chemical states of emerging rules in elementary CA rules 110, 30 and 250 are shown in Fig. S47 with chemical states, CS=0 shown in light blue and CS=1 shown in dark blue. All the three implemented rules implemented as chemical states mapped perfectly with the in-silico elementary CA rules. As discussed in the dynamic feedback model, elementary CA rules can be predicted *in-silico*, with the direct one-to-one mapping as the information loops through chemical and electronic domains. To bypass this limitation, we updated the implementation of elementary CA rules by introducing asymmetric coupling between the nearest neighbouring cells to create novel rules which cannot be predicted directly.

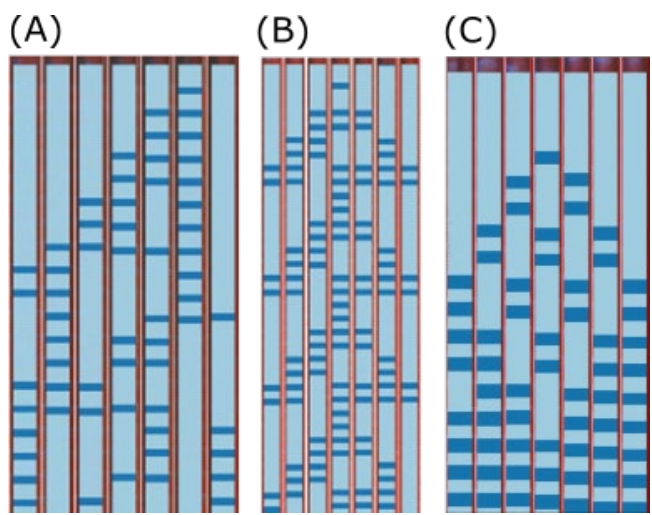

**Fig. S47:** Implementation of three elementary cellular automata. (A) Rule 110, (B) Rule 30 and (C) Rule 250 were implemented on the one-dimensional experimental platform. The figure shows two different cellular automata states where light blue is equivalent to chemical state $CS=0$ and dark blue corresponds to chemical state $CS=1$.

We also extended the definition of the elementary CA and implemented novel CA rules using our experimental platform. These novel rules are defined as 1D-CCA rule, where we use the dynamic feedback loop scheme as shown in Fig. S42(B), see full implementation details in the next section. The novel rules emerge by expanding the state space using chemical states as well as PWM states. Fig. S48(A) shows an experimental implementation of elementary CA rule 30 and (B) shows modified 1D-CCA rule with novel states emerging due to asymmetric interactions between nearest neighbouring cells. In the logic of the asymmetric variation, the interface is only turned on when the left neighbour is at $CS=1$ whilst the cell is at $CS=0$ (see Section 5.3).

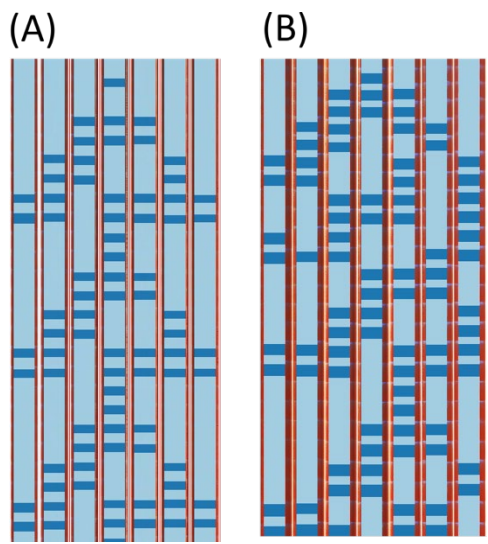

**Fig. S48:** Implementation cellular automata rule 30 together with modified form. (A) Rule 30 with symmetric interactions on both sides of each cell. (B) Rule 30 asymmetric interactions on both sides which creates a novel cellular automata rule.

# 5 1D-CCA: Configuration space and novel 1D-CCA rules

## 5.1 Estimation of BZ input and chemical state space

Consider an experimental set up with $n \times n$ interconnected network of cells, with cell and interfacial stirrers which can be addressed independently. If we assume $p$ possible cell stirrer states and $q$ possible interfacial stirrer states, the total number of possible combinations (Input States: $IS$) are given by $IS = p^{n.n} q^{2n(n-1)}$. The first term $p^{n.n}$ corresponds to all the operations on the cell stirrers where $n.n$ corresponds to the total number of stirrers in the two-dimensional experimental platform. The second term $q^{2n(n-1)}$ corresponds a total number of possible combinations of interfacial stirrers which are present in both horizontal and vertical connections. All combinations of cell stirrer and interfacial stirrer states are a subset of the complete input space. Similar to the input space, we can also estimate the total number of chemical states which can be interpreted from the chemical oscillation patterns by classification using CNN as discussed in previous sections. Assuming, $k$ different possible chemical states on each cell, the total number of possible global chemical states is given by $CS = k^{n.n}$. Depending on the type of operations on our computational platform, we explore through the subset of this input space. Various operations include Chemical Cellular Automata, solving combinatorial optimization problems based on chemical computation logic etc. Each experimental rule or logic creates a trajectory in both input and chemical state space. Fig. S49 shows scaling of total input and chemical states for a two-dimensional set up with the number of cells.

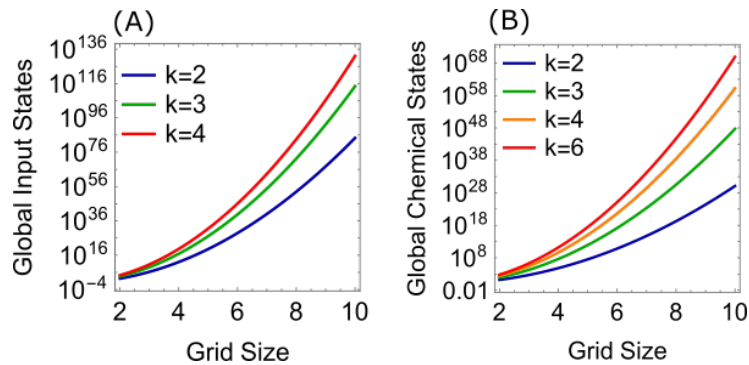

**Fig. S49:** Global input and chemical states in the computational platform. (A) Global inputs states *vs.* grid size with $k$ possible states on each cell stirrer and two different states on each interfacial stirrer (B) Global chemical states *vs.* grid size with $k$ possible detectable chemical states on each oscillating cell.

For the implemented computational architecture with 7×7 array with four distinct PWM states for cell stirrers (as used in two-dimensional chemical cellular automata) and two PWM states for interfacial stirrers, the total number of input states are given by $4^{7\times7} 2^{14\times6} = 6.12 \times 10^{54}$. The total number of possible chemical states are given by $2^{7\times7} = 5.6 \times 10^{14}$.

In the presence of one-to-one mapping between the input and chemical states, the total number of input and chemical states are equal. Hence, the information loops between input and chemical states without any additional benefit towards efficient computation. However, in the current case with many-to-many mapping and for the given initial chemical state, same new chemical state can be observed with various input state configuration in a probabilistic manner. These novel paths leading to the same chemical state is an outcome of the increase in the connectivity in the configuration space in the hybrid electronic-chemical probabilistic computation. In the one- and two-dimensional chemical cellular automata and computation as described in the later sections, the hybrid chemical-electronic state machine uses these probabilistic pathways for the novel computation. This state expansion of many-to-many as compared to one-to-one mappings can be simply quantified by the ratio of input states (many-to-many) and chemical states (equivalent to input states in one-to-one) which is given by $p^{n.n} q^{2n(n-1)} k^{-n.n}$. For the n×n array, with two possible PWM levels for each central and interfacial stirrer such that $p, q, k = 2$, the state expansion ratio is $2^{84} = 1.9 \times 10^{25}$.

## 5.2 Phenomenological model for 1D Chemical Cellular Automata rules

Inspired from the elementary CA rules, we also developed a family of One-dimensional Chemical Cellular Automata(1D-CCA) rules based on a closed-loop state machine utilizing PWM and chemical states. As described in the previous section, a CCA rule can be defined on an experimental platform where based on the observed chemical states, PWM levels of interfacial and cell stirrers selected. In the absence of interfacial stirrers, we can recreate elementary CA rules within a closed feedback loop due to one-to-one mapping between PWM states of cell stirrers and observed chemical states. However, when the interfacial stirrers are active, due to hydrodynamic coupling between the neighbouring cells and coupled hysteresis effects, one-to-one mapping between PWM and chemical states does not exist and novel patterns can emerge out.

Here, we describe the 1D-CCA state machines and phenomenological model which was then used to simulate the emergence of patterns in a one-dimensional geometry. 1D-CCA rules which define the state machine to act on stirrer based on chemical states comprises of two different values, **{Rule_A}-{Rule_B}**. **Rule_A** updates the central cell stirrer PWM state based on chemical states of nearest neighbouring cells ($C_i^t, C_{i-1}^t, C_{i+1}^t$) similar to the elementary CA rule table. **Rule_B** updates the PWM states of the two interfacial stirrers based on the chemical states of the two connecting cells ($C_{i-1}^t, C_i^t$) and ($C_i^t, C_{i+1}^t$). Under this scheme, there are $2^3 = 8$ possible chemical states for three neighbouring cells. Similarly, to the elementary CA rules, there are $2^8 = 256$ possible rules for **Rule_A**. There are $2^2$ possible chemical states for two neighbouring connecting cells at the interfacial stirrer, hence total $2^2 2^2 = 16$ rules are possible for **Rule_B** for both interfacial stirrer. The total number of 1D-CCA rules possible is $2^8 2^4 = 4096$.

Fig. S50 (A) shows rule tables for rules defined by 30-{i} where i is between 1 and 16, as defined by Rule_B. In each of the rule table, the top row (in black and white) are the two chemical states (CS$_0$: black/CS$_1$: white) and the bottom row corresponds to the PWM values of the central cell stirrer and the two interfacial stirrers. There are two possible PWM states for central cell stirrers (drawn as Red and Green) as well as for interfacial stirrers (drawn as Red and Blue).

As an example, the eighth rule table as shown in Fig. S50 (A) is equivalent to,

Cell Stirrers

$\{\{1,1,1\} \to 0, \{1,1,0\} \to 0, \{1,0,1\} \to 0, \{1,0,0\} \to 1, \{0,1,1\} \to 1, \{0,1,0\} \to 1,$

$\{0,0,1\} \to 1, \{0,0,0\} \to 0\}$

Interfacial Stirrers

$\{\{1,1,1\} \to \{1,1\}, \{1,1,0\} \to \{1,1\}, \{1,0,1\} \to \{1,1\}, \{1,0,0\} \to \{1,0\}, \{0,1,1\} \to \{1,1\},$

$\{0,1,0\} \to \{1,1\}, \{0,0,1\} \to \{0,1\}, \{0,0,0\} \to \{0,0\}\}$

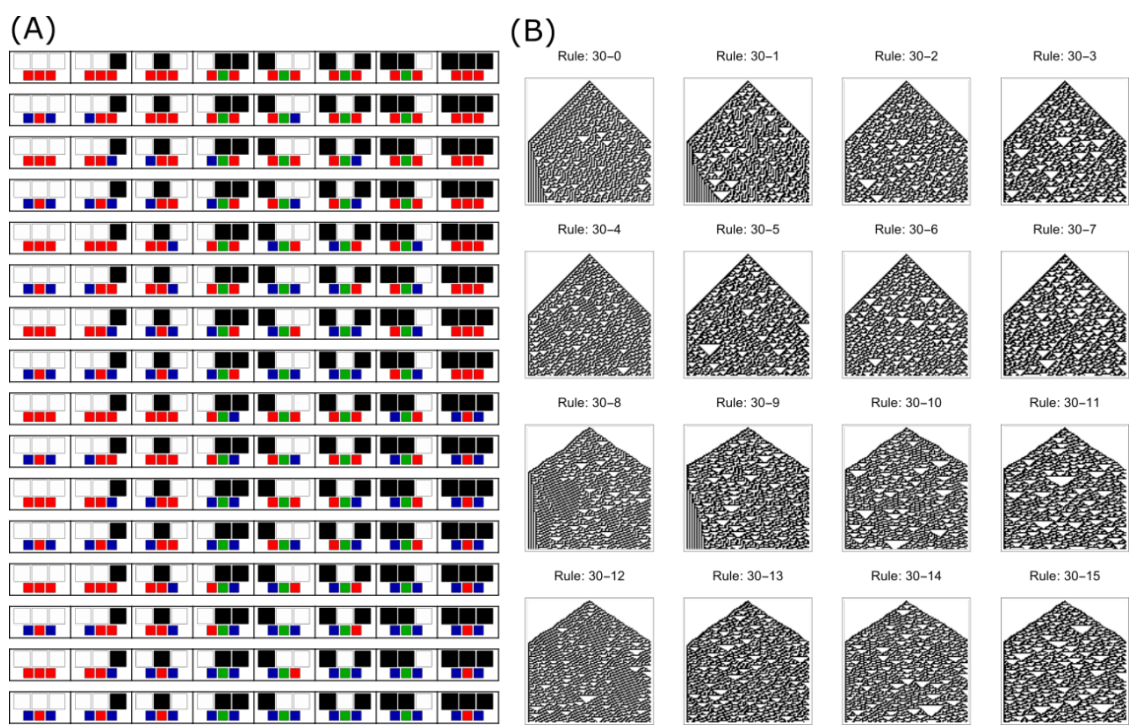

**Fig. S50:** Rule table and evolution of 1D-CCA rule 30-{i} with i ∈ [1, 16]. (A) shows rule table for all 16 rules for interfacial stirrers with CA rule 30 for cell stirrers and (B) Simulated 1D-CCA rule 30-{i} with i ∈ [1, 16] using a phenomenological model.

To simulate the 1D-CCA, we used a simple phenomenological model based on probabiluistic outcomes which updates the chemical states of all the cells based on the PWM values of cell and interfacial stirrers. Here, we define $S_i^t$ as the PWM state of the cell stirrer, $I_{i,i+1}^t$ as the PWM state of interfacial stirrer between $i^{th}$ and $i + 1^{th}$ cells. Here, we define the chemical state of the $i^{th}$ cell is given by $CS_i^{t+1}$. Assuming two different PWM states for both cell and interfacial stirrers, and two different chemical states, we can define the probability for the chemical states purely based on PWM levels of cells and interfacial stirrers. The probabilities for the high chemical state (1) at different conditions are,

$$\begin{cases} S_i^t == 1 \text{ and } S_{i-1}^t == 1/0 \text{ and } S_{i+1}^t == 1/0 \text{ and} \\ I_{i,i+1}^t == 1/0 \text{ and } I_{i,i-1}^t == 1/0 \end{cases} : CS_i^{t+1} \to \mathcal{P}(1) = 1$$

$$\begin{cases} S_i^t == 0 \text{ and } S_{i-1}^t == 0 \text{ and } S_{i+1}^t == 0 \text{ and} \\ I_{i,i+1}^t == 1/0 \text{ and } I_{i,i-1}^t == 1/0 \end{cases} : CS_i^{t+1} \to \mathcal{P}(1) = 0$$

$$\begin{cases} S_i^t == 0 \text{ and } S_{i-1}^t == 1/0 \text{ and } S_{i+1}^t == 1/0 \text{ and} \\ I_{i,i+1}^t == 0 \text{ and } I_{i,i-1}^t == 0 \end{cases} : CS_i^{t+1} \to \mathcal{P}(1) = 0$$

$$\begin{cases} S_i^t == 0 \text{ and } S_{i-1}^t == 1 \text{ and } S_{i+1}^t == 1 \text{ and} \\ I_{i,i+1}^t == 1 \text{ and } I_{i,i-1}^t == 1 \end{cases} : CS_i^{t+1} \to \mathcal{P}(1) = 0.8$$

$$\begin{cases} S_i^t == 0 \text{ and } S_{i-1}^t == 1 \text{ and } S_{i+1}^t == 1/0 \text{ and} \\ I_{i,i+1}^t == 0 \text{ and } I_{i,i-1}^t == 1 \end{cases} : CS_i^{t+1} \to \mathcal{P}(1) = 0.5$$

$$\begin{cases} S_i^t == 0 \text{ and } S_{i-1}^t == 1/0 \text{ and } S_{i+1}^t == 1 \text{ and} \\ I_{i,i+1}^t == 1 \text{ and } I_{i,i-1}^t == 0 \end{cases} : CS_i^{t+1} \to \mathcal{P}(1) = 0.5$$

$$\begin{cases} S_i^t == 0, S_{i-1}^t == 1, S_{i+1}^t == 0, \\ I_{i,i+1}^t == 1/0, I_{i,i-1}^t == 1 \end{cases} : CS_i^{t+1} \to \mathcal{P}(1) = 0.5$$

$$\begin{cases} S_i^t == 0, S_{i-1}^t == 0, S_{i+1}^t == 1, \\ I_{i,i+1}^t == 1, I_{i,i-1}^t == 1/0 \end{cases} : CS_i^{t+1} \to \mathcal{P}(1) = 0.5$$

These probabilities were assigned based on our knowledge on how the experimental setup behaves on the actuation of cell and interfacial stirrers. These probabilities were assigned based on the expected outcome such as, in the first case if the cell stirrer of the central cell is active (PWM state $S_i^t$ = 1) independent of the other controls, the probability of the occurrence of the high chemical state is 1 as due to the stirrer action the chemical state of the cell will be high. If the central cell stirrer is inactive, however, both the neighbouring cells and the interfacial stirrers between them are active (see condition 4 above), the probability of occurrence of the high chemical state was chosen to be 0.8. This high probability is due to the combined interaction from both neighbours together when the interfacial stirrers are also active. Similarly, instead of two neighbouring cells if only one neighbouring cell stirrer is active together with the active interfacial stirrers, the probability for the occurrence of high chemical state was chosen to be 0.5. It is important to note that, these assign probabilities are based on the certain values of PWM levels of cells and interfacial stirrers. We can tune these probabilities by selecting another set of PWM values.

The single step of the simulation consists of running both 1D-CCA state machine for a given rule and phenomenological state machine to update the new chemical states as described by the two-state machines $\boldsymbol{D}$ and $\boldsymbol{C}$.

$$\{S_i^t, I_{i,i-1}^t, I_{i,i+1}^t\} \leftarrow \boldsymbol{D}\,(CS_i^t, CS_{i-1}^t, CS_{i+1}^t)$$

$$CS_i^{t+1} \leftarrow \boldsymbol{C}\,(S_i^t, S_{i-1}^t, S_{i+1}^t, I_{i,i+1}^t, I_{i,i-1}^t)$$

where $\boldsymbol{D}$ is the 1D-CCA rule and $\boldsymbol{C}$ is the probabilistic state machine as defined above. We implemented the 1D-CCA simulation in Mathematica 12 (Wolfram Ltd.) where based on the probabilities, new chemical states are calculated using random choice between the states (0/1) based on weights defined by the probabilities. An example code for 1D-CCA is available on GitHub. Fig. S50 B shows a simulation of 1D-CCA rule 30-{i} with i ∈ [1, 16], where the first rule is similar to elementary CA rule 30 and shows the perfect one-to-one mapping between chemical states and PWM states of cell stirrers. Significant deviations from the basic rule can be observed by introducing the effect of interfacial stirrers, which gives rise to new behaviours. However, it is important to mention that 1D-CCA rules are not completely deterministic. Due to strong hysteresis effects and hydrodynamic coupling, there is associated stochasticity for switching between chemical states. 1D-CCA allows us to program the stochastic effects in the system by selecting PWM levels for cell stirrers as well as interfacial stirrers. We have also demonstrated an experimental example of the emergence of new chemical states by running elementary cellular automata rule 30 with symmetric and antisymmetric couplings (see Fig. S48). Additional 1D CCA examples are shown in Fig. S51 and Fig. S52.

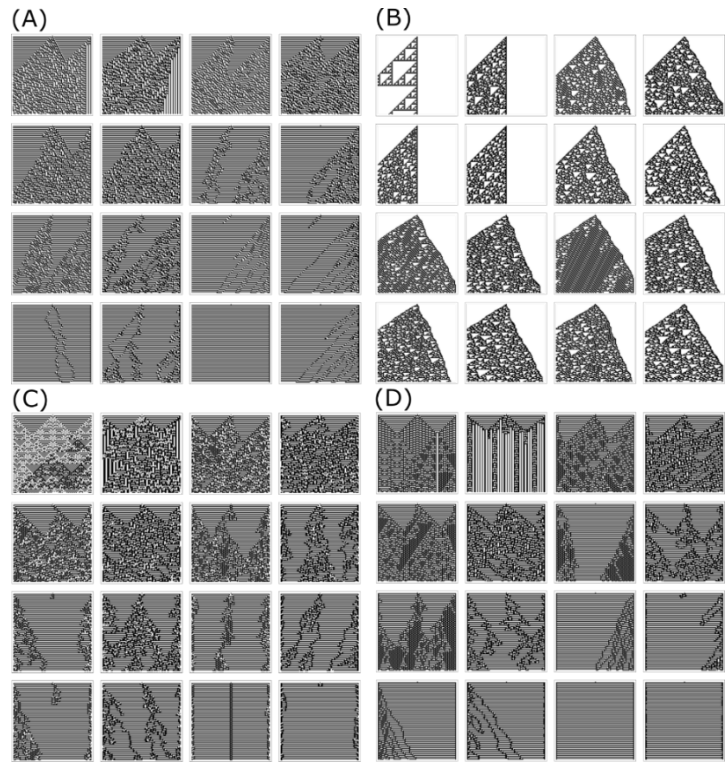

**Fig. S51:** One-dimensional 1D-CCA rule Examples 1-4. (A) 1D-CCA Rule 101-{1-16} (B) 1D-CCA Rule 110-{1-16} (C) 1D-CCA Rule 105-{1-16} (D) 1D-CCA Rule 109-{1-16}. Here, {1-16} means set of 16 different rules.

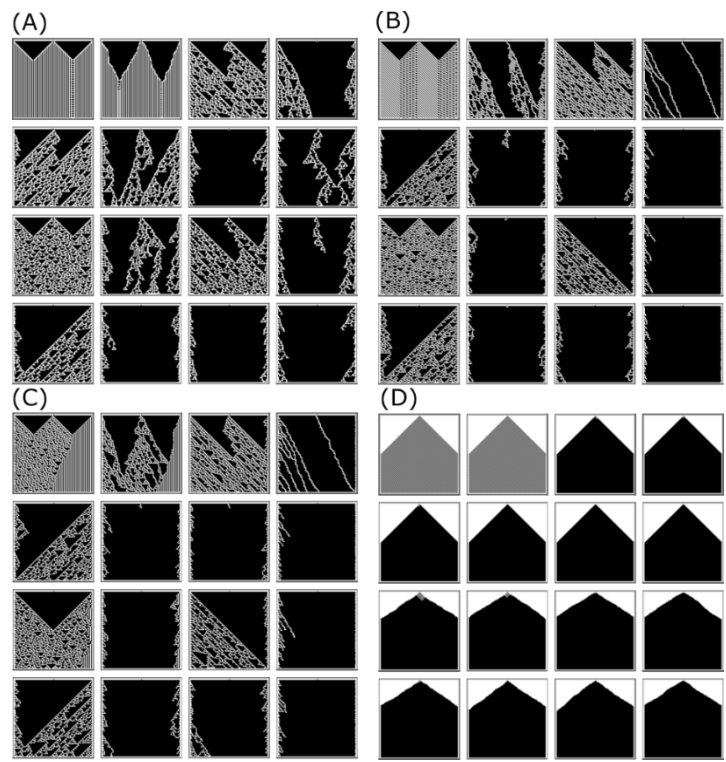

**Fig. S52:** One-dimensional 1D-CCA rule Examples 5-8. (A) 1D-CCA Rule 133-{1-16} (B) 1D-CCA Rule 145-{1-16} (C) 1D-CCA Rule 149-{1-16} (D) 1D-CCA Rule 250-{1-16}. Here, {1-16} means set of 16 different rules.

# 6 The Chemits: 2D hybrid electronic Chemical Automata

Two-Dimensional Chemical Cellular Automata (2D-CCA) is a zero-player two-dimensional Cellular Automata with chemical entities (Chemits) driven by the Chemical Cellular Automata (CCA), which are triggered by chemical oscillations. The key feature is that the Chemit is extended and comprised of 5 cells, with one central cell and its four neighbours. The central cell is the Chemit's core which is essential for existence. The four surrounding cells are used to interact with the fluctuating environment. These cells are not crucial for the existence of the Chemit and they can disappear and appear again during propagation, replication and competition events. Further details of events are described in the later subsections.

In the 2D-CCA experiments, we used four different PWM levels of the cell stirrers defined as $PWM_1 = 0$ (maximum PWM value is 255) for no interaction such that chemical state of the cell returns to the lower state, $PWM_2 = 22$ to create random fluctuations in the environment as BZ chemical oscillator is a damped oscillator, $PWM_3 = 50$ to create the core cell of the Chemit, $PWM_4 = 30$ for the nearest neighbours of the core cell of the Chemit which is used for interaction with other Chemits and random fluctuations. The experimental loop consists of applying PWM states to the cells together with clocking logic, readout the chemical states as described in the previous sections. We use the 2D-CCA State Machine to update the new PWM states of all the cell stirrers. We give a detailed description of the 2D-CCA State Machine in the next subsection.

## 6.1 2D-CCA state machine

The pseudo-code of the state machine which reads in the observed chemical states and updated the PWM levels of all the stirrers is described as follows,

    **Input:**
        **Distance(x, y):** Returns the geometrical distance between x and y.
        **PWMs** with stirring speed for every cell, in which the element can be:
            0 (**OFF**),
            1 (**RANDOM FLUCTUATION**),
            2 (**NEAREST NEIGHBOUR**),
            3 (**CENTER OF THE CHEMIT**).

    **CCA** is the chemical cellular automata states for every cell, in which the element can be **blue** or **lightblue**.

**Output**:
**PWM_new** with updated stirring speed for every cell.
**PWM_interfacial** with stirring speed for interfacial stirrers, which can be
    1 (**HIGH** speed) or
    0 (**LOW** speed).

Initialize element ← 0 in **Freeze**.
(**Freeze** is an array recording if we can change a specific cell (1) or not (0))
Initialize element ← 0 in **PWMs_new**.
Initialize element ← 0 in **PWMs_interfacial**.

**For** *cell* satisfying **PWM**(*cell*) == 3:
    **If SUM(CCA_neighbor(*cell*) == blue) == 0**:
        pass
    **Else**:
        *cell_p*←RandomSelection(*cell* satisfying **CCA_neighbor**(*cell*) == blue)

        **If** distance(*cell_p*, *cell*) == 1 **AND PWM**(*cell_p*) == 3:
        (competition event)
            Probability (**PWM_new**(cell) ← 1) = 0.5
            Probability (**PWM_new**(cell) ← 3) = 0.5

        **If** distance(*cell_p*, *cell*) = 1 **AND PWM**(*cell_p*) ≠ 3:
            (propagation event)
            **PWM_new**(*cell*) ← 1
            **Freeze**(*cell*) ← 1
            **PWM_new**(*cell_p*) ← 3

        **If** distance (*cell_p*, *cell*)>1 **AND PWM**(*cell_p*) ≠3:
        (replication event)
            **PWM_new**(*cell_p*) ← 3

**For** *cell* satisfying **PWM_new**(*cell*)==3:
    **PWM_intefacial**(*cell*) ← On
    **For** *cell_p* satisfying Distance(*cell_p*←*cell*)==1 **AND Freeze**(*cell_p*)==0:
        **PWM_new**(*cell_p*) ← 2
        **Freeze**(*cell_p*) ← 1
    **Freeze**(*cell*) ← 1

**For** *cell* satisfying **Freeze**(*cell*) =0,
    **random**(**PWM_new**(*cell*) ←1)

## 6.2 Phenomenological model

To simulate the dynamics of the Chemits emerging in the 2D-CCA, we developed a simple model of the evolution of chemical states based on the PWM actuation of the cell stirrers. We avoid details of physical models describing hydrodynamic interactions between and cells and its coupling to BZ oscillator kinetics. Computational models of BZ reaction coupled with hydrodynamic interactions are extremely complex to simulate and computationally intensive. Our phenomenological model comprises of two different state machines,

1. **2D-CCA State Machine ($D$)**: Same state machine as used in experiments which read in chemical states and update the PWM states.
2. **Phenomenological State Machine ($C$)**: Based on the observed phenomena in 2D-CCA experiments, it reads in PWM states and previous chemical states and outputs the new chemical states.

The time-stepping the simulation occurs in two steps, at each step first the new PWM states of all the cells were calculated from the state machine $D$,

$$PWM_{ij}^t = D(PWM_{ij}^t, PWM_{pq}^t, CS_{ij}^t, CS_{pq}^t, CS_{mn}^t)$$

where $CS_{ij}^t$ is the chemical state of $ij^{th}$ cell, $CS_{pq}^t, CS_{mn}^t$ are the chemical states of the nearest and next-nearest neighbours and $PWM_{ij}^t, PWM_{pq}^t$ are the PWM states of the central and the neighbouring cells. Once the PWM states were updated, new chemical states were calculated from the second phenomenological state machine $C$,

$$CS_{ij}^{t+1} = C(PWM_{ij}^t, PWM_{pq}^t, CS_{ij}^t, CS_{pq}^t)$$

where $CS_{ij}^t$ defines the chemical state of the $ij^{th}$ cell and $CS_{pq}^t, CS_{mn}^t$ describes of chemical states of nearest neighbours and the next-nearest neighbours. To describe the state machine $C$, we define a list of probabilities for switching chemical states for different scenarios of previous chemical states and the applied PWM levels. Out of the four different PWM levels, for nearest neighbouring interactions, we only consider the effect of $PWM_3$ and $PWM_4$.

**nPWM$_j$:** Number of neighbouring cells with PWM value j.

If (**nPWM$_3$ ≥ 3**): $C_{ij}$ = **P$_1$**

Else If (**nPWM$_3$ ≥ 1 and nPWM$_3$ < 3**): $C_{ij}$ = **P$_2$**

Else If (**nPWM$_4$ ≥ 3 and nPWM$_3$ == 0**): $C_{ij}$ = **P$_3$**

Else If (**nPWM₄ ≥ 1 and nPWM₁ ≤ 3 and nPWM₃**): $C_{ij}$ = **P₄**

Else: $C_{ij}$ = **0**

We defined additional parameter based on the PWM state of the central cell given by $Q$ which can take three different values $\{Q_1, Q_2, Q_4\}$ for PWM level of the central cell $\{PWM_1, PWM_2, PWM_4\}$ respectively. When the PWM of the central cell is $PWM_3$, in that case, there is a high probability that the next chemical state will be High (1), so the effect of neighbours is irrelevant. We also introduced an additional factor $K$ which introduces the effect of the current chemical state due to the hysteresis effect, whether $CS_{ij}^t == 0$ or 1. We defined $K$ for ij$^{th}$ cell as,

$$K_{ij} = \begin{cases} 0.7 & if\ CS_{ij}^t == 0 \\ 1 & if\ CS_{ij}^t == 1 \end{cases}$$

Using these parameters, for a given cell the probability of the high chemical state (1) is given by,

$$P_{ij}(1) = K_{ij} \begin{cases} Q_1 C_{ij} & if\ PWM_{ij} == 1 \\ Q_2 C_{ij} & if\ PWM_{ij} == 2 \\ Q_3 & if\ PWM_{ij} == 3 \\ Q_4 C_{ij} & if\ PWM_{ij} == 4 \end{cases}$$

By estimating the probabilities for all the cells, new chemical states were updated by randomly selecting chemical states **{0/1}** based on weights given by $\{1 - P_{ij}(1), P_{ij}(1)\}$. The simulation requires eight different variables for estimating the probabilities {**P₁, P₂, P₃, P₄, Q₁, Q₂, Q₃, Q₄, K**}. The simulations were performed in Mathematica 12 (Wolfram Ltd.) with basic implementation and source code available on < https://github.com/croningp/BZComputation >.

The simulation parameters we chose are,

{**P₁** = 0.5, **P₂** = 0.3, **P₃** = 0.25, **P₄** = 0.1, **Q₁** = 0.0, **Q₂** = 0.1, **Q₃** = 0.5, **Q₄** = 0.5}

The simulation parameters were chosen similarly as described in the phenomenological model for one-dimensional CCA. As discussed previously, the probabilities can be easily modified by selecting different PWM levels. A single partial step of the simulation for a two-dimensional geometry with 50 × 50 cells to update the PWM states from chemical states using the phenomenological state machine is shown in Fig. S53.

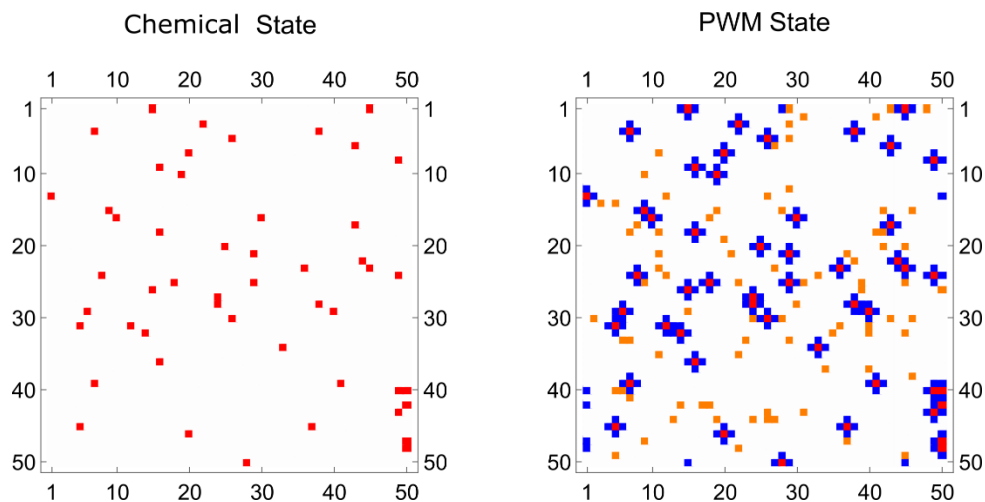

**Fig. S53:** A single simulated step of 2D-CCA using a phenomenological model on 50×50 grid. Left shows Chemical states (RED: 1 and WHITE: 0) which goes as input in the phenomenological state machine. Right shows new PWM states as an output of the state machine. (WHITE: $PWM_1$, ORANGE: $PWM_2$, RED: $PWM_3$, BLUE: $PWM_4$)

As a first step, we tested the outcomes 2D-CCA state machine to test the emergence of propagation, replication and competition events. We created the initial Chemit defined by PWM values $PWM_3$ (core) and $PWM_4$ (interacting nearest neighbours) on 5 × 5 grid. We created the chemical states ourselves to monitor the emergence of new PWM states showing the expected behaviour.

**Propagation Event**

As a first example, to demonstrate the chemical state we created a high chemical state at one of the nearest neighbours (right cell) as shown in Fig. S54. Using the old PWM states and the chemical states, we used the 2D-CCA state machine as described above and create the new PWM states. As shown in the figure, the Chemit propagated in the expected direction. It is important to note that the position of the Chemit can be precisely defined by the position of $PWM_3$ core cell. The surrounding interaction cells can appear or disappear and are not considered as a necessary feature of the Chemit. In the right figure, the core cell of the Chemit propagated to right with some additional random fluctuations ($PWM_2$) occurring as an outcome of the 2D-CCA state machine.

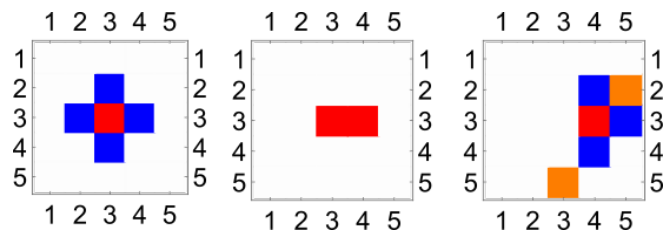

**Fig. S54:** Demonstration of propagation event in simulation over the 5×5 grid. Left shows the initial state of the Chemit described by PWM levels. The middle shows input chemical state which could lead to propagation event (High chemical state at one of the nearest neighbours). Right shows updated PWM state using the 2D-CCA state machine showing the propagation of Chemit.

**Replication Event**

Similar to the propagation event, with initial PWM levels describing the position of the Chemit, we created the high chemical state at one of the next nearest neighbour (bottom right) and used the 2D-CCA state machine to get the new PWM state as shown in Fig. S55. In our construction of state machine, we chose replication events possible when the chemical state at the next nearest neighbour is high to limit the replication events relative to propagation events. We observed replication event as expected in the 5 × 5 grid, see Fig. S55.

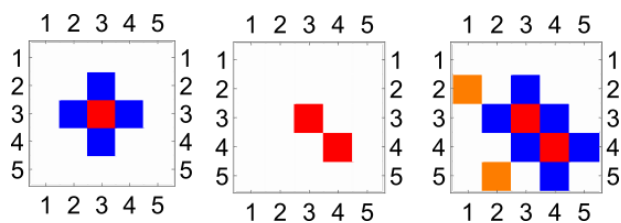

**Fig. S55:** Demonstration of replication event in simulation over the 5×5 grid. Left shows the initial state of the Chemit described by PWM levels. The middle shows input chemical state which could lead to replication event (High chemical state at one of the next nearest neighbours). Right shows updated PWM state using the 2D-CCA state machine showing replication of Chemit.

**Competition Events**

When the density of Chemits increases due to the replication event, two Chemits can meet with each other and a competition event may happen. In the competition event, the Chemit observes a high chemical state which is already the core of another Chemit and interact with it. There is a 50% chance to survive. Since we loop through all the Chemits, if two Chemits are competing with each other, the probability for (i) both of them survive, (ii) both of them die and (iii) one of them survives are 25%, 25% and 50% respectively. Fig. S56 shows three different scenarios, in the first case both Chemits survive, in the second case both Chemits annihilate each other, and in the third case a single Chemit survives.

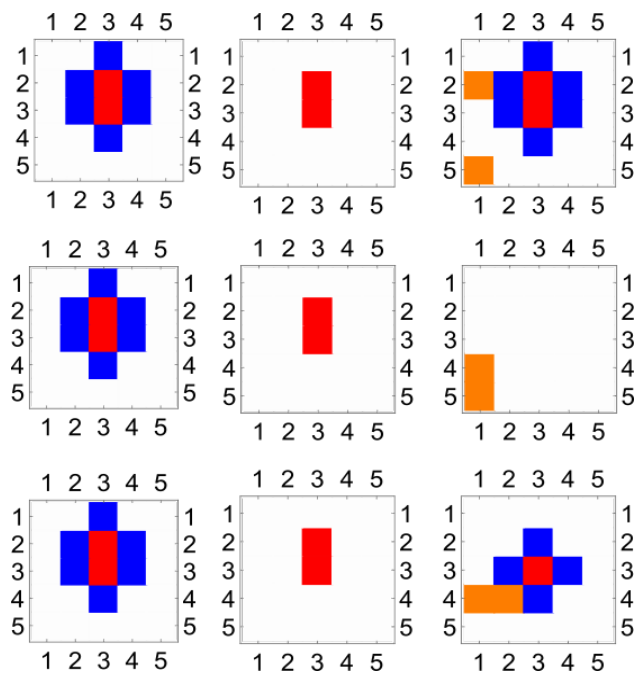

**Fig. S56:** Demonstration of competition events in simulation over the 5×5 grid. The left column shows the initial state of the Chemit described by PWM levels. The middle shows input chemical states with different possibilities of competition events which are same for all the three cases. Right shows updated PWM states using the 2D-CCA showing survival of both, none of the two and single species.

**Random selection among Multiple Events**

In the propagation and replication events, only one of the nearest or the next nearest neighbour was at high chemical state. However, in the experiments, we observed the occurrence of multiple chemical high chemical states simultaneously due to strong coupling among the nearest neighbours and hysteresis in chemical oscillations. In this case, among all the high chemical states, we randomly chose one cell among nearest and next-nearest neighbours, which define competition between possible propagation and replication events. Fig. S57 shows three different examples of multiple events showing propagation in the first and the third case, replication in the second case.

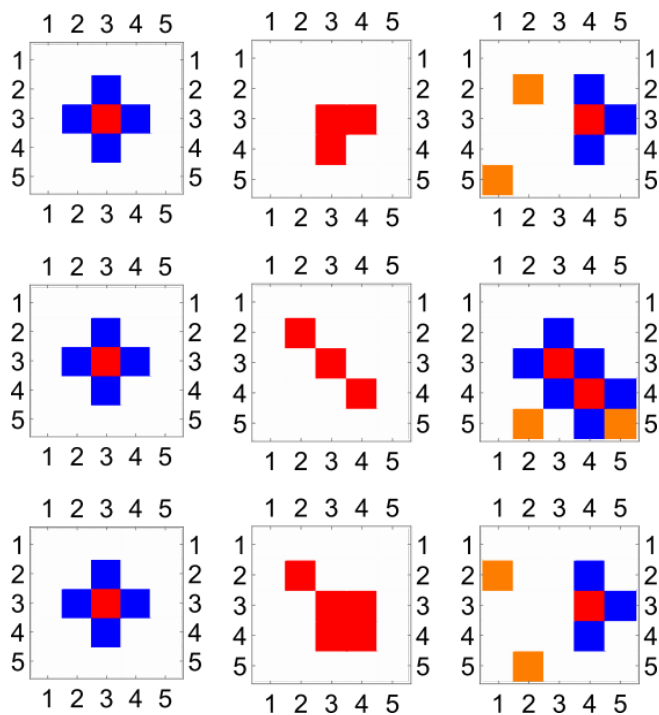

**Fig. S57:** Demonstration of random selection during multiple events in simulation over the 5×5 grid. The left column shows the initial state of the Chemit described by PWM levels. The middle shows input chemical states with different possibilities of competition events (1) two high chemical state nearest neighbours, (2) two high state next-nearest neighbours, and (3) presence of nearest and next-nearest neighbours simultaneously. Right shows updated PWM states using the 2D-CCA state machine showing propagation or replication of Chemits depending on the random selection between the various high chemical states.

As a next step, we tested the simulation step by coupling phenomenological state machine together with 2D-CCA state machine as shown in Fig. S58. For a single step, we start with initial PWM states created using randomly placed Chemits over 50 × 50 array. Fig. S58 (left) shows all four PWM states present in the initial state. Using the phenomenological state machine, we updated the chemical states which are shown in Fig. S58 (middle). It is easy to observe a high density of chemical states at the positions of $PWM_3$ core cells and their neighbours. As a second step, we update the PWM states from the chemical states using 2D-CCA state machine as shown in Fig. S58 (right). We used two-dimensional Periodic Boundary Conditions (PBC) for the Chemits if they cross the boundary cells. Comparing the initial and final PWM states, we can observe propagation, replication and competition events at a different location on the two-dimensional grid.

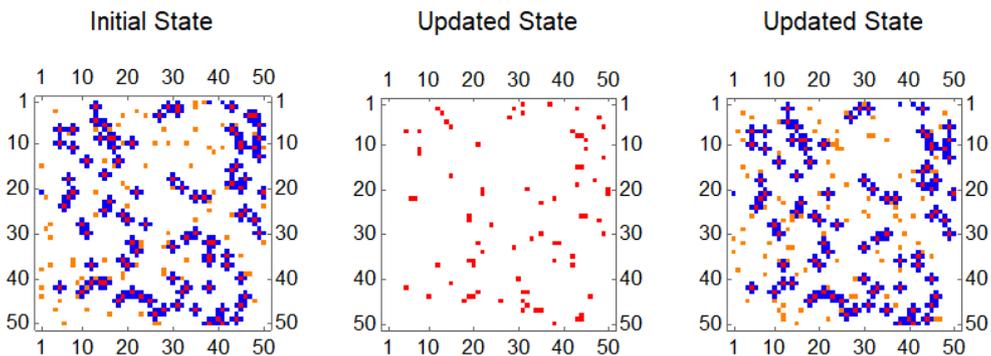

**Fig. S58:** Single simulation time step over the 50×50 array. The left figure shows the initial state of the Chemits described by PWM levels with all four PWM values (Red pixels shows the position of the Chemit core). The middle figure shows updated chemical states using phenomenological state machine using initial PWM states as inputs. Right shows updated PWM states using 2D-CCA state machine completing a full single step of the simulation.

### 6.3 Simulations and results

Using the simulation scheme as described in the previous subsection, we ran simulations over 100 × 100 grid and up to 15000 simulation steps with various initial conditions and parameters. Due to the stochastic nature of the simulations, we ran the same simulation 25 times and estimated the average properties such as the population of Chemits, propagation, replication events over the total simulation time. An example code written in Mathematica 12 notebook is available to download from <https://github.com/croningp/BZComputation>. We ran up to 8 simulations in parallel using Mathematica's parallel kernels. Once the simulation is completed, chemical states and PWM states at each time step were stored and dumped and Mathematica matrix object files (*.mx). A separate script also based on Mathematica was used to analyse the data and estimate the useful properties. In the next subsections, we will show results from simulations with variations in input parameters such as initial populations, size of the experimental space, frequency of random events.

### 6.3.1 Variable initial population

We ran simulations with various numbers of initial Chemits {1, 10, 100, 1000} over a 100 × 100 experimental grids for 15000 steps with 500 random actuation events at each time step. With the different initial numbers of Chemits, we ran simulations 25 times for each case. A complete movie of the simulations is available as Supplementary Video 5. From the simulation data, we estimate the number of Chemits by estimating the number of $PWM_3$ states at each time step. We also quantified propagation, replication and annihilation events throughout the

S78

simulation. To estimate these quantities, we first create a list of positions of all Chemits using $PWM_3$ states. At each time step, as a first step, we calculate the number of still states by finding the intersection between $PWM_3$ states at time t and t−1. The disappeared states between two consecutive steps occur due to propagation and annihilation events. The new appearing states between two consecutive steps occur due to propagation and replication events. For each of the new species appeared, we calculated the list of nearest neighbour and next-nearest neighbours. Based on the positions of new states and neighbours from the old states, we estimate the propagation and replication events. We estimate the annihilation events by subtracting the difference between new and old states from the replication events.

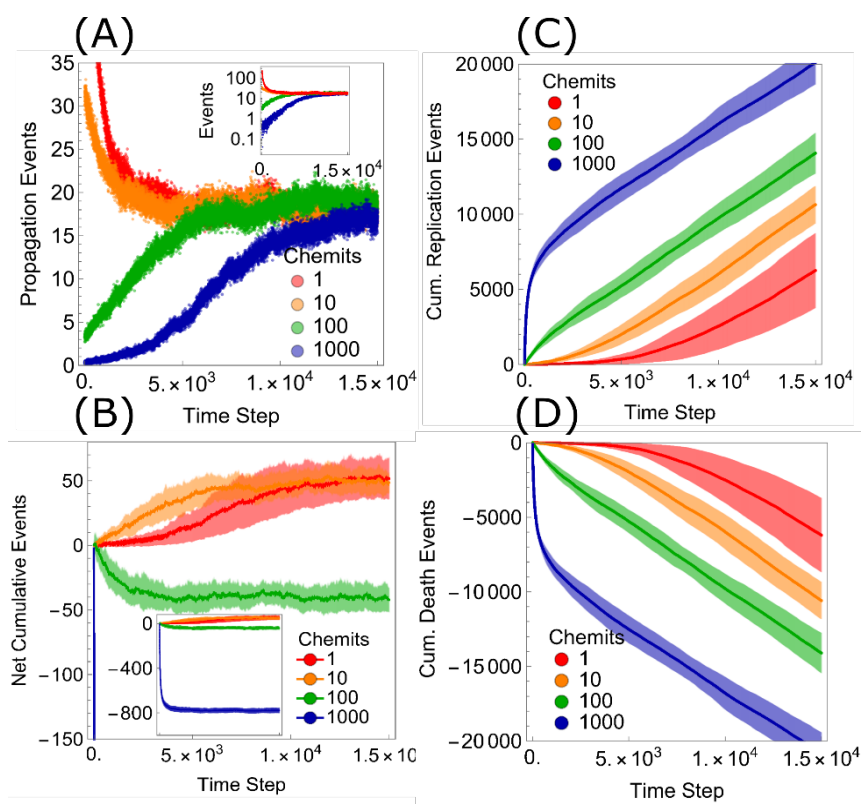

**Fig. S59:** Overall propagation and replication kinetics with different iunitial number of chemits. (A) shows overall propagation events, (B) net cumulative replication events, (C) cumulative replication events, (D) cumulative annihilation (death) events, with different initial number of chemits 1, 10, 100, 1000.

A Mathematica Notebook to analyse the simulation data is available on <https://github.com/croningp/BZComputation>. Fig. S59 (A, B) shows propagation and net cumulative replication events defined as a difference between replication and annhilation events averaged over 25 runs with different initial conditions. Fig. S59 (C, D) shows cumulative replication and annihilation events over the total time scale of the simulation. We observe with the single Chemit, as shown by , replication and propagation kinetics is slow at the start of the simulation. Once the number of species has crossed a threshold value, the growth rate, propagation and replication events start accelerating and reach steady-state value near the end of the simulation.

### 6.3.2 Variable total cell grid size

We investigated the effect of overall space available for Chemits to proliferate by changing the total cell grid size which can be 10 × 10, 20 × 20, 50 × 50, 100 × 100, 150 × 150. The total simulation steps were 10000 for each case and the given set of conditions, each simulation was performed 25 times and averaged properties were estimated.

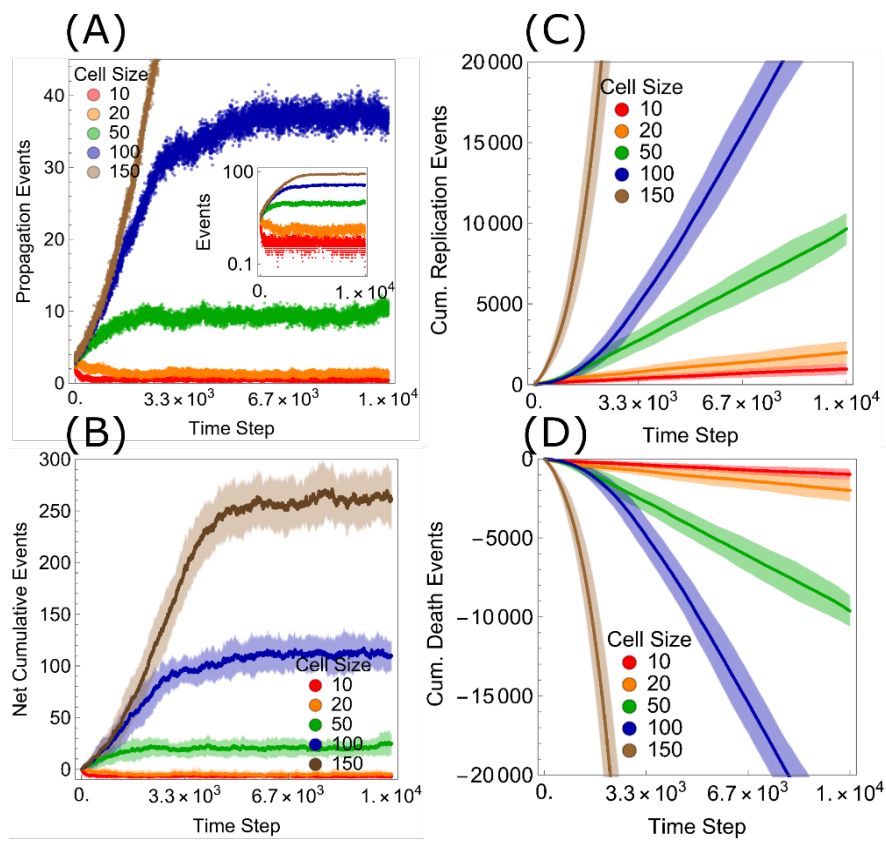

**Fig. S60:** Overall propagation and replication kinetics on a different number of cell grids. (A) shows overall propagation events, (B) cumulative replication events, (C) cumulative replication events, (D) cumulative annihilation (death) events, in five different number of cell grids of 10×10, 20×20, 50×50, 100×100, 150×150.

The initial population of Chemits in each case was kept the same (10 Chemits). The total random events at each time (defined by $PWM_2$) states were chosen based on a fixed ratio of the total number of available cells (1/10). As shown previously in the case of a different number of initial populations, that the Chemits show the tendency to achieve a steady-state population, the number of Chemits at the steady-state is controlled by the total amount of available space. The available space can be seen as an available spatial resource for the Chemits to proliferate or as a parameter which constraints the population of the Chemits and sets an upper bound (for the given set of other conditions such as random fluctuations). The variations in the propagation and replication dynamics at a different number of cell grid size is shown in Fig. S60.

## 7 BZ Computation

### 7.1 Ising model and implementation of combinatorial optimization Problems

Ising model is a mathematical formulation use in statistical mechanics to describe ferromagnetism. It comprises of magnetic spins $s \in (+1, -1)$ defined on a lattice, where each spin can interact with nearest neighbours and external magnetic field. The generalized Hamiltonian for a given configuration of spins $(s_1 \ldots s_n)$ is given by,

$$H(s_1 \ldots s_n) = -\sum_{i=1}^{n} h_i s_i - \sum_{i<j}^{n} h_{ij} s_i s_j$$

The generalized version for quadratic combinatorial optimization model can be described as

$$H(s_1 \ldots s_n) = -\sum_{i=1}^{n} p_i s_i - \sum_{i<j}^{n} q_{ij} s_i s_j$$

where $p_i$ and $q_{ij}$ are the coefficients which can be derived from cost/energy function of the combinatorial optimization problem. This Hamiltonian can be neatly formulated as the Quadratic Unconstrained Binary Optimization (QUBO),

$$H(x_1 \ldots x_n) = h^{(0)} + \sum_{i=1}^{n} h_i^{(1)} x_i + \sum_{i,j}^{n} h_{ij}^{(2)} x_i x_j$$

$h^{(0)}$ is the offset value, $h_i^{(1)}$ is the linear term, $h_{ij}^{(2)}$ is a second rank tensor describing pairwise interaction between the two variables $x_i$ and $x_j$ and $x \in (+1, 0)$. Here, we describe three different combinatorial optimization problems and defines the corresponding Hamiltonians for each problem.

**Number Partitioning Problem**

The number partition problem is defined as, for a given set of $N$ positive numbers P=$\{n_1, n_2, n_3 ... n_N\}$, find a partition, if possible, to two disjoint subsets R and P-R such that sum of all the numbers in both sets is equal. As an example, consider a set P = $\{1, 4, 6, 9\}$, the two disjoint subsets with sum of all elements equal are $\{1, 9\}, \{4, 6\}$.

The Hamiltonian for Number Partitioning for n-variables is given by *(34)*,

$$H = A\left(\sum_{i=1}^{N} n_i s_i\right)^2$$

where $n_i$ is the i$^{\text{th}}$ number in the set and $s_i$ is the associated spin variable. Here, we define Hamiltonians for two different sets of numbers with 4 and 6 numbers respectively, which can be solved in a digital computer using stochastic gradient descent, Fig. S61.

**Four number example**

Consider a set $S = \{1, 3, 4, 8\}$ which needs to be partitioned into two disjoint subsets $(S_1, S_2)$ such that sum of all the elements in both sets is equal. For the given set $S$, the Hamiltonian $H$ in the QUBO formulation using Ising to QUBO transformation $s_i = 2 x_i + 1$ is given by,

$$H = 256 - 64x_1 + 4x_1^2 - 192x_2 + 24x_1x_2 + 36x_2^2 - 256x_3 + 32x_1x_3 + 96x_2x_3 + 64x_3^2 \\ - 512x_4 + 64x_1x_4 + 192x_2x_4 + 256x_3x_4 + 256x_4^2$$

where $\{x_1, x_2, x_3, x_4\}$ are the associated QUBO variables mapped to chemical states of the experimental setup. As in the QUBO model, we have $x_i = x_i^2$, the Hamiltonian can be reformulated as,

$$H = 256 - 60x_1 - 156x_2 + 24x_1x_2 - 192x_3 + 32x_1x_3 + 96x_2x_3 - 256x_4 + 64x_1x_4 \\ + 192x_2x_4 + 256x_3x_4$$

The various coefficients of the generalized Hamiltonian corresponding to offset value, linear and coupling terms are given by,

$$h^{(0)} = 256$$

$$h^{(1)} = (-60, -156, -192, -256)$$

$$h^{(2)} = \begin{pmatrix} 0 & 12 & 16 & 32 \\ 12 & 0 & 48 & 96 \\ 16 & 48 & 0 & 128 \\ 32 & 96 & 128 & 0 \end{pmatrix}$$

To test the Hamiltonian formulation, we solved the problem by minimizing the energy using Stochastic Gradient Descent method.

**Six number example**

Similar to the previous case, consider a six-number set $S = \{1, 3, 4, 6, 5, 1\}$, Hamiltonian is given by,

$$\begin{aligned} H = {} & 400 - 80x_1 + 4x_1^2 - 240x_2 + 24x_1x_2 + 36x_2^2 - 320x_3 + 32x_1x_3 + 96x_2x_3 + 64x_3^2 \\ & - 480x_4 + 48x_1x_4 + 144x_2x_4 + 192x_3x_4 + 144x_4^2 - 400x_5 + 40x_1x_5 \\ & + 120x_2x_5 + 160x_3x_5 + 240x_4x_5 + 100x_5^2 - 80x_6 + 8x_1x_6 + 24x_2x_6 \\ & + 32x_3x_6 + 48x_4x_6 + 40x_5x_6 + 4x_6^2 \end{aligned}$$

where $\{x_1, x_2, x_3, x_4, x_5, x_6\}$ are the associated QUBO variables mapped to chemical states of the experimental setup. As in the QUBO model, we have $x_i = x_i^2$, the Hamiltonian can be reformulated as,

$$\begin{aligned} H = {} & 400 - 76x_1 - 204x_2 + 24x_1x_2 - 256x_3 + 32x_1x_3 + 96x_2x_3 - 336x_4 + 48x_1x_4 \\ & + 144x_2x_4 + 192x_3x_4 - 300x_5 + 40x_1x_5 + 120x_2x_5 + 160x_3x_5 \\ & + 240x_4x_5 - 76x_6 + 8x_1x_6 + 24x_2x_6 + 32x_3x_6 + 48x_4x_6 + 40x_5x_6 \end{aligned}$$

The coefficients of the Hamiltonian are given by,

$$h^{(0)} = 400$$

$$h^{(1)} = (-76, -204, -256, -336, -200, -76)$$

$$h^{(2)} = \begin{pmatrix} 0 & 12 & 16 & 24 & 20 & 4 \\ 12 & 0 & 48 & 72 & 60 & 12 \\ 16 & 48 & 0 & 96 & 80 & 16 \\ 24 & 72 & 96 & 0 & 120 & 24 \\ 20 & 60 & 80 & 120 & 0 & 20 \\ 4 & 12 & 16 & 24 & 20 & 4 \end{pmatrix}$$

To test the Hamiltonian formulation, we solved the problem using Stochastic Gradient Descent method.

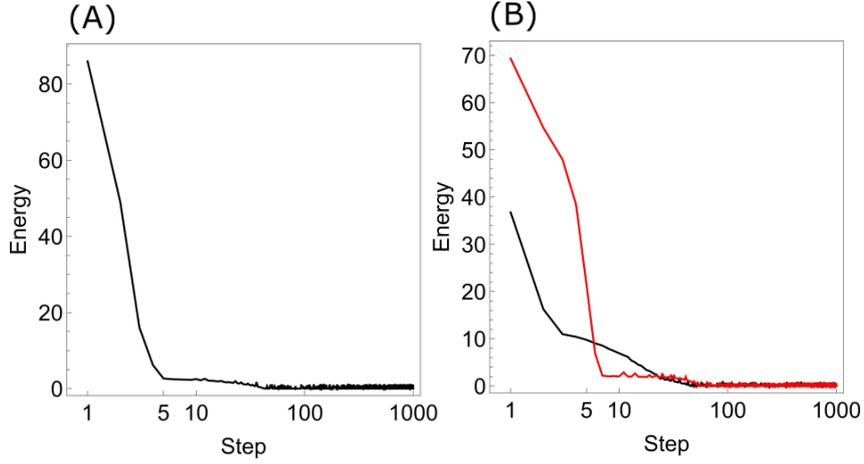

**Fig. S61:** Energy minimization of Number Partitioning Hamiltonian using Stochastic Gradient Descent. (A) shows energy minimization in four-number partitioning problem {1, 3, 4, 8} with the solution as {1, 3, 4} and {8}. (B) shows energy minimization for 6 number partitioning problem {1, 3, 4, 6, 5, 1} where we achieved two different solutions, {1, 3, 5, 1} shown in black and {4, 6} as well as {1, 3, 6}, {4, 5, 1} shown in red.

**Boolean Satisfiability problem**

The generalized Hamiltonian for the 2-SAT Boolean Satisfiability problem is given by,

$$H = A \sum_i^N \prod_j^2 (1 - w_{i,j} s_{i,j})$$

where $w_{i,j} \in \{-1, +1\}$ and $w_{i,j}$ is -1 and +1 for $\overline{s_{i,j}}$ and $s_{i,j}$ respectively and A is a constant. This is based on the genralized definition of k-SAT problem *(35)*. We formulate Hamiltonian for different two different examples.

**Example 1:**

The 2-SAT problem with 4 variables and 3 clauses in the conjunctive normal form is given by,

$$S = (s_1 \vee s_2) \wedge (s_2 \vee \overline{s_4}) \wedge (s_3 \vee s_4)$$

The Hamiltonian corresponding to this problem is given by,

$$H = 8 - 4x_1 - 4x_2 + 4x_1 x_2 - 4x_3 - 4x_2 x_4 + 4x_3 x_4$$

The coefficients of the Hamiltonian including offset, linear and quadratic terms are given by,

$$h^{(0)} = 8$$

$$h^{(1)} = (-4, -4, -4, 0)$$

$$h^{(2)} = \begin{pmatrix} 0 & 2 & 0 & 0 \\ 2 & 0 & 0 & -2 \\ 0 & 0 & 0 & 2 \\ 0 & -2 & 2 & 0 \end{pmatrix}$$

**Example 2:**

Another example of the 2-SAT problem with 4 variables and 6 clauses in the conjunctive normal form is given by,

$$S = (s_1 \lor s_2) \land (s_2 \lor \overline{s_4}) \land (s_3 \lor s_4) \land (s_1 \lor \overline{s_3}) \land (s_1 \lor \overline{s_2}) \land (\overline{s_3} \lor s_4)$$

Similarly to the previous example, the Hamiltonian, offset, linear and quadratic terms are given by,

$$H = 8 - 4x_1 + 4x_3 - 4x_1x_3 - 4x_2x_4$$

$$h^0 = 8$$

$$h^{(1)} = (-4, 0, 4, 0)$$

$$h^{(2)} = \begin{pmatrix} 0 & 0 & -2 & 0 \\ 0 & 0 & 0 & -2 \\ -2 & 0 & 0 & 0 \\ 0 & -2 & 0 & 0 \end{pmatrix}$$

**Travelling Salesman Problem**

The Hamiltonian for n-cities Travelling Salesman Problem is given by *(34)*,

$$H = A \sum_i^N (1 - \sum_j^N x_{i,j})^2 + A \sum_j^N (1 - \sum_i^N x_{i,j})^2 + A \sum_i^N \sum_j^N x_{i,j} x_{i+1,j}$$

where if i+1 is larger than N, i+1 is replaced by 1 and A is a constant.

We consider a 4-city travelling salesman problem, with four cities coordinates are given by City A: $\{0, 0\}$, City B: $\{1, 0\}$, City C: $\{3, 3\}$, City D: $\{0, 10\}$. The Ising Hamiltonian expression for these cities is given by,

$$H = 8. - 4. x_{1,1} + 2. x_{1,1}^2 - 4. x_{1,2} + 2. x_{1,1} x_{1,2} + 2. x_{1,2}^2 - 4. x_{1,3} + 2. x_{1,1} x_{1,3} + 2. x_{1,2} x_{1,3} + 2. x_{1,3}^2 - 4. x_{1,4} + 2. x_{1,1} x_{1,4} + 2. x_{1,2} x_{1,4} + 2. x_{1,3} x_{1,4} +$$
$$2. x_{1,4}^2 - 4. x_{2,1} + 2. x_{1,1} x_{2,1} + 0.01 x_{1,2} x_{2,1} + 0.0422 x_{1,3} x_{2,1} + 0.0995 x_{1,4} x_{2,1} + 2. x_{2,1}^2 - 4. x_{2,2} + 0.01 x_{1,1} x_{2,2} + 2. x_{1,2} x_{2,2} +$$
$$0.0359 x_{1,3} x_{2,2} + 0.1 x_{1,4} x_{2,2} + 2. x_{2,1} x_{2,2} + 2. x_{2,2}^2 - 4. x_{2,3} + 0.0422 x_{1,1} x_{2,3} + 0.0359 x_{1,2} x_{2,3} + 2. x_{1,3} x_{2,3} + 0.0758 x_{1,4} x_{2,3} +$$
$$2. x_{2,1} x_{2,3} + 2. x_{2,2} x_{2,3} + 2. x_{2,3}^2 - 4. x_{2,4} + 0.0995 x_{1,1} x_{2,4} + 0.1 x_{1,2} x_{2,4} + 0.0758 x_{1,3} x_{2,4} + 2. x_{1,4} x_{2,4} + 2. x_{2,1} x_{2,4} + 2. x_{2,2} x_{2,4} +$$
$$2. x_{2,3} x_{2,4} + 2. x_{2,4}^2 - 4. x_{3,1} + 2. x_{1,1} x_{3,1} + 2. x_{2,1} x_{3,1} + 0.01 x_{2,2} x_{3,1} + 0.0422 x_{2,3} x_{3,1} + 0.0995 x_{2,4} x_{3,1} + 2. x_{3,1}^2 - 4. x_{3,2} +$$
$$2. x_{1,2} x_{3,2} + 0.01 x_{2,1} x_{3,2} + 2. x_{2,2} x_{3,2} + 0.0359 x_{2,3} x_{3,2} + 0.1 x_{2,4} x_{3,2} + 2. x_{3,1} x_{3,2} + 2. x_{3,2}^2 - 4. x_{3,3} + 2. x_{1,3} x_{3,3} + 0.0422 x_{2,1} x_{3,3} +$$
$$0.0359 x_{2,2} x_{3,3} + 2. x_{2,3} x_{3,3} + 0.0758 x_{2,4} x_{3,3} + 2. x_{3,1} x_{3,3} + 2. x_{3,2} x_{3,3} + 2. x_{3,3}^2 - 4. x_{3,4} + 2. x_{1,4} x_{3,4} + 0.0995 x_{2,1} x_{3,4} + 0.1 x_{2,2} x_{3,4} +$$
$$0.0758 x_{2,3} x_{3,4} + 2. x_{2,4} x_{3,4} + 2. x_{3,1} x_{3,4} + 2. x_{3,2} x_{3,4} + 2. x_{3,3} x_{3,4} + 2. x_{3,4}^2 - 4. x_{4,1} + 2. x_{1,1} x_{4,1} + 0.01 x_{1,2} x_{4,1} + 0.0422 x_{1,3} x_{4,1} +$$
$$0.0995 x_{1,4} x_{4,1} + 2. x_{2,1} x_{4,1} + 2. x_{3,1} x_{4,1} + 0.01 x_{3,2} x_{4,1} + 0.0422 x_{3,3} x_{4,1} + 0.0995 x_{3,4} x_{4,1} + 2. x_{4,1}^2 - 4. x_{4,2} + 0.01 x_{1,1} x_{4,2} +$$
$$2. x_{1,2} x_{4,2} + 0.0359 x_{1,3} x_{4,2} + 0.1 x_{1,4} x_{4,2} + 2. x_{2,2} x_{4,2} + 0.01 x_{3,1} x_{4,2} + 2. x_{3,2} x_{4,2} + 0.0359 x_{3,3} x_{4,2} + 0.1 x_{3,4} x_{4,2} + 2. x_{4,1} x_{4,2} +$$
$$2. x_{4,2}^2 - 4. x_{4,3} + 0.0422 x_{1,1} x_{4,3} + 0.0359 x_{1,2} x_{4,3} + 2. x_{1,3} x_{4,3} + 0.0758 x_{1,4} x_{4,3} + 2. x_{2,3} x_{4,3} + 0.0422 x_{3,1} x_{4,3} + 0.0359 x_{3,2} x_{4,3} +$$
$$2. x_{3,3} x_{4,3} + 0.0758 x_{3,4} x_{4,3} + 2. x_{4,1} x_{4,3} + 2. x_{4,2} x_{4,3} + 2. x_{4,3}^2 - 4. x_{4,4} + 0.0995 x_{1,1} x_{4,4} + 0.1 x_{1,2} x_{4,4} + 0.0758 x_{1,3} x_{4,4} +$$
$$2. x_{1,4} x_{4,4} + 2. x_{2,4} x_{4,4} + 0.0995 x_{3,1} x_{4,4} + 0.1 x_{3,2} x_{4,4} + 0.0758 x_{3,3} x_{4,4} + 2. x_{3,4} x_{4,4} + 2. x_{4,1} x_{4,4} + 2. x_{4,2} x_{4,4} + 2. x_{4,3} x_{4,4} + 2. x_{4,4}^2$$

As in the QUBO model, we have $x_i = x_i^2$, the Hamiltonian can be reformulated as,

$$H = 8. - 2 x_{1,1} - 2 x_{1,2} + 2 x_{1,1} x_{1,2} - 2 x_{1,3} + 2 x_{1,1} x_{1,3} + 2 x_{1,2} x_{1,3} - 2 x_{1,4} + 2 x_{1,1} x_{1,4} + 2 x_{1,2} x_{1,4} + 2 x_{1,3} x_{1,4} - 2 x_{2,1} + 2 x_{1,1} x_{2,1} + 0.01 x_{1,2} x_{2,1} +$$
$$0.0422 x_{1,3} x_{2,1} + 0.0995 x_{1,4} x_{2,1} - 2 x_{2,2} + 0.01 x_{1,1} x_{2,2} + 2 x_{1,2} x_{2,2} + 0.0359 x_{1,3} x_{2,2} + 0.1 x_{1,4} x_{2,2} + 2 x_{2,1} x_{2,2} - 2 x_{2,3} + 0.0422 x_{1,1} x_{2,3} +$$
$$0.0359 x_{1,2} x_{2,3} + 2 x_{1,3} x_{2,3} + 0.0758 x_{1,4} x_{2,3} + 2 x_{2,1} x_{2,3} + 2 x_{2,2} x_{2,3} - 2 x_{2,4} + 0.0995 x_{1,1} x_{2,4} + 0.1 x_{1,2} x_{2,4} + 0.0758 x_{1,3} x_{2,4} + 2 x_{1,4} x_{2,4} +$$
$$2 x_{2,1} x_{2,4} + 2 x_{2,2} x_{2,4} + 2 x_{2,3} x_{2,4} - 2 x_{3,1} + 2 x_{1,1} x_{3,1} + 2 x_{2,1} x_{3,1} + 0.01 x_{2,2} x_{3,1} + 0.0422 x_{2,3} x_{3,1} + 0.0995 x_{2,4} x_{3,1} - 2 x_{3,2} + 2 x_{1,2} x_{3,2} +$$
$$0.01 x_{2,1} x_{3,2} + 2 x_{2,2} x_{3,2} + 0.0359 x_{2,3} x_{3,2} + 0.1 x_{2,4} x_{3,2} + 2 x_{3,1} x_{3,2} - 2 x_{3,3} + 2 x_{1,3} x_{3,3} + 0.0422 x_{2,1} x_{3,3} + 0.0359 x_{2,2} x_{3,3} + 2 x_{2,3} x_{3,3} +$$
$$0.0758 x_{2,4} x_{3,3} + 2 x_{3,1} x_{3,3} + 2 x_{3,2} x_{3,3} - 2 x_{3,4} + 2 x_{1,4} x_{3,4} + 0.0995 x_{2,1} x_{3,4} + 0.1 x_{2,2} x_{3,4} + 0.0758 x_{2,3} x_{3,4} + 2 x_{2,4} x_{3,4} + 2 x_{3,1} x_{3,4} +$$
$$2 x_{3,2} x_{3,4} + 2 x_{3,3} x_{3,4} - 2 x_{4,1} + 2 x_{1,1} x_{4,1} + 0.01 x_{1,2} x_{4,1} + 0.0422 x_{1,3} x_{4,1} + 0.0995 x_{1,4} x_{4,1} + 2 x_{2,1} x_{4,1} + 2 x_{3,1} x_{4,1} + 0.01 x_{3,2} x_{4,1} +$$
$$0.0422 x_{3,3} x_{4,1} + 0.0995 x_{3,4} x_{4,1} - 2 x_{4,2} + 0.01 x_{1,1} x_{4,2} + 2 x_{1,2} x_{4,2} + 0.0359 x_{1,3} x_{4,2} + 0.1 x_{1,4} x_{4,2} + 2 x_{2,2} x_{4,2} + 0.01 x_{3,1} x_{4,2} +$$
$$2 x_{3,2} x_{4,2} + 0.0359 x_{3,3} x_{4,2} + 0.1 x_{3,4} x_{4,2} + 2 x_{4,1} x_{4,2} - 2 x_{4,3} + 0.0422 x_{1,1} x_{4,3} + 0.0359 x_{1,2} x_{4,3} + 2 x_{1,3} x_{4,3} + 0.0758 x_{1,4} x_{4,3} + 2 x_{2,3} x_{4,3} +$$
$$0.0422 x_{3,1} x_{4,3} + 0.0359 x_{3,2} x_{4,3} + 2 x_{3,3} x_{4,3} + 0.0758 x_{3,4} x_{4,3} + 2 x_{4,1} x_{4,3} + 2 x_{4,2} x_{4,3} - 2 x_{4,4} + 0.0995 x_{1,1} x_{4,4} + 0.1 x_{1,2} x_{4,4} +$$
$$0.0758 x_{1,3} x_{4,4} + 2 x_{1,4} x_{4,4} + 2 x_{2,4} x_{4,4} + 0.0995 x_{3,1} x_{4,4} + 0.1 x_{3,2} x_{4,4} + 0.0758 x_{3,3} x_{4,4} + 2 x_{3,4} x_{4,4} + 2 x_{4,1} x_{4,4} + 2 x_{4,2} x_{4,4} + 2 x_{4,3} x_{4,4}$$

The coefficients of the Hamiltonian are given by,

$$h^0 = 8$$

$$h^{(1)} = (-2, -2, -2, -2, -2, -2, -2, -2, -2, -2, -2, -2, -2, -2, -2, -2)$$

$$h^{(2)} = \begin{pmatrix}
0 & 2 & 2 & 2 & 2 & 0.01 & 0.0422 & 0.0995 & 2 & 0 & 0 & 0 & 2 & 0.01 & 0.0422 & 0.0995 \\
2 & 0 & 2 & 2 & 0.01 & 2 & 0.0359 & 0.1 & 0 & 2 & 0 & 0 & 0.01 & 2 & 0.0359 & 0.1 \\
2 & 2 & 0 & 2 & 0.0422 & 0.0359 & 2 & 0.0758 & 0 & 0 & 2 & 0 & 0.0422 & 0.0359 & 2 & 0.0758 \\
2 & 2 & 2 & 0 & 0.0995 & 0.1 & 0.0758 & 2 & 0 & 0 & 0 & 2 & 0.0995 & 0.1 & 0.0758 & 2 \\
2 & 0.01 & 0.0422 & 0.0995 & 0 & 2 & 2 & 2 & 2 & 0.01 & 0.0422 & 0.0995 & 2 & 0 & 0 & 0 \\
0.01 & 2 & 0.0359 & 0.1 & 2 & 0 & 2 & 2 & 0.01 & 2 & 0.0359 & 0.1 & 0 & 2 & 0 & 0 \\
0.0422 & 0.0359 & 2 & 0.0758 & 2 & 2 & 0 & 2 & 0.0422 & 0.0359 & 2 & 0.0758 & 0 & 0 & 2 & 0 \\
0.0995 & 0.1 & 0.0758 & 2 & 2 & 2 & 2 & 0 & 0.0995 & 0.1 & 0.0758 & 2 & 0 & 0 & 0 & 2 \\
2 & 0 & 0 & 0 & 2 & 0.01 & 0.0422 & 0.0995 & 0 & 2 & 2 & 2 & 2 & 0.01 & 0.0422 & 0.0995 \\
0 & 2 & 0 & 0 & 0.01 & 2 & 0.0359 & 0.1 & 2 & 0 & 2 & 2 & 0.01 & 2 & 0.0359 & 0.1 \\
0 & 0 & 2 & 0 & 0.0422 & 0.0359 & 2 & 0.0758 & 2 & 2 & 0 & 2 & 0.0422 & 0.0359 & 2 & 0.0758 \\
0 & 0 & 0 & 2 & 0.0995 & 0.1 & 0.0758 & 2 & 2 & 2 & 2 & 0 & 0.0995 & 0.1 & 0.0758 & 2 \\
2 & 0.01 & 0.0422 & 0.0995 & 2 & 0 & 0 & 0 & 2 & 0.01 & 0.0422 & 0.0995 & 0 & 2 & 2 & 2 \\
0.01 & 2 & 0.0359 & 0.1 & 0 & 2 & 0 & 0 & 0.01 & 2 & 0.0359 & 0.1 & 2 & 0 & 2 & 2 \\
0.0422 & 0.0359 & 2 & 0.0758 & 0 & 0 & 2 & 0 & 0.0422 & 0.0359 & 2 & 0.0758 & 2 & 2 & 0 & 2 \\
0.0995 & 0.1 & 0.0758 & 2 & 0 & 0 & 0 & 2 & 0.0995 & 0.1 & 0.0758 & 2 & 2 & 2 & 2 & 0
\end{pmatrix}$$

**Implementation of Hybrid Electronic-chemical Logic**

Similar to those in 2D-CCA, we implemented a Finite State Machine logic *in silico* which reads in chemical states and updates the PWM states which are applied to the cell and interfacial stirrers. The evolution of chemical states is dominated by physicochemical principles which leads to the emergence of short-term memory or hysteresis effects in the system. In the current strategies to implement hybrid electronic-chemical computation logic, the combinatorial optimization problem is initialized by formulating an Ising Hamiltonian as a function of spin variables. Depending on the number of pairwise couplings, the computational problem was mapped to the two-dimensional experimental setup such that all the pairwise coupling variables in the Hamiltonian can exist as nearest neighbours. In the two-dimensional geometry, if all the couplings cannot be described by nearest neighbours, multiple instantiations of the same variables were created to map all the necessary pairs. These additional cells were called as auxiliary cells, and they flow the same logic as the primary cell associated with the given variable. Either the chemical states or PWM states were interpreted as Ising equivalent spins. At the start, all the spins are initialized randomly making sure that the auxiliary spins have the same states as the primary spins using. The emerging chemical states were read into the hybrid state machine, which updates the new PWM states. This information processing loop between digital and chemical domains was applied until the minimal energy configuration was achieved. Once the minimal energy configuration was achieved, the state of spin variables was interpreted as the solution to the combinatorial optimization problem.

**Electronic-Chemical State Machine Type 1**

In this hybrid electronic-chemical state machine, we used chemical states of the cells as the representation of Ising equivalent spins, where high chemical state $CS_1$ corresponds to +1 state and low chemical state $CS_0$ corresponds to -1. The Hamiltonian function was created *in-silico* and the system was initialized to random spins (chemical states). The basic pseudocode is defined as,

1. Initialize **PWM states** to randomly assign the **Chemical States** to all the cell associated with Hamiltonian variables.

2. Assign the **lowest energy** to be infinity.

3. While **True**:

    a. Randomly flip the **PWM state** of a cell and observe the emergence of all the new **Chemical States**.

    b. Calculate **current Energy** based on the **Chemical States**.

c. If **current Energy** ≤ **lowest energy**:

**lowest energy** = **current Energy**

d. **Accept** or **Reject** the flipping based on the estimated energy difference and update the **lowest energy**. The probability of accepting the change is:

$$p_{accept} = Max(e^{\frac{-\Delta E}{k}}, 1)$$

where $\Delta E$ is the energy difference after flipping the spin and we set $k = 5$.

e. If the **lowest energy cannot be further reduced (termination)**:

Save current spins and interpret them as a solution.

**Break**

This is the simplistic approach, where most of the information processing occurs in the digital domain, however the information loops through the chemical and digital domain. There is also inherent hysteresis in the system from the physicochemical processes which also influences the chemical states and prevents the system to get stuck in local minima. This step is crucial towards implementing more advanced logic and computational algorithms. Using this approach, we solved three different combinatorial optimization problems, whose associated Hamiltonians have been described in the previous subsection.

**Example 1**: 4 number partition Problem: {1, 3, 4, 8}

The stepwise spins, associated energy and minimum energy of the system are given by:

1. [-1, -1, -1, -1], {E = 256.0, $E_{min}$=256.0}
2. [-1, 1, -1, -1], {E = 100.0, $E_{min}$=100.0}
3. [1, 1, -1, -1], {E = 64.0, $E_{min}$=64.0}
4. [-1, 1, -1, 1], {E = 36.0, $E_{min}$=36.0}
5. [-1, 1, -1, -1], {E = 100.0, $E_{min}$=36.0}
6. [-1, 1, 1, 1], {E = 196.0, $E_{min}$=36.0}
7. [-1, 1, -1, 1], {E = 36.0, $E_{min}$=36.0}
8. [1., 1, -1, 1], {E = 64.0, $E_{min}$=36.0}
9. [-1, -1, -1, -1], {E = 256.0, $E_{min}$=36.0}
10. [-1, -1, -1, 1], {E = 0.0, $E_{min}$=0.0}

The final configuration of spins [-1, -1, -1, 1] for the 4 number set {1,3, 4, 8} splits the set into two disjoint subsets with -1 and +1 spins, {1, 3, 4} and {8} as the solution to the problem.

**Example 2**: 6 number partition Problem {1, 3, 4, 6, 5, 1}

Similar to the 4-number partition problem, for 6 number partition problem, the initial spins were assigned as, {-1, -1, -1, -1, -1, -1} with E=400. The hybrid logic was able to minimize the energy with the spin configuration, {-1, -1, 1, 1, -1, -1}. So, based on the achieved spins states, the six-number set splits into two disjoint subsets with -1 and +1 spins, {1, 3, 5, 1} and {4, 6} respectively.

**Example 3**: Travelling Salesman Problem: Four cities coordinate City A: {0, 0}, City B: {1, 0}, City C: {3, 3}, City D: {0, 10}. For 4 cities Travelling Salesman Problem, we initialized the system with 16 spins for pairwise interaction between the city connections as shown in the Hamiltonian in the previous subsection. The system was initialized with spins,

{{-1, -1, 1, -1}, {1, -1, -1, 1}, {1, -1, 1, 1}, {-1, -1, -1, -1}} with E=8.492. The hybrid logic was able to minize the energy with the final spin configuration,

{{-1, -1, -1, 1}, {1, -1, -1, -1}, {-1, 1, -1, -1}, {-1, -1, 1, -1}} with E=0.221. This configuration can be interpreted as the solution to the 4-cities Travelling Salesman Problem where the optimal path is defined as linking the +1 spin variables. The indices of spin +1 are {{1, 4},{2, 1}, {3, 2}, {4, 3}}, which defines the links between the cities and the solution to the Travelling Salesman Problem.

**Electronic-Chemical State Machine Type 2 (Extended Approach)**

In this hybrid electronic-chemical state machine, we used PWM states of the cell stirrers to represent Ising spin variables. The Ising spin variables based on all the pairwise coupling terms were mapped into the experimental domain using primary and auxiliary cells. Similar to the previous case, the Hamiltonian function was created *in-silico* and the system was initialized to random spins (PWM states). The hydrodynamic interactions between the neighbouring cells were introduced as a part of the hybrid logic. The quadratic formulation of the Hamiltonian allows writing the energy terms specific to a given cell (both primary and auxiliary) as the summation of interactions between the given cell and neighbouring cells. Similar to the first hybrid state machine, we flipped the PWM state randomly and introduce the effect of local interactions between the cells when we estimate the energy change for the pairwise interactions. To introduce chemical decision making, a comparison was made between the emerging chemical states and the lookup table which describes the ideal emergence of chemical states in the absence of hysteresis and noise effects (see Fig. 6 in the main text). Even though the phenomenological state machine is probabilistic, we can still estimate the most probable results from the local interactions logic. The overall energy change was calculated pair-wise and while calculating for a specific pair, if the observation of new chemical state is inconsistent with the lookup table, we change the sign of the energy change of this pair. The overall energy changed is accumulated pair-wisely and, similar to that in the first strategy, a decision is made to accept or reject the change of the PWM state using a Finite State Machine. The basic pseudocode is defined as,

1. Initialize **PWM states** randomly for all Ising spin variables associated with the problem Hamiltonian.

2. While **True**:

    a. Randomly flip the **PWM state** of the primary cell and create auxiliary cells related to all pairwise interactions.

    b. Set **energy_benefit** = 0

    c. For all possible pair interaction including **primary cell**:

    Calculate **pairwise energy change** after flipping the PWM state

    Create the auxiliary cell and activate interfacial stirrers for interactions

    If the new **Chemical States consistent** with **lookup table**:

    d_energy = pairwise energy change

    Else:

    d_energy = - pairwise energy change

    **Energy benefit** = Energy benefit + d_energy

    f. **Accept** the flipping if **Energy benefit** <= 0.

    g. If the configuration **energy reached minima (termination)**:

    Save current spins and interpret them as a solution.

**Break**

In this approach, we introduced a hybrid electronic-chemical logic where information processing loops between chemical and electronic domains. Similar to the previous logic, calculation of energy occurs in the electronic domain, however, energy benefit for each step is based on the chemical decision making which occurs due to strong hysteresis, non-linear couplings and noise which are inherent in the system. Using this approach, we solved three different combinatorial optimization problems, whose associated Hamiltonians have been described in the previous subsection.

**Example 1**: 4 number partition Problem: {1, 3, 4, 8}

The stepwise spins, associated energy of the system are given by:

1. [-1, 1, 1, -1], {E = 4.0}
2. [-1, 1, 1, 1], {E = 196.0}
3. [-1, 1, -1, 1], {E = 36.0}
4. [-1, 1, -1, -1], {E = 100.0}
5. [1, 1, -1, -1], {E = 64.0}
6. [1, 1, -1, -1], {E = 64.0}
7. [1, 1, -1, 1], {E = 64.0}
8. [-1, -1, -1, 1], {E = 0.0}
9. [-1, -1, -1, 1], {E = 0.0}

The final configuration of spins [-1, -1, -1, 1] for the 4 number set {1,3, 4, 8} splits the set into two disjoint subsets with -1 and +1 spins, {1, 3, 4} and {8} as the solution to the problem.

**Example 2**: 2 SAT problem (4 variable, 3 clauses)

In conjunctive normal form, the 2-SAT problem with 4 variables and 3 clauses is given by,

$$S = (s_1 \vee s_2) \wedge (s_2 \vee \overline{s_4}) \wedge (s_3 \vee s_4)$$

The system was initialized with spins [-1, -1, -1, -1] with total energy 8.0. The minimized energy configuration reached was [1, -1, 1, -1] with total energy 0.0. On these spin states, the solution is interpreted with +1 corresponds to 1 and -1 corresponds to 0. Hence, the solution becomes [1, 0, 1, 0], where the solution matches correctly ((1 or 0) and (0 and not 0), (1 and 0)).

**Example 3**: 2 SAT problem (4 variable, 6 clauses)

In conjunctive normal form, the 2-SAT problem with 4 variables and 6 clauses is given by,

$$S = (s_1 \vee s_2) \wedge (s_2 \vee \overline{s_4}) \wedge (s_3 \vee s_4) \wedge (s_1 \vee \overline{s_3}) \wedge (s_1 \vee \overline{s_2}) \wedge (\overline{s_3} \vee s_4)$$

The system was initialized with spins [-1, -1, -1, -1] with total energy 8.0. The minimized energy configuration reached was [1, 1, -1, 1] with total energy 0.0. On these spin states, the solution is interpreted with +1 corresponds to 1 and -1 corresponds to 0. Hence, the solution corresponds to minimum energy (0.0) is [1, 1, 0, 1].

## 7.2 Qualification of chemical computation

In this section, we describe the role of chemistry in the computational logic as implemented in the proposed hybrid electronic-chemical computer. To demonstrate the computation in hybrid electronic-chemical architecture, we redefine the question as follows:

1. Do one- & two-dimensional Chemical Cellular Automata and solving combinatorial optimization problems using the computational architecture give the complete picture towards hybrid computation?
2. Is there a one-to-one mapping between the electronic and chemical domain, such that digital logic controls all the phenomena and information just loops between electronic and chemical domain without any additional benefit?
3. What does the mapping between input PWM states and emerging chemical states look like?
4. Is it possible to simulate the complete electronic-chemical computational model in-silico?

In the following sections, we will describe our computational hybrid state machine in context to one- and two-dimensional Chemical Cellular Automata (CCA) as well as for chemical computation logic for combinatorial optimization problems. Based on our description, we aim to answer all four proposed questions. In the implemented hybrid computational logic, at each step the information loops between digital and chemical domains, where different parts of computation take place. To describe them clearly, as a first step we distinguish two different aspects of our hybrid computational architecture including information transfer and processing.

1. **Information Transfer and Time Stepping**
   The information transfer from the electronic to the chemical domain and vice versa is based on time-stepping logic which is essential for the proposed computational architecture. As the hybrid logic uses state machines in both digital and chemical domains, dynamic communication between them is essential and is considered as a separate process from Information Processing. The two information communication processes between electronic and chemical domains are described below.

a. **Electronic to Chemical:** Information transfer form electronic domain to chemical domain occurs by mapping the output of digital Finite State Machine to the PWM levels of mechanical stirrers, which influences the emergence of chemical states.

b. **Chemical to Electronic:** Information transfer from the chemical state to the electronic state occurs through imaging and a recognition state machine based on a neural network (CNN) that converts analogue chemical states ($\underline{CS}$) to the digital equivalent of chemical states ($CS$).

c. **Time Stepping:** Time stepping is an essential feature to synchronize chemical and electronic logic and is defined by the chemical oscillation clock in the chemical domain.

**2. Information Processing and Computational Operations**

The information processing logic and the implemented algorithms utilise both electronic and chemical processes. These include the digital and chemical states as well as the various state machines which act on chemical and digital states. Here, we define all states and state machines,

1. $\underline{CS_i^t}$ : defines the analogue chemical state of the i$^{th}$ cell at time step t which is equivalent to the chemical oscillation.
2. $CS_i^t$ : defines the digital chemical state of the i$^{th}$ cell at time step t which is the outcome of the recognition finite state machine (rFSM) based on a neural network ($CNN$) applied on the analogue chemical state.
3. $S_i^t$ : defines the digital PWM state of the cell stirrer of the i$^{th}$ cell at time t.
4. $I_{i,j}^t$ : defines the digital PWM state of the interfacial stirrer placed between the i$^{th}$ and j$^{th}$ cell at time t, where i and j are neighbouring cells.
5. $T(CNN)$: defines the state machine which acts on the analogue chemical state ($\underline{CS_i^t}$) and converts it into the digital chemical state ($CS_i^t$).
6. $D$ : defines the digital Finite State Machine which acts on $CS_i^t$ and updates cell stirrer states ($S_i^t$) and the interfacial stirrer states ($I_{i,j}^t$).
7. $P$ : defines the function which reads the PWM states and brings their effect to chemical analogue states in a physical world.

8. **$C$**: defines the chemical state machine which reads in the previous analogue chemical state ($CS_i^t$) and analogue equivalent physical interaction of the stirrer using function $P$. Here, we include the $T(CNN)$ state machine within $C$ for simplicity and hence directly updates the digital chemical state $CS_i^t$. It is important to note that even CNN occurs in the digital medium, we still consider it as a part of $C$ as it links two different versions of the chemical state.

9. **$K$**: defines the hybrid state machine which includes all the state machines ($T$, $C$, $P$, $D$), such that it reads digital chemical state and outputs directly the new chemical state.

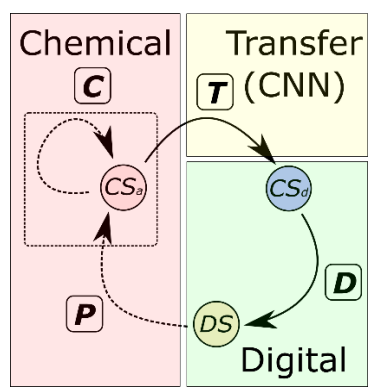

**Fig. S62:** Hybrid Electronic-Chemical Computational State Machine. The figure shows the information flow between the electronic $DS \equiv \{S_i^t, I_{i,j}^t\}$ and chemical domains ($CS_a \equiv CS_i^t$ and $CS_d \equiv CS_i^t$). We show transfer ($T$, which is *CNN*) as a separate unit from $C$ only because it occurs in the digital domain. However, it is included in $C$ in the state machine logic.

The complete representation of the hybrid electronic-chemical state machine is shown in Fig. S62, where information loops between the chemical and the digital domain. Based on the definition of various state variables and machines, we describe the electronic and chemical operations in details.

**Electronic Computation** comprises the digital Finite State Machine (**$D$**) which takes the digital chemical states of the cell and its neighbours as inputs and updates the new PWM states to be applied on the stirrers. Assuming only nearest neighbours in one-dimensional CCA, the new cell stirrer state and the two interfacial stirrers states can be defined as,

$$\{S_i^t, I_{i,i-1}^t, I_{i,i+1}^t\} = D\ (CS_i^t, CS_{i-1}^t, CS_{i+1}^t)$$

This general formulation of the state machine can be defined for any dimensions and higher-order neighbours as well. The digital state machine is deterministic and activates by the chemical clock signal.

**Chemical Computation** comprises a combination of two different state machines $P$ and $C$, hence reads the digital PWM states and updates the new chemical states based on the physical phenomena which include temporal BZ oscillation chemistry coupled with hydrodynamic interactions from the stirrers.

<u>It is important to note that the emergence of a new chemical state has both implicit and explicit dependence on the previous chemical state. By explicit, we mean the previous states determine the PWM level explicitly through $P$ and this PWM level influences the emergence of CS, while the hysteresis effect from previous operations is implicit.</u>

Furthermore, the emergence of a new digital chemical state is a complex function of the previous chemical state and multiple physical interactions which include neighbouring interactions and interaction of stirrers on the ongoing chemical oscillation. So, for one-dimensional CCA, we can write a new digital chemical state as,

$$CS_i^{t+1} = f\left(CS_i^t, S_i^t, S_{i-1}^t, S_{i+1}^t, I_{i,i+1}^t, I_{i,i-1}^t\right)$$

where we assume there is no direct dependence on the chemical states of the neighbouring cells and $f$ is a function that takes into account all the physical effects related to the given variables. We start with simplifying the function by excluding the neighbouring interactions and considering the central cell chemical state and PWM state of the cell stirrers. The functional relationship between current and previous chemical state is defined by (which has both hybrid electronic and chemical interaction components)

$$CS_i^{t+1} = f\left(CS_i^t, S_i^t\right)$$

Even we expect $f(x)$ to be a non-linear function, we split it into two different parts to distinguish chemical and electronic processing by assuming linear combinations of explicit (digital) and implicit (chemical) dependence of the previous chemical state with $\alpha$ is the weighting factor.

$$CS_i^{t+1} = \alpha\left(\underline{CS_i^t}\right) + (1-\alpha)\,P(S_i^t) = \alpha\left(\underline{CS_i^t}\right) + (1-\alpha)\,P(D(CS_i^t))$$

It is important to understand the difference between functions which involves $CS_i^t$ and $\underline{CS_i^t}$ even when both represent different aspects of the same chemical state. The two terms in the equation clearly show the separation of chemical and electronic processes in hybrid computation.

1. $\boldsymbol{\alpha}\left(\underline{CS_i^t}\right)$: occurs in a **chemical medium** and relates to the Hysteresis effect and noise.
2. $(1 - \boldsymbol{\alpha})\, \boldsymbol{P}(S_i^t)$: occurs in the **chemical medium** and is a physical process that includes hydrodynamic interaction between existing chemical oscillation that occurs over a single clocking step.
3. $S_i^t = \boldsymbol{D}(\boldsymbol{T}(CS_i^t))$: occurs in the **digital medium** using a Finite State Machine ($\boldsymbol{D}$).

**Term 1 and 2 corresponds to chemical information processing and are probabilistic while Term 3 corresponds to digital information processing which is deterministic.** In our experimental design, the parameter $\boldsymbol{\alpha}$ also tunable by careful selection of PWM levels. This flexibility in our computational architecture allows us to switch between deterministic and probabilistic domains for the efficient implementation of hybrid computation algorithms.

As a next step, we extend our formulation by introducing the effect of nearest neighbours. In the case of one-dimensional CCA, as discussed previously the functional relationship can be defined as,

$$CS_i^{t+1} = \boldsymbol{f}\,(CS_i^t, S_i^t, S_{i-1}^t, S_{i+1}^t, I_{i,i+1}^t, I_{i,i-1}^t)$$

Similarly, we formulate $\boldsymbol{f}$ as a linear combination of different interactions which can be described as

$$CS_i^{t+1} = \boldsymbol{\alpha}\left(\underline{CS_i^t}\right) + \boldsymbol{\beta}\,\boldsymbol{P_C}(S_i^t) + (1 - \boldsymbol{\alpha} - \boldsymbol{\beta})\boldsymbol{P}(S_i^t, S_{i-1}^t, S_{i+1}^t, I_{i,i-1}^t, I_{i,i+1}^t)$$

Here, we introduced two different types of transfer functions for different coupling interactions

$\boldsymbol{P_C} \equiv \boldsymbol{P_C}(S_i^t)$ describes the physical effect of the central cell stirrer on the chemical oscillation.

$P_I \equiv P_I(S_i^t, S_{i-1}^t, S_{i+1}^t, I_{i,i-1}^t, I_{i,i+1}^t)$ describes the physical effect of the interactions between neighbouring cells. This is complex phenomena due to the coupling vortices of three neighbouring cells via interfacial stirrers.

The various electronic operations which use the digital finite state machine $D$ are defined as,

$$S_i^t \equiv S_i^t(CS_i^t, CS_{i+1}^t, CS_{i-1}^t)$$

$$I_{i,i-1}^t \equiv I_{i,i-1}^t(CS_i^t, CS_{i-1}^t)$$

$$I_{i,i+1}^t \equiv I_{i,i+1}^t(CS_i^t, CS_{i+1}^t)$$

$$S_i^t, I_{i,i-1}^t, I_{i,i+1}^t = D(CS_i^t, CS_{i+1}^t, CS_{i-1}^t)$$

In this formulation, the role of digital and chemical information processing can be separated and can be described by the following,

1. $\alpha\left(CS_i^t\right)$: occurs in a **chemical medium** and relates to the Hysteresis effect and noise
2. $\beta\, P_C(S_i^t)$: occurs in a **chemical medium** describes the physical effect of the central cell stirrer on the oscillations.
3. $(1 - \alpha - \beta)P_I(S_i^t, S_{i-1}^t, S_{i+1}^t, I_{i,i-1}^t, I_{i,i+1}^t)$ : occurs in the **chemical medium** describes the physical effects of hydrodynamic coupling between neighbouring cells on the existing chemical oscillations.
4. $\{S_i^t, I_{i,i-1}^t, I_{i,i+1}^t\}$: occurs in the **electronic domain** controlled by the digital Finite State Machine which is defined as $S_i^t, I_{i,i-1}^t, I_{i,i+1}^t = D(CS_i^t, CS_{i+1}^t, CS_{i-1}^t)$

### 7.2.1 Time stepping in hybrid electronic-chemical logic

Using the description of state machine and variables which distributes information processing in electronic and chemical domains, we now describe the complete hybrid state machine $K$, and formulate time-stepping logic where at each step the information loops through chemical and electronic domains. The hybrid state machine uses digital processing ($D$) and physical and chemical processing state machines ($P$, $C$). To distinguish the initial state from the generalized state at time step t, we describe an additional state machine $C_0$ which represents the initial condition. The hybrid chemical-digital state machine $K$ acting on a chemical state is defined as

$$K^t(CS_i) \equiv C(P(D(CS_i, CS_j)), \underline{CS_i})$$

such that, we can describe time-stepping of the digital chemical state as,

$$CS_i^{t+1} = K^t(CS_i^t, CS_j)$$

where $CS_i^t$ represents the digital chemical state of the central cell and $CS_j$ describes the combined digital chemical state of all the neighbouring cells. At the initial state, for any CCA, the chemical state and PWM state of the underlying stirrer has one-to-one mapping which is controlled by the initial chemical state machine $C_0$ without any surrounding hysteresis effects and noise effects from chemical fluctuations.

The initial condition is well-defined by digital chemical states or equivalent stirrer operations and at this point, chemical interactions start.

$$IC = CS_i^0 = C_0(S_i^0, I_{i,j}^0)$$

$$IC = (S_i^0, I_{i,j}^0) = D(CS_i^0)$$

The temporal evolution is given by the interaction of the cell with its neighbouring cells together with hysteresis, hydrodynamic coupling and noise effects defined by chemical state machines $C$ and $P$ with the given initial conditions,

**Step 1:**

$$CS_i^1 = C\left(P(S_i^0, I_{i,j}^0), \underline{CS_i^0}\right) = C(IC)$$

$$(S_i^1, I_{i,j}^1) = D(CS_i^1, CS_j^1)$$

For the next step, to update the digital chemical state $CS_i^2$, we can write **Step 2** as:

**Step 2:**

$$CS_i^2 = C\left(P(S_i^1, I_{i,j}^1), \underline{CS_i^1}\right) = C(P(D(CS_i^1, CS_j^1)), \underline{CS_i^1}) = K^1(C(IC))$$

$$(S_i^2, I_{i,j}^2) = D(CS_i^2, CS_j^2)$$

where the hybrid state machine can be expanded as,

$$K^1(C(IC)) = C(P(D(C(IC))), C(IC))$$

**Step 3:**

$$CS_i^3 = C\left((S_i^2, I_{i,j}^2), \underline{CS_i^2}\right) = C\left(P(D(CS_i^2, CS_j^2)), \underline{CS_i^2}\right) = K^2(K^1(C(IC)))$$

$$(S_i^3, I_{i,j}^3) = D(CS_i^3, CS_j^3)$$

**Step N:**

$$CS_i^N = C\left(P(S_i^{N-1}, I_{i,j}^{N-1}), \underline{CS_i^{N-1}}\right) = C\left(P(D(CS_i^{N-1}, CS_j^{N-1})), \underline{CS_i^{N-1}}\right)$$
$$= K^{N-1}K^{N-2}\ldots K^1(C(IC))$$

$$(S_i^N, I_{i,j}^N) = D(CS_i^N, CS_j^N)$$

Hence, the general formulation for the digital chemical state at the t$^{th}$ time step to a hybrid chemical-electronic state machine is defined as,

$$CS_i^t = K^{t-1}K^{t-2}\ldots K^1(C(IC))$$

$$K^t(CS_i) \equiv C(P(D(CS_i^t, CS_j^t)), \underline{CS_i^t})$$

So, a hybrid chemical-electronic state machine $K$ comprised of three different operations which occur in the digital and analogue domain defined by three state machines $D, P$ and $C$. Based on this formulation of the hybrid chemical-electronic state machine, we can write our state machine in computation and "display screen" mode can be defined,

$$K_{comp}^t(CS_i^t) \equiv C(P(D(CS_i^t, CS_j^t)), \underline{CS_i^t})$$

$$K_{disp}^t(CS_i^t) \equiv C(P(D(CS_i^t, CS_j^t))) = H(D(CS_i^t, CS_j^t) - p)$$

$$H(x) = \begin{cases} 0, & x < 0 \\ 1, & x \geq 0 \end{cases}$$

where, $K_{comp}^t$ is the hybrid state machine in computation mode, $K_{disp}^t$ is the hybrid state machine in a fully deterministic display screen mode, $H(x)$ is a piecewise function and $p$ is the threshold stirrer level at which the digital chemical state changes. In the display screen mode, there are no hysteresis, physical coupling and noise effects, so that there exists a one-to-one mapping in the digital PWM states and chemical states. The pictorial representation of the hybrid chemical-electronic state machine in computation mode and in "display screen" mode is shown in **Fig. S63** and **Fig. S64**.

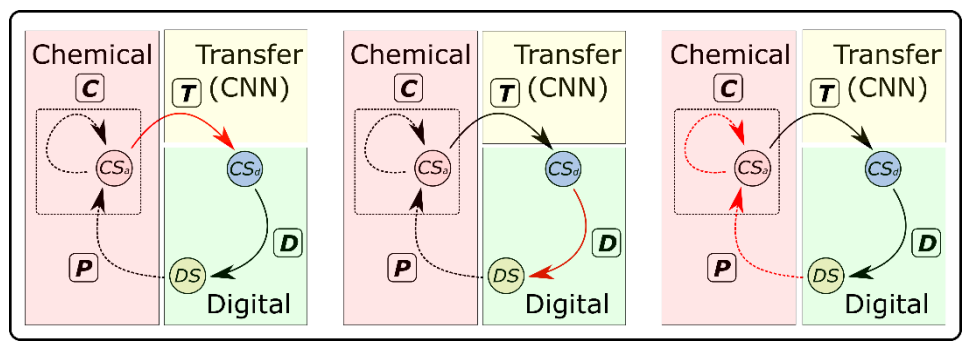

**Fig. S63:** Hybrid Electronic-Chemical Computational State Machine. (A-C) shows the concept of the proposed hybrid state machine showing information flow in chemical and electronic domains. The chemical domain comprises of two state machines $P$ and $C$ which computes in a probabilistic manner (shown in dotted) and the digital state machine $D$ computes in the digital domain in the deterministic manner (shown in continuous). The information transfer domain $T(CNN)$ converts analogue chemical state ($CS_i^t$) into the digital equivalent of a chemical state ($CS_i^t$).

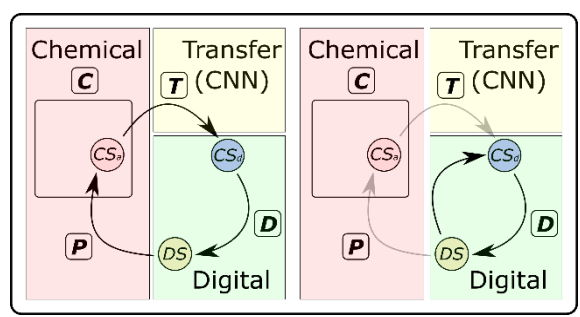

**Fig. S64:** Hybrid Electronic-Chemical "Display Screen" State Machine showing one-to-one mapping. (A, B) shows the concept of the proposed hybrid state machine in "display screen" mode where information loops between chemical and digital domains in the absence of hysteresis and probabilistic effects. Due to a complete deterministic information loop with one-to-one mapping, the information flow is equivalent to processing in a purely digital domain as shown in B.

**Emergence of Probabilistic Behaviour in Hybrid Computational Architecture**

In this section, we will describe the emergence of the probabilistic nature of computation in our proposed computational architecture. Given an initial state, we define response as the effect on the initial state due to the added stimulus of control input. Starting from the digital PWM states which are the output of the digital state machine, the response of the PWM value on the stirrer (motor) can be assumed to be linear. If we define emerging states based on only the response of the stirrer, we can distinct two output states as shown in Fig. S65 A and hence, well-defined deterministic output. Chemical response to stirrer input becomes non-linear, with a region where the stirrer response on emerging chemical states is strong (see Fig. S65 B). We have demonstrated similar behaviour experimentally (see SI Section 2.4, Fig. S26), where the peak oscillation amplitude rises strongly with the change in PWM values by creating a step function to increase the PWM values from 20 to 80.

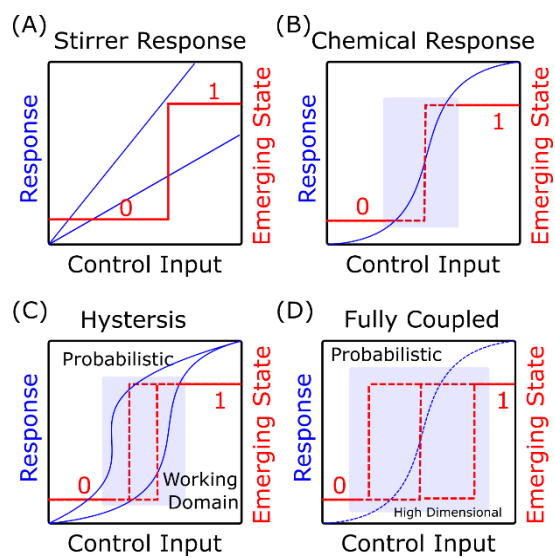

**Fig. S65:** Representation of probabilistic outcomes in hybrid computational architecture. (A) shows the simple linear response of the PWM value on the stirrer which is classified to deterministic digital states. (B) Chemical response on stirrer control input creating a non-linear response. (C) Hysteresis effects creating probabilistic outputs of the chemical states due to complex hydrodynamic interaction with the ongoing chemical oscillations. (D) Expanded probabilistic outcome space due to the non-linear response of PWM input, hysteresis effects and hydrodynamic couplings between neighbours.

Due to short term memory in the chemical system (forced-damped oscillator), the system shows strong hysteresis behaviour which leads to a complex interaction between the previous oscillatory state and non-linear chemical response from the stirrer. These emergent analogue chemical states when classified with CNN leads to probabilistic chemical states Fig. S65 C. However, during the actual computation, the interaction is not limited to the previous chemical oscillatory states and stirrer response, but also with the interactions of nearest neighbours and response from the multiple interacting cells and interfacial stirrers. The coupled oscillatory chemical states with high-dimensional nearest neighbours, hysteresis and stirrer interactions expand the probabilistic space, which when gets classified by CNN leads to probabilistic outcomes as shown in Fig. S65 D.

Based on the defined hybrid chemical-electronic state machine above, we further extend our formulation towards computation in one- and two-dimensional Chemical Cellular Automata (CCA) and identifying the role of chemistry in the computational architecture.

**Example 1: One-dimensional Chemical Cellular Automata**

In the one-dimensional CCA implementation, consider a 1D-CCA rule defined by $\boldsymbol{DS}$-$\boldsymbol{DI}$ where the finite state machine $\boldsymbol{DS}$ updates the central cell stirrer and $\boldsymbol{DI}$ updates the interfacial cell stirrers based on readout of digital chemical states. In this case, the transfer function state machine takes the output from two digital finite state machines ($\boldsymbol{DS}$ and $\boldsymbol{DI}$) and transfers both cell and interfacial stirrer operations into the chemical analogue domain. Hence, the generalized 1D-CCA hybrid chemical-electronic state machine is defined as,

$$\boldsymbol{K}^t(CS_i^t) \equiv \boldsymbol{C}\left(\boldsymbol{P}(\{\boldsymbol{DS}(CS_i^t, CS_j^t), \boldsymbol{DI}(CS_i^t, CS_j^t)\}), \underline{CS_i^t}\right)$$

The various implemented 1D CCA rules utilize the generalized hybrid state machine, where digital state machines ($\boldsymbol{DS}$ and $\boldsymbol{DI}$) were defined in a deterministic way and chemical/physical state machines ($\boldsymbol{D}, \boldsymbol{C}$) was implemented using a phenomenological probabilistic model. The emergence of novel patterns occurs due to increased connectivity in configuration space due to chemical information processing defined by ($\boldsymbol{D}, \boldsymbol{C}$).

Similarly, the implemented elementary CA rule state machine ("display screen mode") can be defined from the generalized state machine as,

$$\boldsymbol{K}_{el}^t(CS_i^t) \equiv \boldsymbol{C}\left(\boldsymbol{P}(\{\boldsymbol{DS}(CS_i^t, CS_j^t)\})\right) = H(\boldsymbol{DS}(CS_i^t, CS_j^t) - p)$$

which is equivalent to the 1D-CCA rule $\boldsymbol{DS}$-**0**, where there is no interaction between the neighbouring cells as interfacial stirrers are off and direct one-to-one mapping on the central cell for the given rule defined by $\boldsymbol{DS}$.

**Example 2: Two-dimensional Chemical Cellular Automata**

In the 2D CCA, the digital finite state machine that leads to propagation, replication, competition and multiple events of chemical entities (Chemits) is defined by $\boldsymbol{D_C}(CS_{i,j}^t, CS_{m,n}^t)$, which reads the local digital chemical states between the nearest and next-nearest neighbours where $CS_{i,j}^t$ represents the central cell and $CS_{m,n}^t$ represents all the neighbours collectively.

$$\boldsymbol{K}_C^t(CS_{i,j}^t, CS_{m,n}^t) \equiv \boldsymbol{C}(\boldsymbol{P}(\boldsymbol{D_C}(CS_{i,j}^t, CS_{m,n}^t)), (\underline{CS_{i,j}^t}, \underline{CS_{m,n}^t}))$$

The finite state machine $\boldsymbol{D_C}(CS_{i,j}^t, CS_{m,n}^t)$ for Chemits comprises two different parts which output four different PWM states of cell stirrers and two for interfacial stirrers,

1. $\boldsymbol{D_1}(CS_{i,j}^t, CS_{i,j}^t) \rightarrow \{1_{i,j}^t, 3_{i,j}^t, 4_{i,j}^t\}$ with 1, 3 and 4 are PWM levels corresponding to inactive, cell core and interaction stirrers.

2. $\boldsymbol{D_2}(CS_{i,j}^t) \rightarrow \{1_{i,j}^t, 2_{i,j}^t\}$ with 1 and 2 are PWM levels corresponding to inactive and random fluctuations stirrer to create surrounding weak oscillations.

The implementation of Chemical Entities (**Chemits**) in both experiments and simulations demonstrated emergent behaviour in the hybrid chemical-electronic state machine, where the Chemits interact with the fluctuating medium.

### 7.2.2 Chemical computation towards combinatorial optimization problems

In the previous section, we defined a hybrid state machine where both chemical and electronic state machines take part in the probabilistic computation. The computation in the two-dimensional Chemical Cellular Automata (CCA) shared between the chemical and digital domain, where the emergence of the chemical state drives the digital state machine. In this section, we extend the qualification of chemical computation by demonstrating the role of chemistry in solving combinatorial optimization problems.

The introduction of chemical logic into the hybrid algorithm increases the number of the trajectories to connect different configurations in an Ising system in BZ computation, and we use a discrete-time Markov chain to understand and analyze it. To emphasize the role of chemistry, we started to introduce a greedy search algorithm and later combined it with chemical states which form our chemical computation strategy.

For a system with $N$ spins, the total number of the configurations $C$ is $2^N$, where a configuration $c$ is defined as a set of specific spins values. The transition between configurations forms the key to our optimization. If in two configurations, there is only 1 spin with different values and the rest of the spins are the same, they are neighbouring configurations. In a greedy algorithm, only the transition to neighbouring configurations with lower energy is allowed. Although it helps with fast convergence, this algorithm can be easily trapped in a local minimum and starting with some configurations, it will never reach the global minimum due to the limited number of configuration connections.

Chemistry is probabilistic in the sense even though we define the current PWM level, the emerged CS is not purely dependent on the current PWM values but a probabilistic distribution. Although this distribution depends on the accumulation of all the previous operations, we can control the PWM values so that this distribution is not completely random but quite consistent with our look-up table with a probability $p_{chem} \pm \Delta p$, where $\Delta p$ is a small value counting the effects from previous states. We use a discrete-time Markov chain to analyze the optimization process given $\Delta p$ is small.

First, the set of configurations are defined as $\boldsymbol{C} = \{c_1, c_2, \ldots, c_{2^N}\}$ with that the individual configuration is a representation of the spins. Probability distribution in this configuration set is further defined with the initial condition $\boldsymbol{P_0} = \{p_{1,0}, p_{2,0}, \ldots, p_{2^N,0}\}$. The probability distribution at time step t is represented by $\boldsymbol{P_t} = \{p_{1,t}, p_{2,t}, \ldots, p_{2^N,t}\}$ with $\boldsymbol{P_t} = \boldsymbol{P_0} \boldsymbol{T}^t$, where $\boldsymbol{T}$ is the transition matrix with a size of $2^N \times 2^N$ and the summation within a row to be 1.

The ($i^{th}$, $j^{th}$) element in the transition matrix described the probability of configuration transition from $c_i$ to $c_j$. It is important to note that only the transition between neighbouring configurations is allowed. Now the problem remains to compare the energy of two neighbouring configurations $c_i$ and $c_j$ which defines the ($i^{th}$, $j^{th}$) element in $\boldsymbol{T}$. If the energy $c_j$ is smaller than $c_i$, we change the configuration from $c_i$ to $c_j$, else the configuration is still $c_i$. $\boldsymbol{P_t}$ will be converged given t is large enough.

We only need to consider the transition of neighbouring configurations and those that are non-neighbouring are 0. Let us assume we flip the spin $s_{h1}$ to $s_{h2}$. Then the energy difference can be calculated as:

$$\Delta E = (s_{h2} - s_{h1}) \sum_{i \neq h} H_{obs,(h,i)} K_{h,i} s_i$$

Note the summation goes through all the $2^N$ spins except for spin h and $K_{h,i}$ is the coefficients of the spin interactions. $\boldsymbol{H_{obs,(h,i)}}$ is from the comparison between observation and the look-up table, which is the crucial step for us to introduce probabilistic effects from chemistry. If they are consistent, $H_{obs,(h,i)} = +1$ else $H_{obs,(h,i)} = -1$.

Since $\Delta E$ is calculated from individual interaction terms, the chance of $H_{obs,(h,i)}$ being 1 is $p_{chem}$ for one term, which makes the overall distribution of $\Delta E$ being polynomial. If $\Delta E > 0$, this transition is rejected and $\Delta E \leq 0$ is accepted, which contributes to the probability in the diagonal (no change) and ($i^{th}$, $j^{th}$) element in the transition matrix respectively.

If $p_{chem}$ is 1, our algorithm reduces to the greedy search purely in the digital domain. If $p_{chem}$ is 0.5, regardless of the spins, the probability of $\Delta E \leq 0$ is 0.5 where the system loses the tendency to minimize energy and become ergodic. By programming the chemical computer, we tune the value of $p_{chem}$ and introduce the ergodicity from chemistry while enabling its tendency to minimize energy.

**Implementation via Chemical and Digital State Machines**

1. A set of spins $S = \{s_1, s_2, \ldots, s_N\}$ is randomized and stored in the digital computer.
2. The HIGH or LOW PWM values in the corresponding cells are applied according to the spin state. (**The digital finite state machine**)
3. One spin $s_h$ is randomly flipped from $s_{h1}$ to $s_{h2}$ and the pair-wise energy change was calculated in the digital computer. (Which is $(s_{h2} - s_{h1})K_{h,i}s_i$ for every $i$ spins) (**The digital finite state machine**)
4. Update the PWM values to the latest spin values, and the corresponding $H_{obs,(h,i)}$ was calculated from the chemical finite state machine. If the observation is consistent with the look-up table, $H_{obs,(h,i)} = +1$ else $H_{obs,(h,i)} = -1$. (**The chemical finite state machine**)
5. The overall energy change is calculated via $\Delta E = (s_{h2} - s_{h1})\sum_{i \neq h} H_{obs,(h,i)}K_{h,i}s_i$ and if $\Delta E < 0$, the change of the spin is accepted otherwise rejected.

**Example: Number Partitioning Problem**

Consider a number set $S = \{1, 3, 4, 9, 3, 5, 3, 6\}$ which needs to be partitioned into two disjoint subsets $(S_1, S_2)$ such as the sum of all the elements of $S_1$ and $S_2$ is equal. We compare the performance of a purely digital and hybrid strategy. We started with setting $p_{chem} = 1$, which creates a purely digital greedy algorithm. In this case, some initial configurations are already trapped in a local minimum, or they can lead to a local minimum, which means the probability of reaching the global minimum (thus solve the problem) is not possible. By slightly reducing $p_{chem}$ to 0.99, the ergodicity of the system is enabled and all the configurations can reach the global minimum, by sacrificing the performance of some configurations.

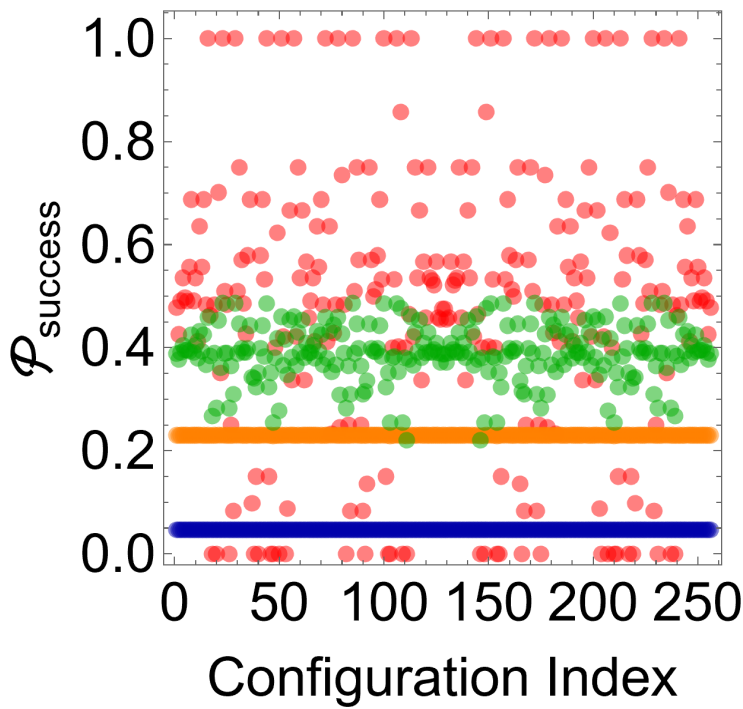

**Fig. S66:** Probability to reach minimum energy configuration using deterministic and probabilistic computational machines for a number partition problem. The figure shows the probability to reach minimal energy configuration starting at different initial configurations at different deterministic indices whose value describes digital, ergodic and hybrid algorithms. A deterministic index equals 1 (shown in red) corresponds to the purely digital deterministic greedy algorithm, and 0.5 corresponds to ergodic (shown in Blue) and 0.99/0.95 (shown in green/orange) corresponds to hybrid logic.

Here we drew the probability of reaching global minimum starting with different configurations under different $p_{chem}$. With decreased $p_{chem}$, this distribution is shrunk which means the preference of selecting a "good" configuration is decreased after introducing the ergodicity (see Fig. S66). As an example, Fig. S67 shows an example of the trajectory in the configurational space towards the solution to the number partitioning problem starting at configuration no. 50 in the configuration space. In the pure deterministic case (index=1), the system stuck into local minima and no part of the trajectory overlaps with solution configuration (red). In a pure ergodic approach (index=0.5), the algorithm spans through configurational space randomly and hence can find the solution but inefficiently. However, in the hybrid approach (index=0.99, 0.95), the trajectory overlaps with the solutions substantially and demonstrates an efficient combination using a probabilistic approach.

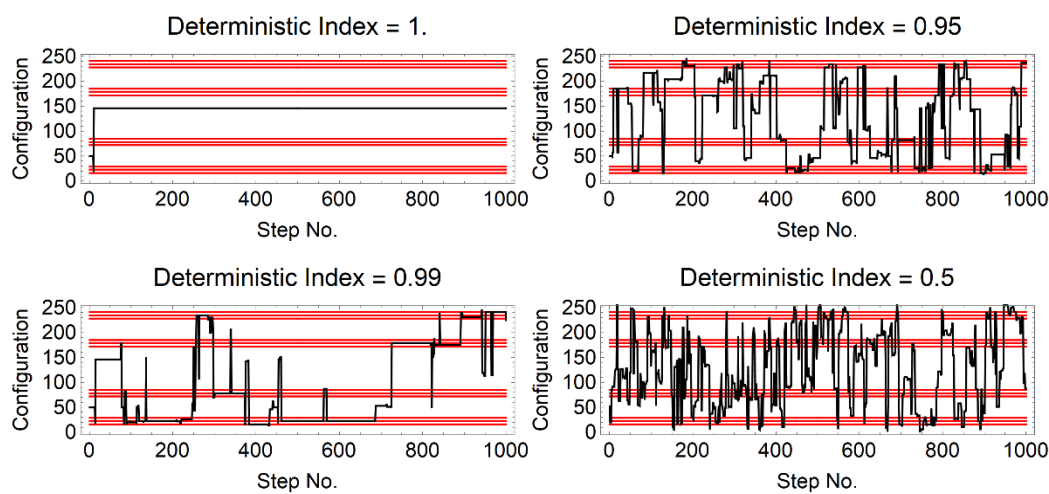

**Fig. S67:** Configuration Trajectories in 8-number partition problem. The figure shows trajectories of finding the solution of the number partitioning problem at different deterministic indices. A deterministic index equals 1 corresponds to the purely deterministic greedy algorithm, and other indices demonstrate the role of chemistry. The black lines are the actual trajectories spanned by the algorithms and the red lines correspond to the solution configurations.

### 7.2.3 Conclusions

In this section, we described the functionality of the proposed hybrid computation architecture and using the state-machine algebra, we demonstrated the role of electronic and chemical logic in constructing algorithms using the hybrid architecture. We posed questions at the start of the document which could be used to qualify our computational architecture towards the probabilistic chemical computer, which were answered in different aspects in various subsections. To answer these questions, we defined state variables and machines to describe information flow between electronic and chemical domains at each step in the hybrid computation with example in one- and two-dimensional CCA and combinatorial optimization problems. The roles of electronics and chemistry were shown by separating digital processing (deterministic finite state machine) from physical and chemical processes (probabilistic state machine). Hence, demonstrating that one-to-one mapping is a subset of probabilistic mapping between electronic and chemical states. Using the one- and two-dimensional CCA, we demonstrated that the computational architecture could work in both deterministic and probabilistic modes, which increases the configurational paths through which the system can evolve. We extended further by demonstrating a hybrid-computation algorithm to solve combinatorial optimization problem mapped to the Ising lattice ground state problem. We demonstrated the power of programmable probabilistic hybrid computation logic over deterministic algorithm by simplifying it to a Markovian process. The description of the probabilistic algorithm as the Markov process was considered due to the simplicity of the model to describe the role of chemical and physical phenomena. As the problem size scales, hybrid logic prevents the system from getting stuck into local minima. For a very large system, we expect a large number of iterations with various initial configurations are required to find the global minima. Instead, using hybrid computation logic, a single run is enough to find the minima as it utilizes the probabilistic logic and has the capacity to explore alternate paths towards convergence with less tendency of getting stuck to local minima.

In a conclusion, we have demonstrated that our computational architecture is probabilistic in nature where the information processing distributes between electronic and chemical domains and the implementation of simple algorithms towards useful computation. The computational architectural design is quite generic and there is a large scope of implementing more efficient algorithms over a diverse range of mathematical problems.

## 7.3 Towards fully Chemical Computation Logic

Based on our current computational approach, using hybrid chemical electronic logic, we can further increase the information processing in the chemical domain and utilize the full potential of massively parallel interactions. This is possible by mapping the energy/cost minimization problem completely into the physical behaviour of the chemical system governed by time evolution. There exists a variety of physicochemical systems capable of spontaneously minimizing energy but it becomes extremely complex to program the system such that mathematical problems can be mapped precisely and solution can be readout. This is a big challenge that exists in molecular computation frameworks. Our computational architecture bypasses this limitation and is capable of utilizing precise digital control for I/O and massively parallel information processing in the chemical domain.

As an example, by increasing the accuracy of the couplings between the cells and finding the input controls to create perfectly symmetric chemical states, the combinatorial optimization problem can be directly mapped to the experimental platform without defining a state machine in the digital domain. Fig. S68 A shows a fully coupled graph with four variables which could be equivalent to 4-number partitioning problem, where each edge defines the couplings coefficients. The direct mapping of this graph to the interconnected chemical oscillator is shown in Fig. S68 B. The four-variables are mapped directly to the chemical states of the active cells. We define four cells to introduce self-interactions and four additional auxiliary cells to complete the connectivity between the diagonally placed spins (x2, x3) and (x1, x4). The cell-cell interface can be activated with the weights proportional to the coupling coefficients. At each step, all neighbouring interactions occur and based on purely chemical decision making, new chemical states should emerge. This emergence of these chemical states can be directly enhanced using digital control, such as PWM states of cell stirrers. By utilizing the natural tendency of the physicochemical system and precise control of couplings using digital control, highly efficient computing hardware can be developed. In this way, complete chemical processing occurs in the chemical domain and digital electronics state machine act as an amplifier to sustain chemical states and precise control of I/O. A viable solution towards a fully chemical computational platform is to switch to electrochemical framework coupled with chemical oscillators, where electrochemical potential or current could be used to precisely define the coupling strength.

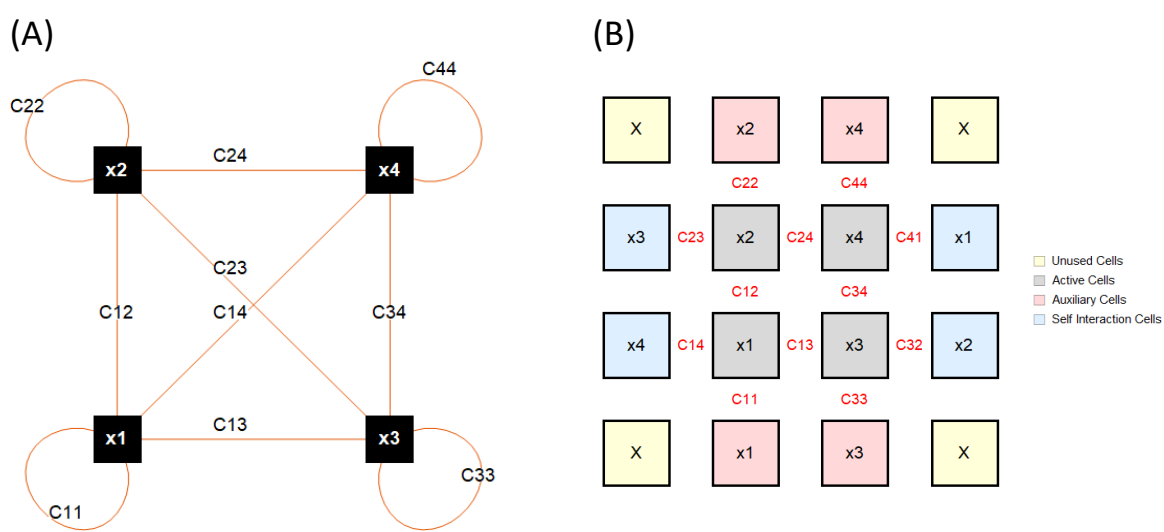

**Fig. S68:** Implementation of fully chemical computation logic. (A) Fully connected graph with 4-variables with edges defining the coupling coefficients for self and pairwise coupling interactions. (B) shows mapping to a two-dimensional computational architecture where couplings between neighbouring cells can be precisely defined and weighted by the coupling coefficients from the fully-connected Hamiltonian.

Due to mapping of fully connected Hamiltonian on the two-dimensional computational architecture, with the increase in the number of variables and coupling coefficients mapping a large-scale combinatorial optimization problem could be inefficient due to requirement of a large number of auxiliary cells to satisfy all the pairwise interactions such as number partitioning problem on large set or Travelling Salesman Problem with a large number of cities. One simpler solution is to use a network of denser grids such as hexagonal, where each cell can interact up to six neighbours as compared to four neighbour interaction for a square grid. This could lead to more efficient mapping of strongly coupled Hamiltonian and reducing the number of auxiliary cells. Here, we propose an alternative approach to couple cells beyond the nearest neighbours inspired from the idea of implementing high dimensional quasispaces *(36)* in lower-dimensional geometries. The idea is to miniaturize the computational platform such that the coupling between cells can be instantiated by evolving travelling waves of the reaction-diffusion system. In this case, the coupling between non nearest neighbouring cells can be achieved by programming the path lengths of the interface connecting the cells. By programming the pathlengths in the interconnected network of cells, waves propagating from one cell can reach nearest and next-nearest neighbours at the same time if the path lengths joining them are same. Fig. S69 A shows our current strategy to map Ising spins on a square grid and use auxiliary cells to instantiate interactions between diagonally placed cells. However, based on the proposed strategy, a fully connected four-cell network is achievable without using auxiliary cells as shown in Fig. S69 (B, C). Here, path lengths between S1-S2, S1-S3 and S1-S4 have been increased by creating meander like structures such that all the connections have same path length. In Fig. S69 B, assuming an equilateral triangle (S2-S3-S4) with length L, the path lengths between S1-S2, S1-S2 and S1-S4 should be increased to L.

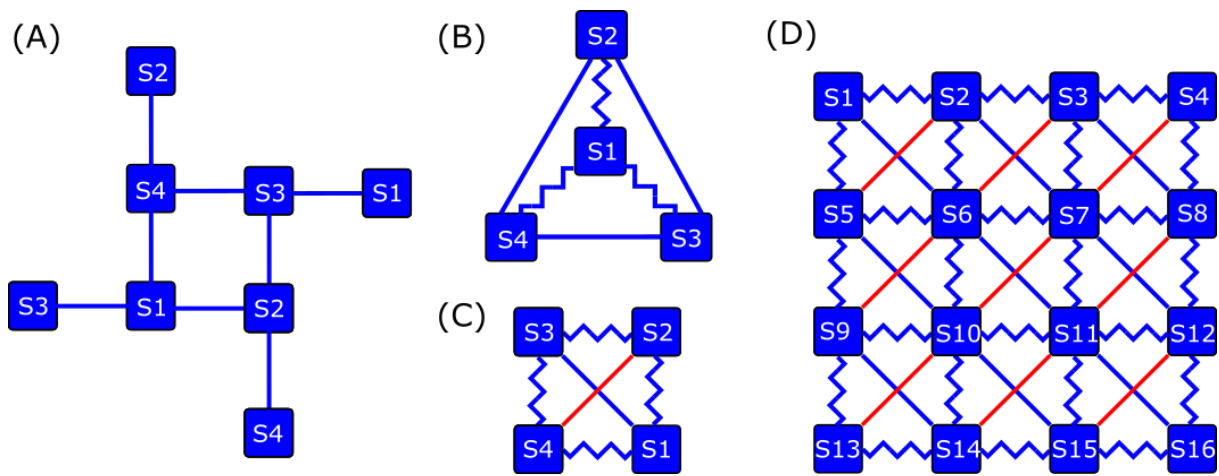

**Fig. S69**: Efficient implementation of non-nearest neighbouring couplings. (A) shows the current approach to map 4 variable full coupled graph on our computational architecture. (B and C) shows efficient mapping by programming path lengths between the cells avoiding the need for auxiliary cells using single layer and multilayer geometries. (C) shows an example of efficient mapping on large scale networks with multi-layered structure and programmed pathlengths for up to next-nearest neighbour couplings.

These mappings can be made more efficient and compact by using multilayer structure as shown in Fig. S69 C, where path lengths between S1-S2, S2-S3, S3-S4 and S4-S1 can be increases to $\sqrt{2}L$ to match the path lengths between diagonally placed elements. Fig. S69 D shows a large scale implementation of the strategy shown in Fig. S69 C to couple next-nearest neighbours. Based on the current approach towards chemical computation together with efficient mappings of non-nearest neighbour couplings, a very efficient computation architecture based on Ising models can be created towards highly efficient computation.

# 8 References

31. O. Steinbock, *et al.*, Oxygen Inhibition of Oscillations in the Belousov-Zhabotinsky Reaction. *J. Phys. Chem. A.* **104**, 6441-6415 (2000)

32. J. M. Parrilla-Gutierrez, *et al.*, A programmable chemical computer with memory and pattern recognition. *Nat. Commun.* **11**, 1442 (2020).

33. P. L. Gentili, J.-C. Micheau, Light and chemical oscillations: Review and perspectives. *J. Photochem. Photobiol. C Photochem. Rev.* **43**, 100321 (2020).

34. A. Lucas, Ising formulations of many NP problems. *Front. Phys.* **2**, 1–14 (2014).

35. S. Y. Guo, *et al.* A molecular computing approach to solving optimization problems via programmable microdroplet arrays. *Matter.* **4**, 1107–1124 (2021).

36. J. S. McCaskill, From quasispecies to quasispaces: coding and cooperation in chemical and electronic systems. *Eur. Biophys. J.* **47**, 459–478 (2018).